\documentclass[lettersize,journal]{IEEEtran}
\usepackage{amsmath,amsfonts}
\usepackage{algorithmic}
\usepackage{algorithm}
\usepackage{array}
\usepackage[caption=false,font=normalsize,labelfont=sf,textfont=sf]{subfig}
\usepackage{textcomp}
\usepackage{stfloats}
\usepackage{url}
\usepackage{verbatim}
\usepackage{graphicx}
\usepackage{cite}
\usepackage[hidelinks]{hyperref}
\usepackage{cleveref}

\usepackage{multicol}
\usepackage{multirow}
\usepackage{color, colortbl}
\usepackage[svgnames]{xcolor}

\usepackage{graphicx}
\usepackage{tabu}                      %
\usepackage{booktabs}                  %

\usepackage{mathptmx}                  %
\usepackage{subfig}
\usepackage{comment}
\usepackage{bm}
\usepackage{setspace}
\usepackage{soul}
\usepackage{enumitem}

\definecolor{NavyBlue}{RGB}{0,0,128}

\renewcommand{\paragraph}[1]{\vspace{4pt}\noindent\textbf{#1:}}
\Crefformat{figure}{#2Fig.~#1#3}
\Crefmultiformat{figure}{Figs.~#2#1#3}{ and~#2#1#3}{, #2#1#3}{ and~#2#1#3}
\Crefformat{section}{#2Sect.~#1#3}
\Crefmultiformat{section}{Sect.~#2#1#3}{ and~#2#1#3}{, #2#1#3}{ and~#2#1#3}

\newcommand{\lnorm}        {\ell}

\definecolor{green}{HTML}{1b9e77}
\definecolor{yellow}{HTML}{e6ab02}
\definecolor{pink}{HTML}{e7298a}
\definecolor{grey}{HTML}{808080}
\definecolor{darkgrey}{HTML}{666666}

\begin{document}

\title{Visual Stenography: Feature Recreation and Preservation in Sketches of Noisy Line Charts}

\author{Rifat Ara Proma, Michael Correll, Ghulam Jilani Quadri, Paul Rosen

\thanks{Rifat Ara Proma is with the University of Utah. E-mail: rifat.proma@utah.edu.}
\thanks{Michael Correll is with Northeastern University. E-mail: m.correll@northeastern.edu.}
\thanks{Ghulam Jilani Quadri is with the University of Oklahoma. E-mail: quadri@ou.edu.}
\thanks{Paul Rosen is with the University of Utah. E-mail: prosen@sci.utah.edu.}}

\markboth{Journal of \LaTeX\ Class Files,~Vol.~14, No.~8, August~2021}%
{Proma \MakeLowercase{\textit{et al.}}: Visual Stenography}

\maketitle

\begin{figure*}[!ht]
    \captionsetup[subfloat]{labelfont=scriptsize,textfont=scriptsize}
    \centering
    \begin{minipage}[t]{0.35\linewidth}
    \vfill
    \vspace{-2.4em}
        \includegraphics[width=0.875\textwidth, angle=90]{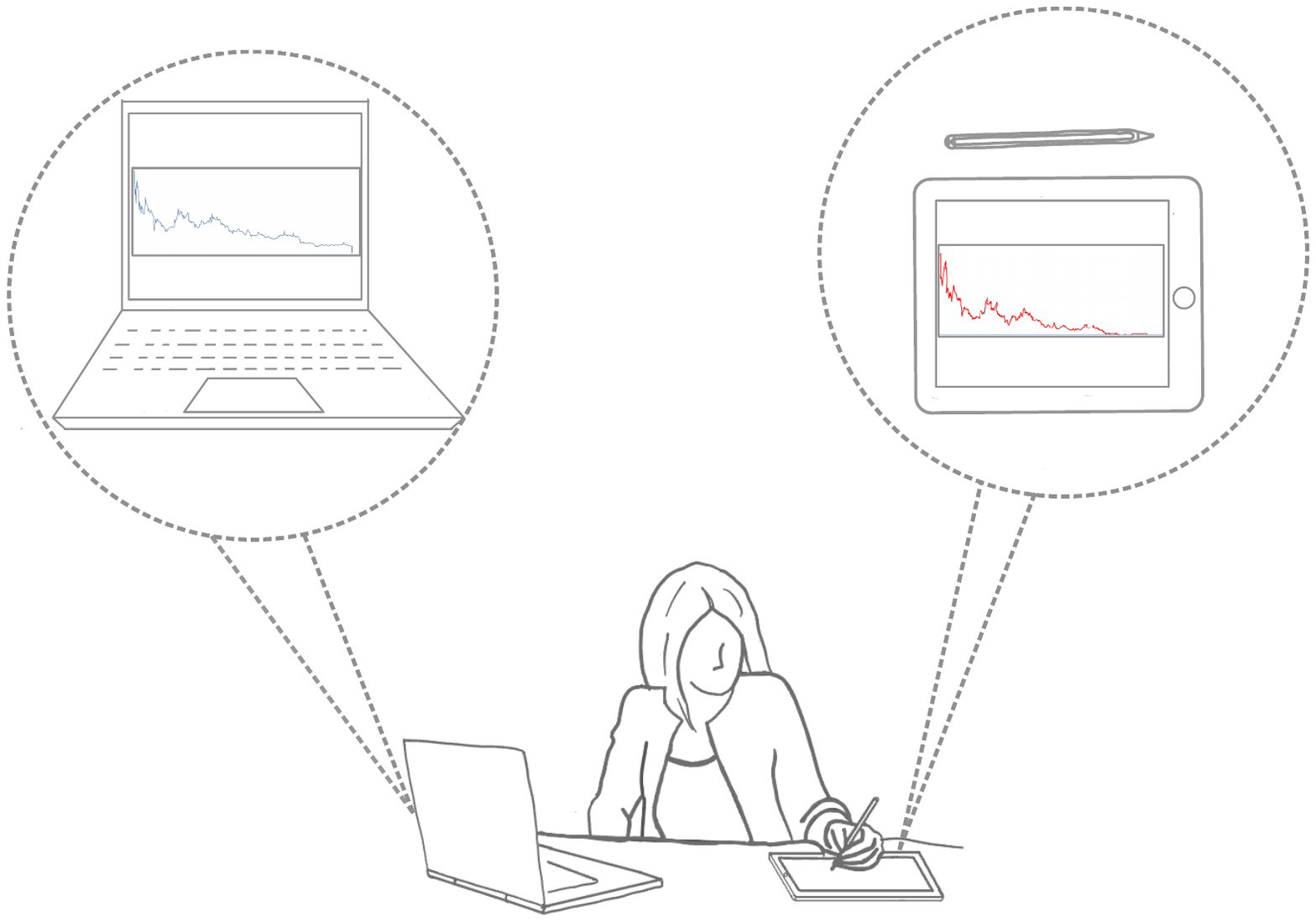}
       
    \end{minipage}
    \hfill %
    \begin{minipage}[t]{0.6\linewidth} %
    
        \setlength{\fboxrule}{2pt}
        \begin{minipage}{\linewidth}
            \hspace{-10pt}
            \begin{minipage}[t]{0.28\linewidth}
                \fcolorbox{green}{white}{
                    \subfloat[\textcolor{green}{\textbf{Replicator}}\label{fig:replicator_teaser}]{\includegraphics[width=\linewidth]{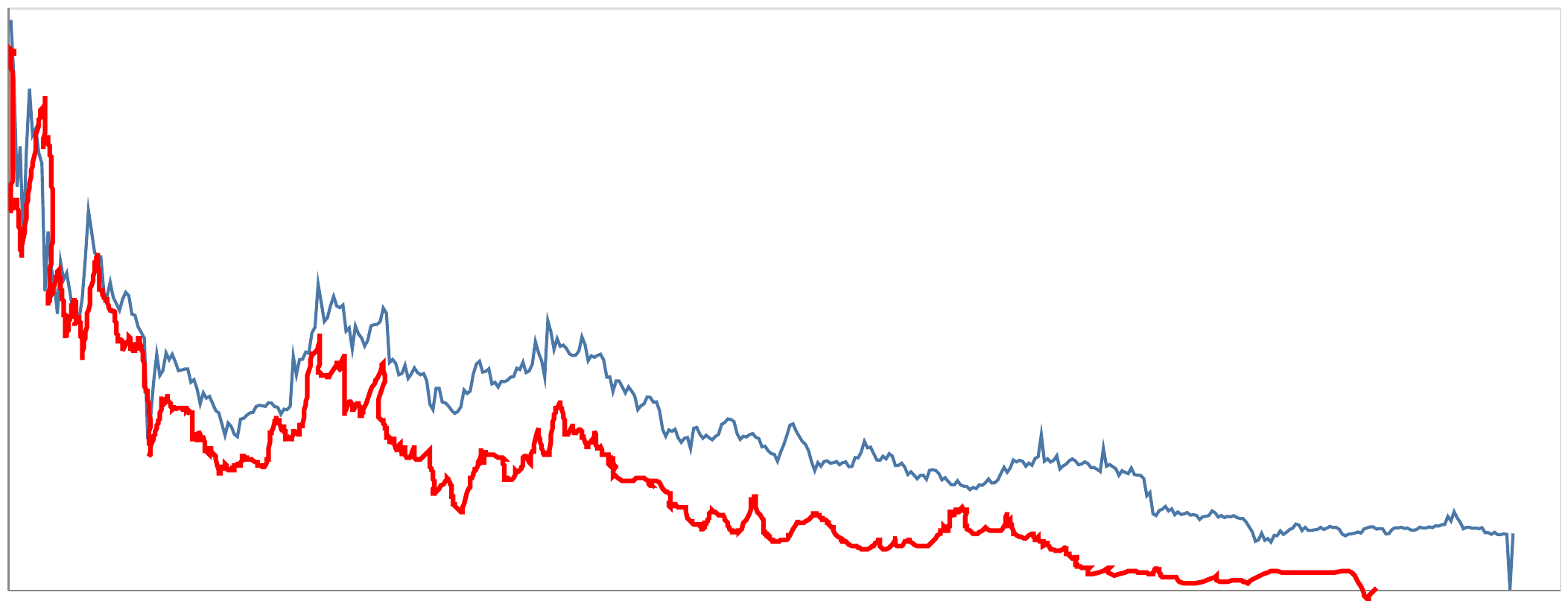}}
                    \hspace{0pt}
                }
                
            \end{minipage}
            \hspace{12pt}
            \begin{minipage}[t]{0.28\linewidth}
                \fcolorbox{yellow}{white}{
                    \subfloat[\textcolor{yellow}{\textbf{Trend Keeper}}\label{fig:trend_keeper_teaser}]{\includegraphics[width=\linewidth]{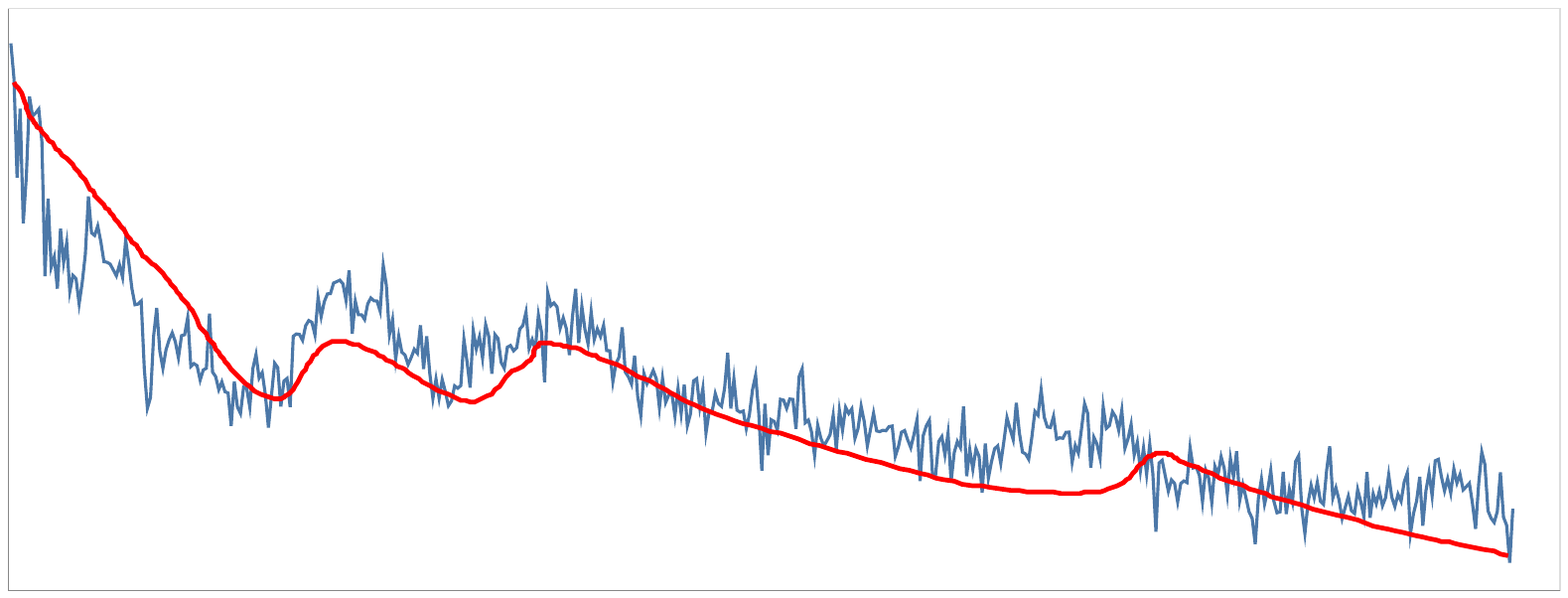}}
                    \hspace{0pt}
                }
               
            \end{minipage}
            \hspace{12pt}
            \begin{minipage}[t]{0.28\linewidth}
                \fcolorbox{pink}{white}{
                    \subfloat[\textcolor{pink}{\textbf{De-noiser}}\label{fig:de-noiser_teaser}]{\includegraphics[width=\linewidth]{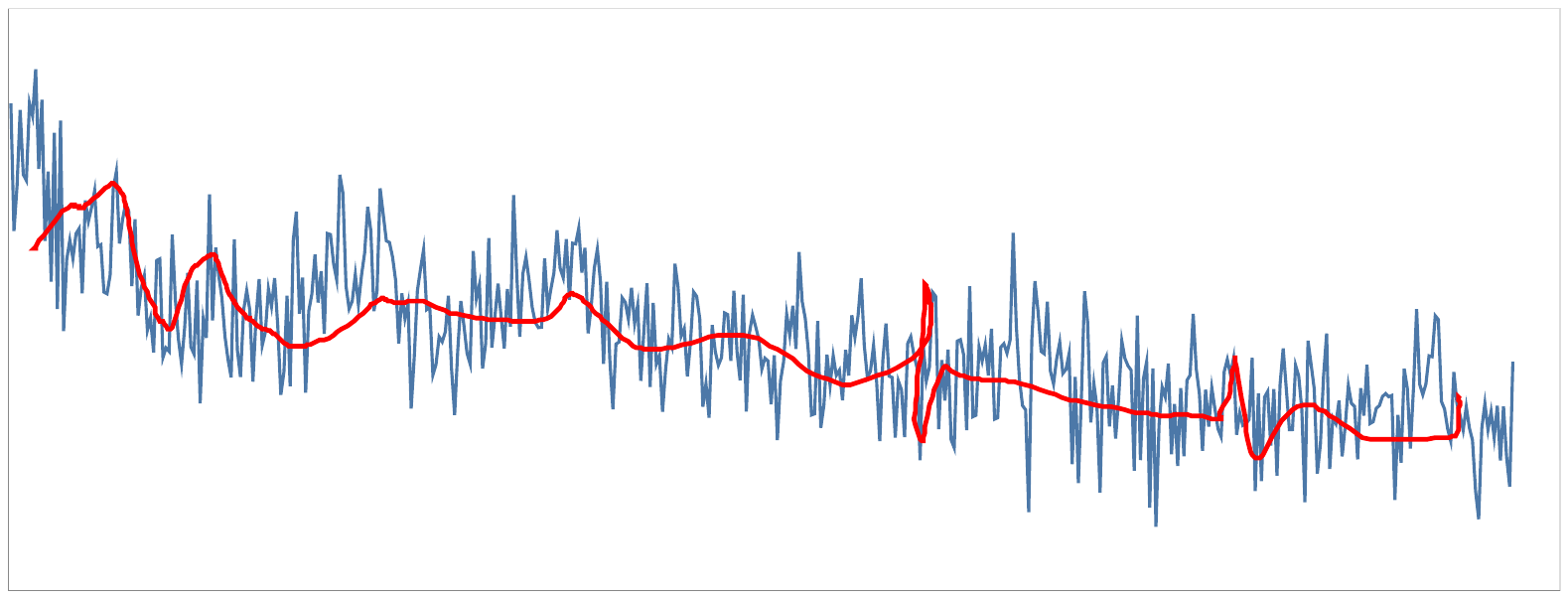}}
                    \hspace{0pt}
                }
               
            \end{minipage}
            
            \vspace{-3pt}
            \begin{center}
                \scriptsize
                Patterns of Behaviors
            \end{center}
        \end{minipage}
    
        \vspace{7pt} 
    
        \begin{minipage}{\linewidth}
            \begin{minipage}[t]{0.28\linewidth}
    
                    \subfloat[\texttt{Astronomy}\label{fig:astro_max_teaser}]{\includegraphics[width=\linewidth]{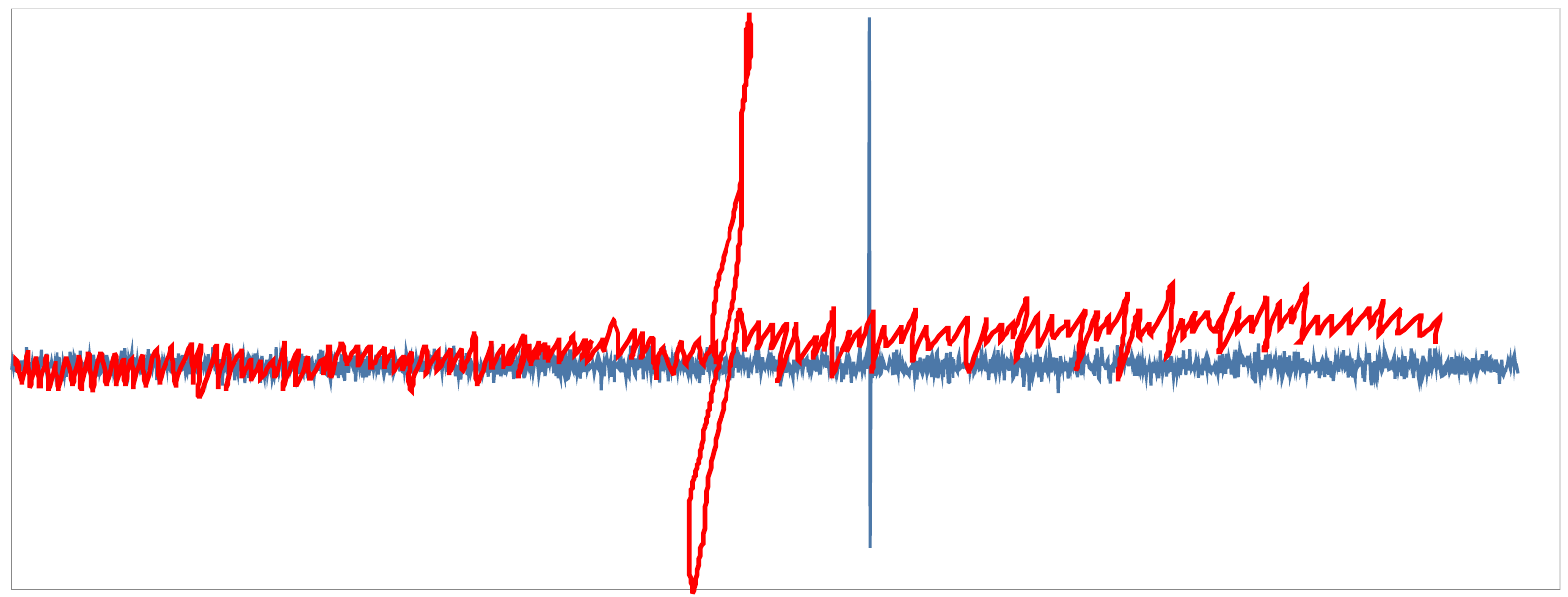}}
    
            \end{minipage}
            \hfill
            \begin{minipage}[t]{0.28\linewidth}
    
                    \subfloat[\texttt{EEG}\label{fig:eeg_max_teaser}]{\includegraphics[width=\linewidth]{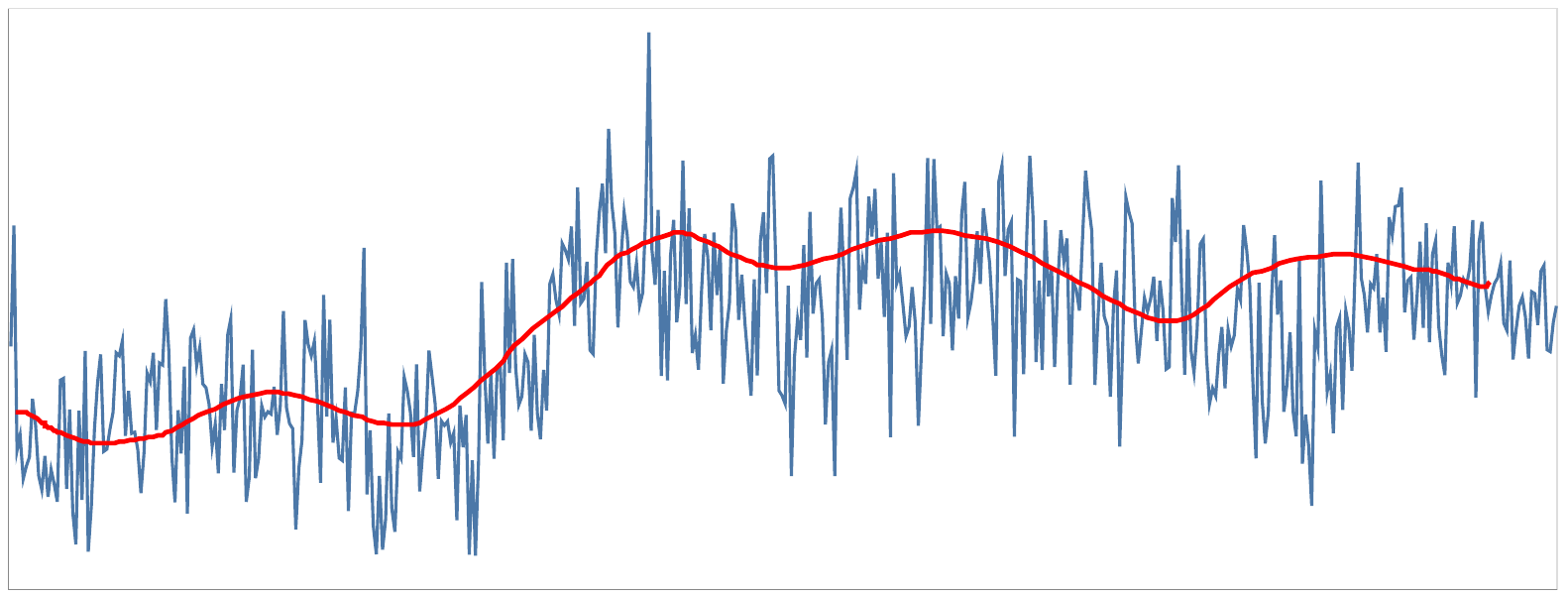}}
         
            \end{minipage}
            \hfill
            \begin{minipage}[t]{0.28\linewidth}
    
                    \subfloat[\texttt{Unemployment}\label{fig:unemployment_max_teaser}]{\includegraphics[width=\linewidth]{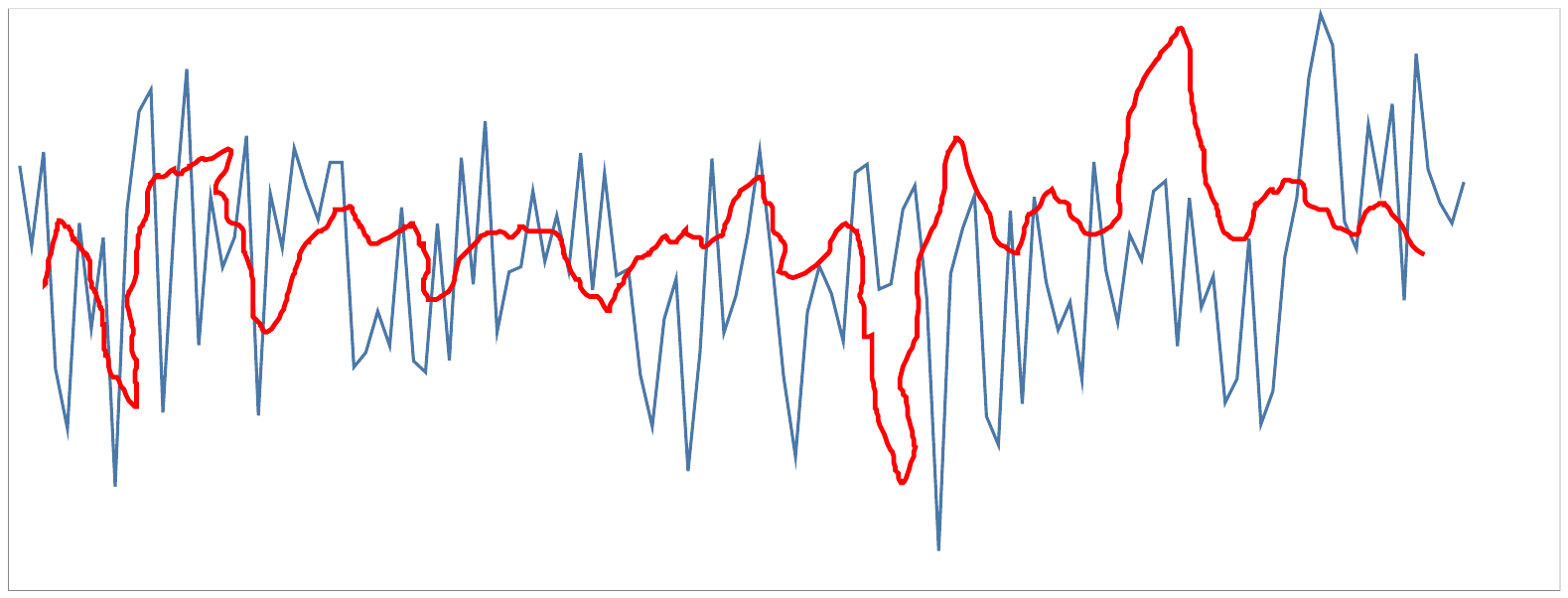}}
    
            \end{minipage}
            \hspace{5pt}
            
            \vspace{-5pt}
            \begin{center}
                \scriptsize
                Robustness to Noise
            \end{center}
        \end{minipage}

        \vspace{7pt}

        \begin{minipage}{\linewidth}
            \begin{minipage}[t]{0.28\linewidth}
    
                     \subfloat[\texttt{Chicago}\label{fig:chi_teaser}]{\includegraphics[width=\linewidth]{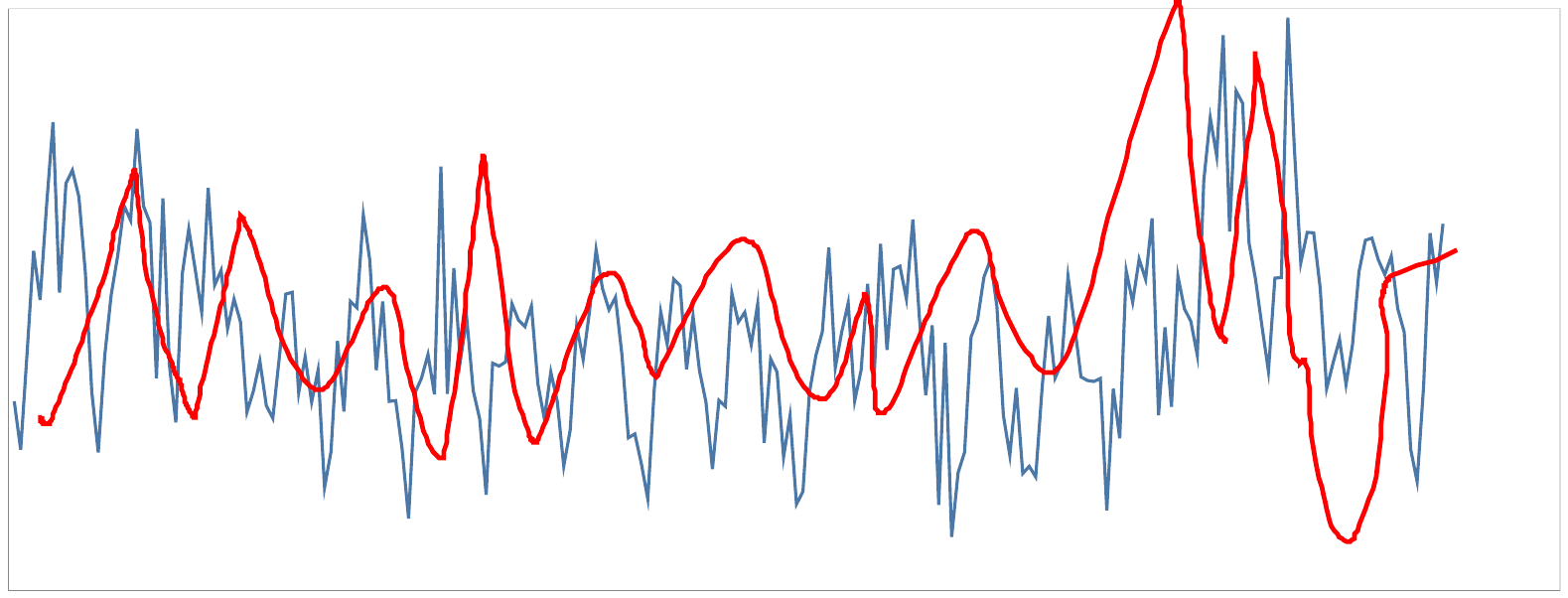}}
    
            \end{minipage}
            \hfill
            \begin{minipage}[t]{0.28\linewidth}
    
                   \subfloat[\texttt{Temperature}\label{fig:climate_teaser}]{\includegraphics[width=\linewidth]{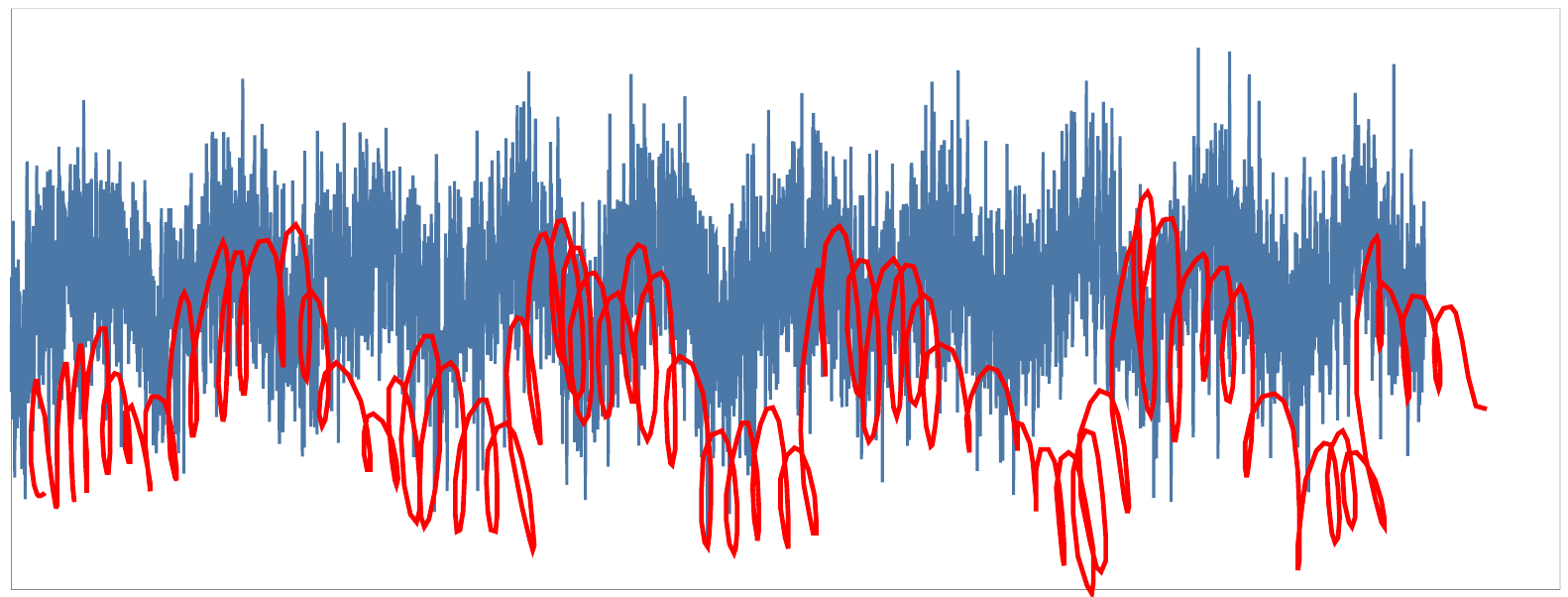}}
         
            \end{minipage}
            \hfill
            \begin{minipage}[t]{0.28\linewidth}
    
                    \subfloat[\texttt{Unemployment}\label{fig:unemployment_teaser}]{\includegraphics[width=\linewidth]{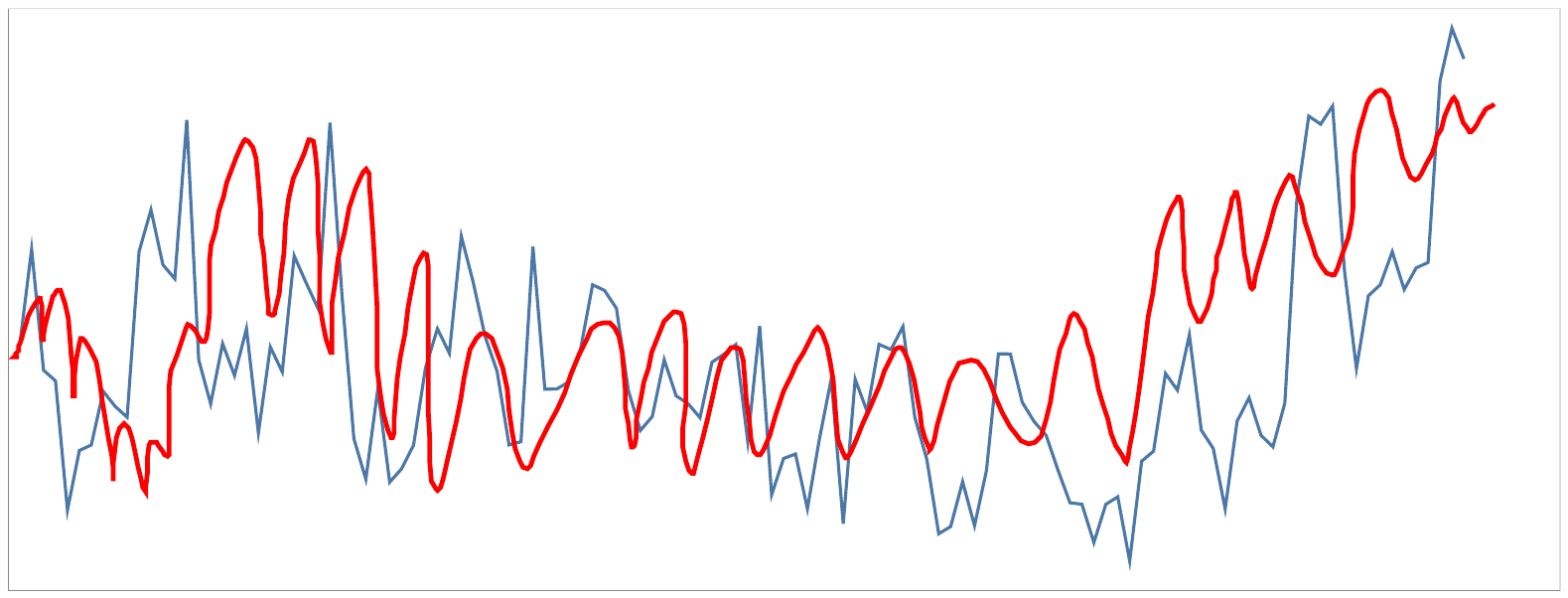}}
    
            \end{minipage}
            \hspace{5pt}
            
            \vspace{-5pt}
            \begin{center}
                \scriptsize
                Representational Semantics
            \end{center}
        \end{minipage}
    
    \end{minipage}

    \caption{The illustration on the left showcases the process of visual stenography, where the participants were shown line charts and were tasked to re-draw them on an iPad. The stimuli are in blue, and the participant sketches are in red. (a-c)~We compared the sketches with the shown stimuli and identified three behavior pattern clusters---\textcolor{green}{Replicator}, \textcolor{yellow}{Trend Keeper}, and \textcolor{pink}{De-noiser}. (d-f)~Further, participants showed general robustness to noise in terms of preserving trends, periodicity, peaks, and valleys in their sketches across different datasets, and (g-i)~they tended to represent the periodicity and noisiness of the stimuli semantically (i.e., conceptually) rather than faithful (i.e., accurately) replication in their sketches.}
    \label{fig:teaser}
\end{figure*}

\begin{abstract}
Line charts surface many features in time series data, from trends to periodicity to peaks and valleys. However, not every potentially important feature in the \textit{data} may correspond to a \textit{visual feature} which readers can detect or prioritize. In this study, we conducted a \textit{visual stenography} task, where participants re-drew line charts to solicit information about the visual features they believed to be important. We systematically varied noise levels (SNR $\approx$ 5–30 dB) across line charts to observe how visual clutter influences which features people prioritize in their sketches. We identified three key strategies that correlated with the noise present in the stimuli: the \textcolor{green}{\textit{Replicator}} attempted to retain all major features of the line chart including noise; the \textcolor{yellow}{\textit{Trend Keeper}} prioritized trends disregarding periodicity and peaks; and the \textcolor{pink}{\textit{De-noiser}} filtered out noise while preserving other features. Further, we found that participants tended to \textit{faithfully} retain trends and peaks and valleys when these features were present, while periodicity and noise were represented in more qualitative or gestural ways: \textit{semantically} rather than accurately. These results suggest a need to consider more flexible and human-centric ways of presenting, summarizing, pre-processing, or clustering time series data.
\end{abstract}


\section{Introduction}

\IEEEPARstart{L}{ine} charts are widely used for visualizing time-series data, enabling viewers to observe key patterns such as trends (e.g., in stock prices~\cite{mussweiler2003goes}), periodicity or cyclical patterns (e.g., in environmental data analysis~\cite{assfalg2009periodic}), or the peaks and valleys (e.g., in COVID-19 pandemic data~\cite{stein2023effect}). However, there can be mismatches between the \textit{statistical} patterns in time series data and the \textit{visual} patterns in line charts. While statistical patterns may be present, how viewers interpret and internalize them remains unclear.

Prior work has explored how viewers match statistical patterns to visual features in charts, examining tasks like mean estimation~\cite{correll2012comparing}, trend detection~\cite{correll2017regression}, and similarity evaluation~\cite{batista2014cid}. Research also shows contextual cues like captions~\cite{kim2021towards} and anchoring effects~\cite{xiong2019curse} influence interpretation. However, little attention has been given to which features (e.g., trends, periodic patterns, or peaks and valleys) naturally draw viewers' attention, especially under varying noise levels, which can impact cognitive load~\cite{vredeveldt2011eyeclosure} and judgement~\cite{hu2023application}. Knowing how people attend to these features is critical to designing visualizations that reflect human understanding of data. 

While many prior works measure people's accuracy for graphical-perception tasks, our goal is instead to reveal how people choose which visual features to preserve when sketching noisy line charts and to characterize the distinct strategies that emerge. To address this, we conducted a human subjects evaluation of 20 participants to identify how individuals represent trends, periodic patterns, and peaks and valleys in line charts given a free-form sketching medium, a valuable method for capturing how individuals perceive visual features~\cite{lee2013sketchstory, xia2018dataink} as it directly captures how users naturally structure and prioritize visual features without imposing explicit constraints on their perceptual judgments~\cite{muthumanickam2016shape}. 
Each participant saw nine line charts and completed what we term a \textit{visual stenography} task--- re-drawing the time series data to capture the noteworthy visual features of the chart (see~\Cref{sec:study_design}). The nine datasets were selected to include a variety of trends, periodic patterns, and notable peaks and valleys~(see~\Cref{sec:study_design:datasets}). We performed both qualitative~(see~\Cref{sec:qualitative}) and quantitative (see~\Cref{sec:quantitative}) analyses of the resulting sketches to assess the retention of visual features under various levels of noise. We explicitly focused on noise as our independent variable because real-world line charts commonly include random fluctuations that often cause visual clutter. By systematically controlling signal-to-noise ratio (SNR), we reveal how increasing visual complexity influences viewers’ mental prioritization and representation of visual features.

We then conducted a validation experiment (\Cref{sec:follow_up_analysis}), where participants selected pre-generated sketches that best-represented reference charts. This allowed us to validate findings from our primary study, particularly given the subjectivity of sketching~\cite{cohen1997can, lee2019you}, and gain insights into their selection rationale.

Our analysis revealed three distinct patterns of behavior surrounding the inclusion and exclusion of feature: (1)~\textcolor{green}{\textit{Replicator}} (see \Cref{fig:replicator_teaser}), preserve all major features, including noise. (2)~\textcolor{yellow}{\textit{Trend Keeper}} (see \Cref{fig:trend_keeper_teaser}) focused primarily on retaining trends in the data while disregarding finer details such as periodic patterns and peaks (3)~\textcolor{pink}{\textit{De-noiser}} (see \Cref{fig:de-noiser_teaser}) filtered out noise while retaining other major features. Noise significantly influenced behavior, with participants more likely to adopt \textcolor{yellow}{Trend Keeper} strategies as noise increased. We also noted similar behavior in how accurately subjects represented features. While trends and peaks and valleys were  often represented \textit{faithfully} (accurately with respect to the stimuli observed), periodicity and noise were represented \textit{semantically}: in ways that qualitatively denote the presence of a feature but may not accurately reflect exact statistical properties of the feature like scale or location. Our results suggest the use of not only \textit{mathematical} but also \textit{human-centric} representations of time series data for goals like designing and annotating visualizations, selecting algorithms for pre-processing data (e.g., smoothing~\cite{rosen2020linesmooth} or down sampling~\cite{steinarsson2013downsampling}), and performing data-related tasks (e.g., visual querying~\cite{correll2016semantics} or summarization~\cite{tufte2006beautiful}).

\vspace{4pt}
\noindent
In summary, the key contributions of this work are:
\vspace{0pt}
\begin{enumerate}[noitemsep,itemsep=4pt]
    \item Introduction of visual stenography as a novel sketch‐based method for externalizing viewers’ subjective prioritization of chart features under noise.
    \item Identification of three behavior groups based on feature inclusion and exclusion patterns; 
    \item A study that verified the robustness of visual feature extraction of trends, periodicity, and peaks and valleys under various noise levels; and
    \item Insights into human-centric feature preservation, showing that trends and peaks are represented more faithfully, while periodicity and noise are represented semantically.
\end{enumerate}

\section{Prior Work}

In this section, we discuss prior work in graphical perception and the utility of sketching. In \Cref{sec:background}, we provide background on visual features in line charts.

\subsection{Graphical Perception}

An extensive body of work on perceptual and cognitive processes underlies human estimations in visualizations from seminal papers like Cleveland \& McGill~\cite{cleveland1984graphical}. More recent work has examined not just extraction or comparison of individual values but graphical perception for more holistic statistical tasks~\cite{franconeri2021science,szafir2016four}. For instance, Gleicher et al.~\cite{gleicher2013perception} found that viewers can reliably estimate average values in scatterplots, even with noise. Other studies confirm that viewers can identify scatterplot trends, even in the presence of noise and outliers~\cite{harvey1997effects,ciccione2021can, hong2021weighted, ciccione2022analyzing}. Furthermore, investigations into the reliability of assumptions individuals rely on when estimating trends from various plots have shed light on viewers' capabilities to accurately assess trends in line charts and other bivariate visualizations~\cite{correll2017regression}. However, noise's impact on visual estimation differs from its effect on statistical analysis, suggesting that visual estimations may not align with statistical summaries~\cite{reimann2021visual}. The estimation of summary statistics from visualizations is influenced by the attributes of the visualization design and underlying data properties~\cite{kim2018assessing, ciccione2023outlier, gogolou2018comparing}. 

Most of this research focuses on scatterplots using synthetic data. In contrast, our study examines whether these findings hold when line charts depict real-world time series augmented with controlled levels of synthetic noise, focusing on how viewers preserve key features as visual clutter increases.

\subsection{Utility of Sketching}
Sketching is an unbounded medium for generating visual content~\cite{eitz2012humans}, allowing individuals to highlight areas of interest~\cite{correll2016semantics}. Therefore, various studies have adopted pen-and-touch interfaces for chart sketching~\cite{lee2013sketchstory, xia2018dataink, mannino2018expressive}, recognizing their adaptability and effectiveness in generating more expressive and customized visual representations. Unlike selection-based methods, free-form sketching avoids priming participants toward predefined visual features, capturing more authentic perceptual priorities~\cite{correll2016semantics,lee2013sketchstory,xia2018dataink}. Although explicitly querying participants about their intentions could clarify perceptual priorities, we intentionally avoided this to preserve the spontaneity and natural authenticity of their sketching behavior. Nonetheless, inherent subjectivity in sketching, especially among non-experts, remains a methodological challenge~\cite{eitz2012humans}.

While visual query systems leverage sketching to represent queries visually, these systems face challenges, including the inherent ambiguity in sketches~\cite{correll2016semantics} and the frequent inability or reluctance of users to depict their intentions accurately~\cite{lee2019you}. Prior work has shown how user input through free‑hand sketches~\cite{muthumanickam2016shape} or example‑based labeling~\cite{lekschas2020peax} can effectively tune a system to human perceptual priorities. Our work complements these efforts by systematically studying which visual features people naturally prioritize for preservation, followed by a validation evaluation where participants chose sketches and explained their reasons to reduce the subjectivity associated with sketching.  

Collectively, these approaches provide valuable insights into feature prioritization and contribute to the development of improved visual query systems.

\section{Background on Visual Features}
\label{sec:background}

Existing literature presents different ways of defining and identifying visual features such as trends, periodicity, peaks and valleys, and noise within time series data. While not exhaustive, these features align with the principles of scagnostics~\cite{wilkinson2005graph, friedman1974projection}. Adapting scagnostics for line charts allows these features to be interpreted as measurable properties.
Additionally, while scagnostics measures do not explicitly define noise, it remains a critical feature to explore due to its impact on perception~\cite{vredeveldt2011eyeclosure}. This section establishes a foundation for the upcoming analysis by reviewing established methods for estimating and quantifying these visual features and explaining their role in the current study.

\subsection{Defining Trend}

Trends are widely used to interpret time series data~\cite{wang2017line}, but their definition varies across fields. In mathematics and statistics, they are associated with monotonic patterns, event rates, and timing ~\cite{lawless2012testing}, while in climate science, they represent systematic variations in parameters like temperature and extreme weather events~\cite{alexander2006global, seregina2019new}.

In the visualization community, trends are generally seen as upward or downward movements over time~\cite{wang2009temporal, robertson2008effectiveness, fan2022annotating}, often depicted using trend lines or regression lines~\cite{ahmed2022semantics, rahman2017ve, filipowicz2023visual, correll2017regression, johnson1991trend}. However, real-world data often lacks clear trends~\cite{wang2017line}. In such scenarios, locally estimated scatterplot smoothing (LOESS) regression~\cite{cleveland1979robust} is useful for capturing the main patterns~\cite{wang2017line}. Likewise, in signal processing, the Fast Fourier Transform (FFT) is often used to analyze time series~\cite{popivanov2002similarity} and detect trends by removing higher frequencies~\cite{musbah2020novel, duda2018detection, denholm1998practical}, based on Duda et al.\cite{duda2018detection}'s definition of trend as the non-periodic component of a time series. Our analysis employed LOESS regression and FFT with a low pass filter to capture the non-linear trends in our real-world data.

\subsection{Detecting Periodicity}
A time series is considered periodic if values repeat at
(almost) regular intervals over time\cite{elfeky2005periodicity}. Identifying periodicity in time series for a number of tasks, including aiding interpretation~\cite{lazcano2022improved}, detecting anomalies~\cite{shehu2023efficient}, and forecasting~\cite{wen2021robustperiod}. 
Several approaches exist for detecting periodicity including FFTs~\cite{ musbah2019identifying, musbah2020novel, wen2023robust}, wavelet transformation~\cite{mohammed2019developing, stratimirovic2018analysis, otazu2002detection}, and dynamic time warping~\cite{elfeky2005warp, boulnemour2018qp, song2022robust}. In this study, we used FFTs to detect prominent periodic patterns by identifying the most prominent peak in the frequency domain as described by Musbah et al.~\cite{musbah2019identifying}.

\subsection{Detecting Peaks and Valleys}
Peaks are local maxima~\cite{suh2019topolines, rosen2020linesmooth} and valleys local minima~\cite{rensink2010perception} in time series data. Multiple studies explore methods for peak and valley detection, such as entropy-based methods~\cite{palshikar2009simple}, wavelet transformation~\cite{zhou2022improved, du2006improved}, and Topological Data Analysis (TDA)~\cite{suh2019topolines}. This paper adopts the TDA approach due to its robustness to noise and its effectiveness in preserving prominent peaks and valleys~\cite{rosen2020linesmooth}.

\subsection{Detecting Noise}

Time series data often contain random fluctuations that mask the underlying signal, referred to as noise~\cite{price1993signals}. Various techniques, such as filtering or smoothing~\cite{rosen2020linesmooth} and downsampling~\cite{steinarsson2013downsampling}, are employed to mitigate noise. Additionally, frequency domain low-pass filters, including the Butterworth~\cite{butterworth1930theory} and Chebyshev~\cite{rhodes1980generalized} filters, are effective in noise reduction. This study adopts the approach of employing an FFT with a high-pass filter to isolate noise components within the signal~\cite{musbah2019identifying, raj2007fft, karar1997apd}. Given that noise can complicate the interpretation of visualizations~\cite{parekh2018investigating}, this research also incorporates the use of Pixel Approximate Entropy (PAE)~\cite{2018entropy} as a metric to estimate the noise level in line charts.

\begin{figure}[!t]
    \centering

    \subfloat[{\scriptsize \texttt{Apple}} \label{fig:datasets:apple}]{
        \includegraphics[width=0.28\linewidth]{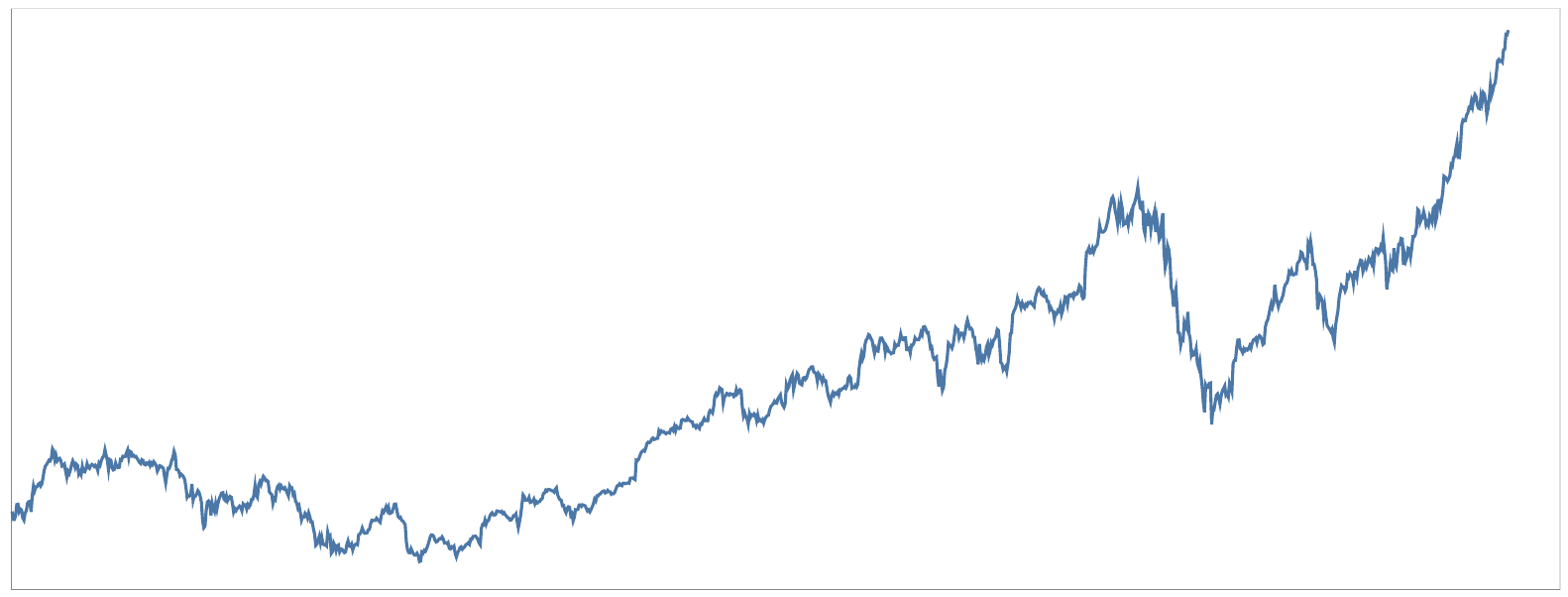}
    }
    \hspace{5pt}
    \subfloat[{\scriptsize\texttt{Astronomy}} \label{fig:datasets:astro}]{
        \includegraphics[width=0.28\linewidth]{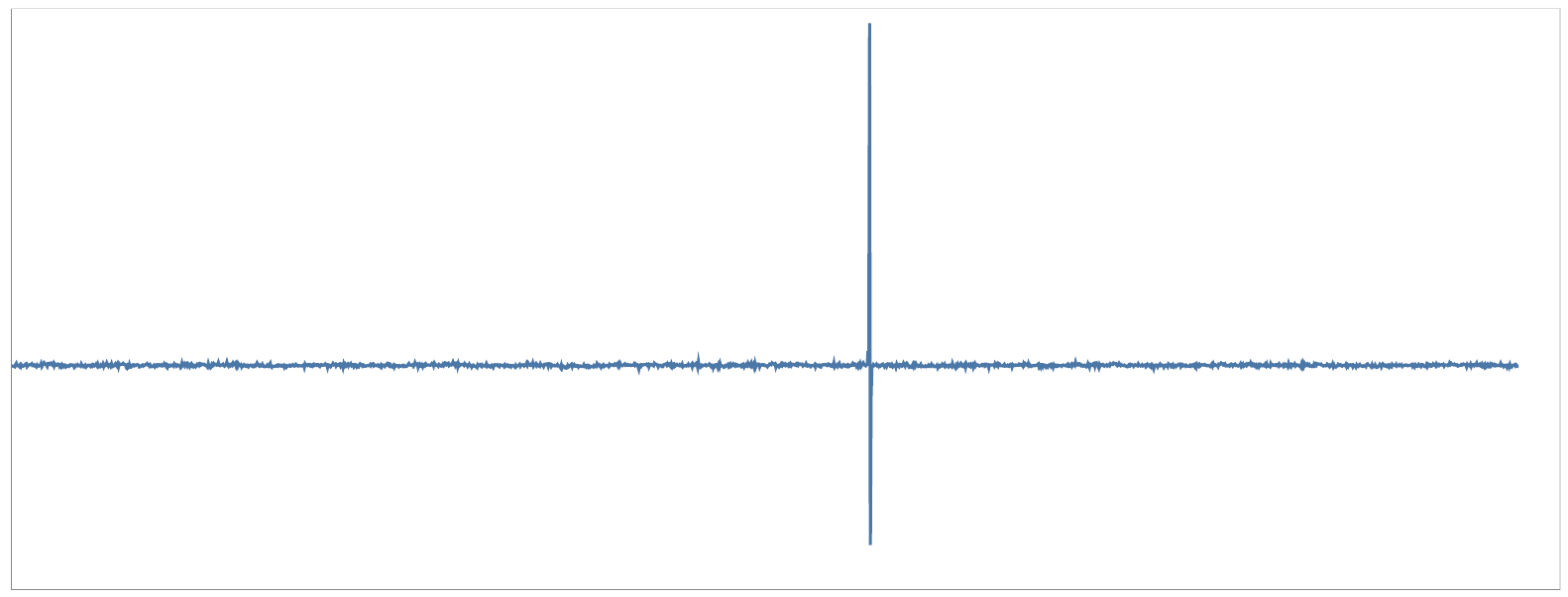}
    }
    \hspace{5pt}
    \subfloat[{\scriptsize \texttt{Chicago}} \label{fig:datasets:chi}]{
        \includegraphics[width=0.28\linewidth]{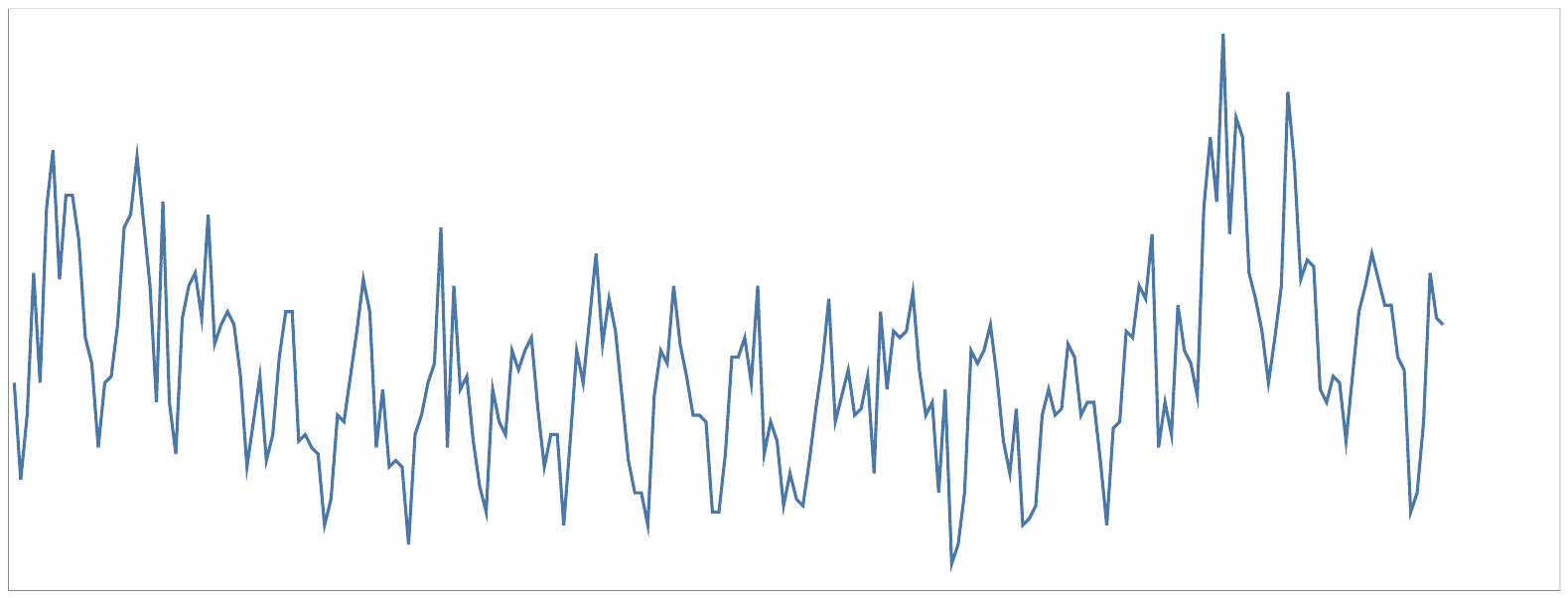}
    }

    \subfloat[{\scriptsize \texttt{Doge}} \label{fig:datasets:doge}]{
        \includegraphics[width=0.28\linewidth]{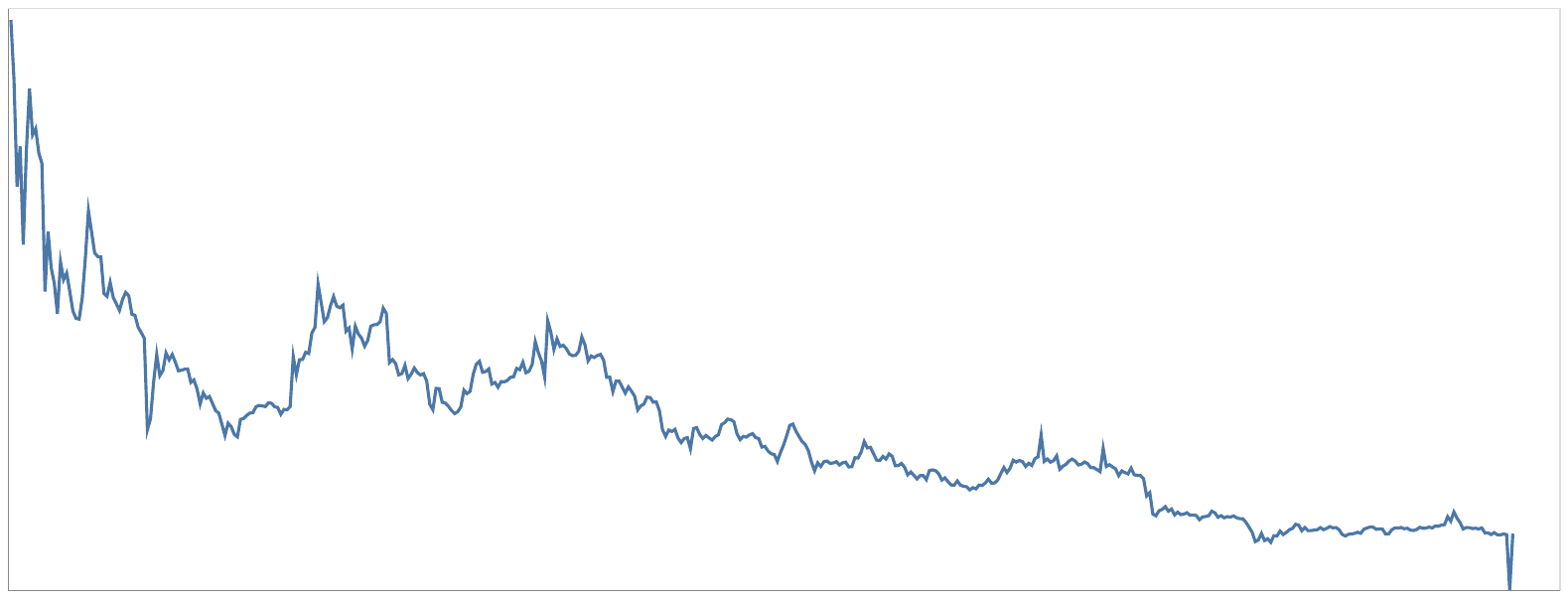}
    }
    \hspace{5pt}
    \subfloat[{\scriptsize \texttt{EEG}} \label{fig:datasets:eeg}]{
        \includegraphics[width=0.28\linewidth]{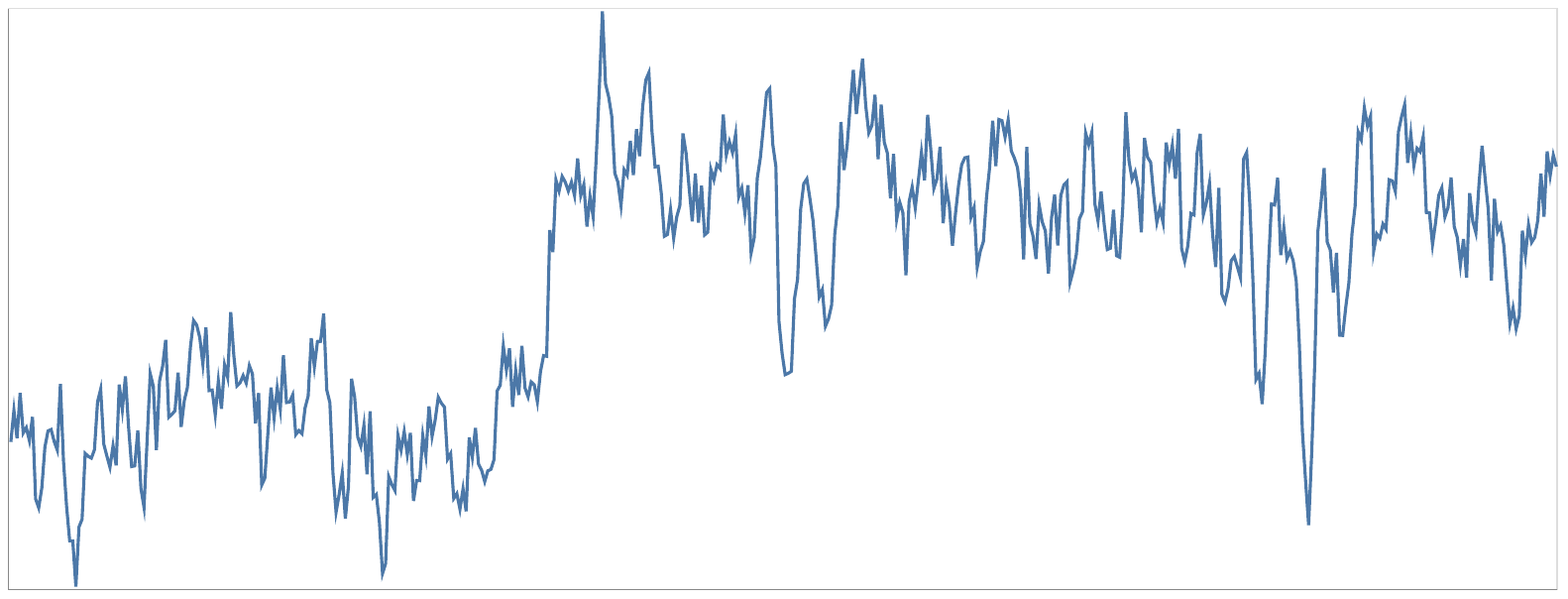}
    }
    \hspace{5pt}
    \subfloat[{\scriptsize \texttt{Flights}} \label{fig:datasets:flights}]{
        \includegraphics[width=0.28\linewidth]{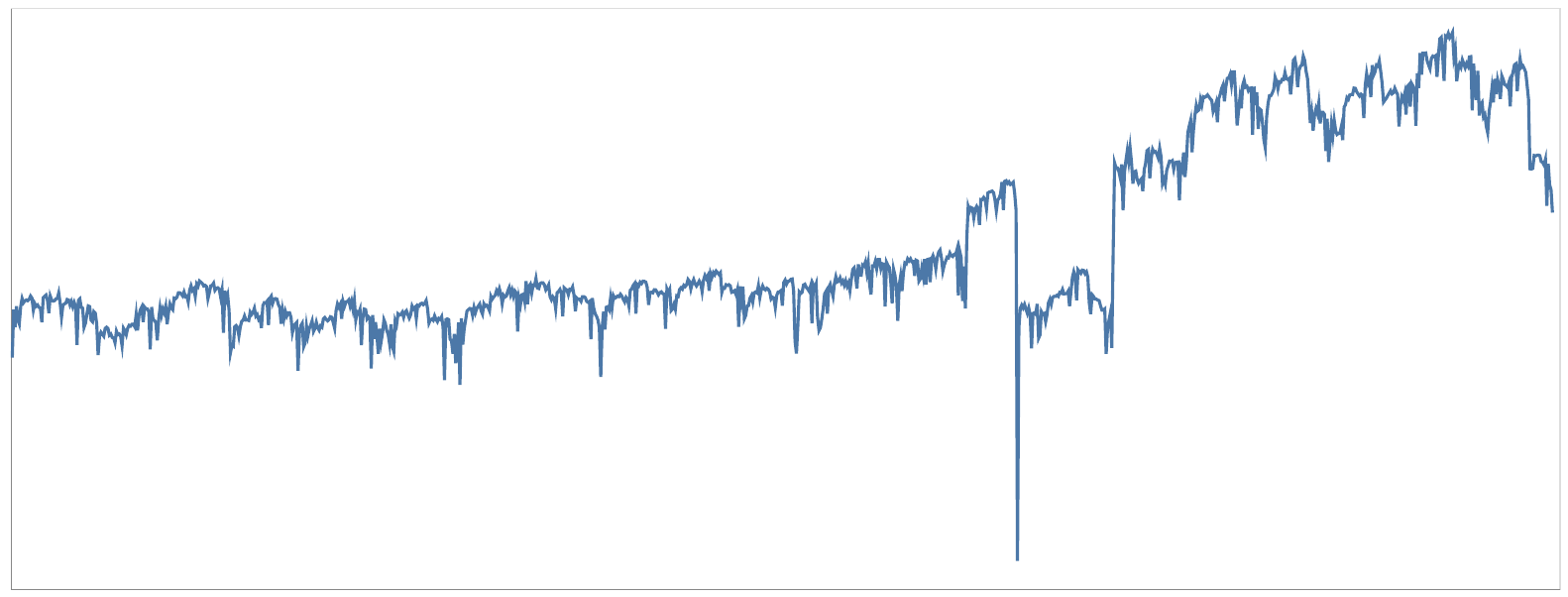}
    }

    \subfloat[{\scriptsize \texttt{Temperature} } \label{fig:datasets:cli}]{
        \includegraphics[width=0.28\linewidth]{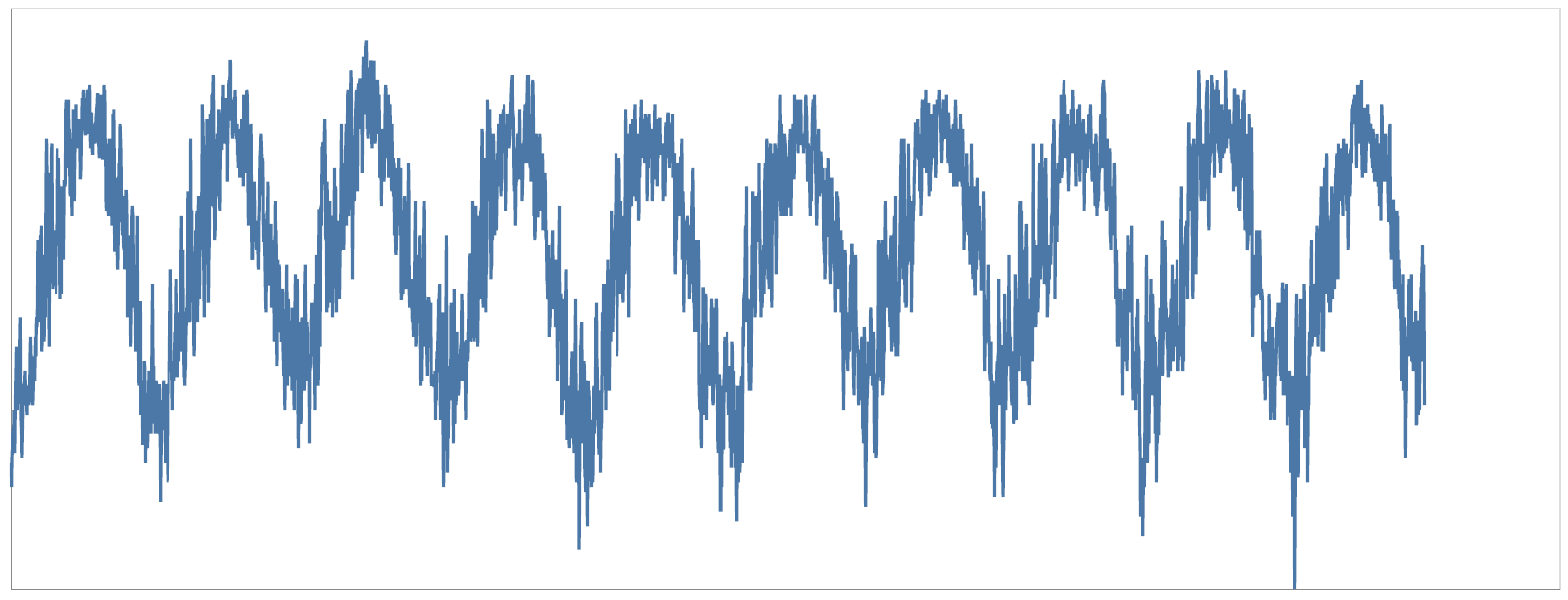}
    }
    \hspace{5pt}
    \subfloat[{\scriptsize \texttt{Tourists}} \label{fig:datasets:nz}]{
        \includegraphics[width=0.28\linewidth]{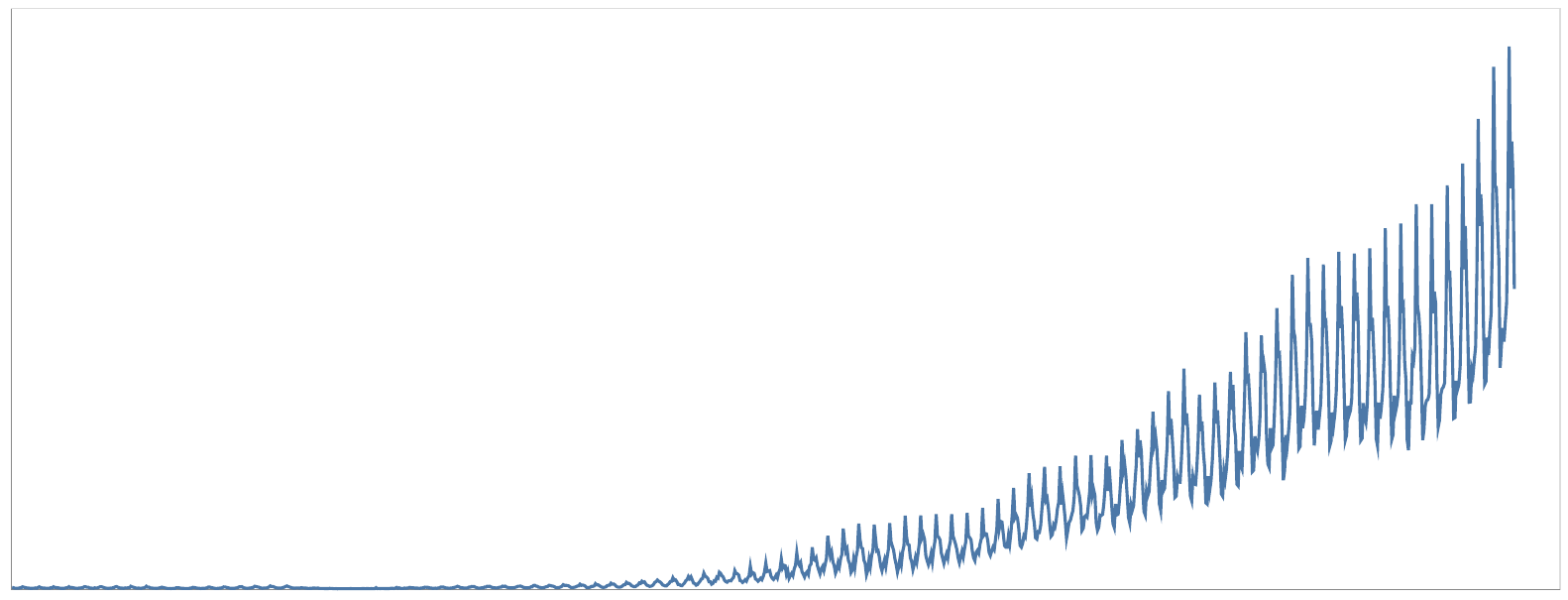}
    }
    \hspace{5pt}
    \subfloat[{\scriptsize \texttt{Unemployment}} \label{fig:datasets:unemployment}]{
        \includegraphics[width=0.28\linewidth]{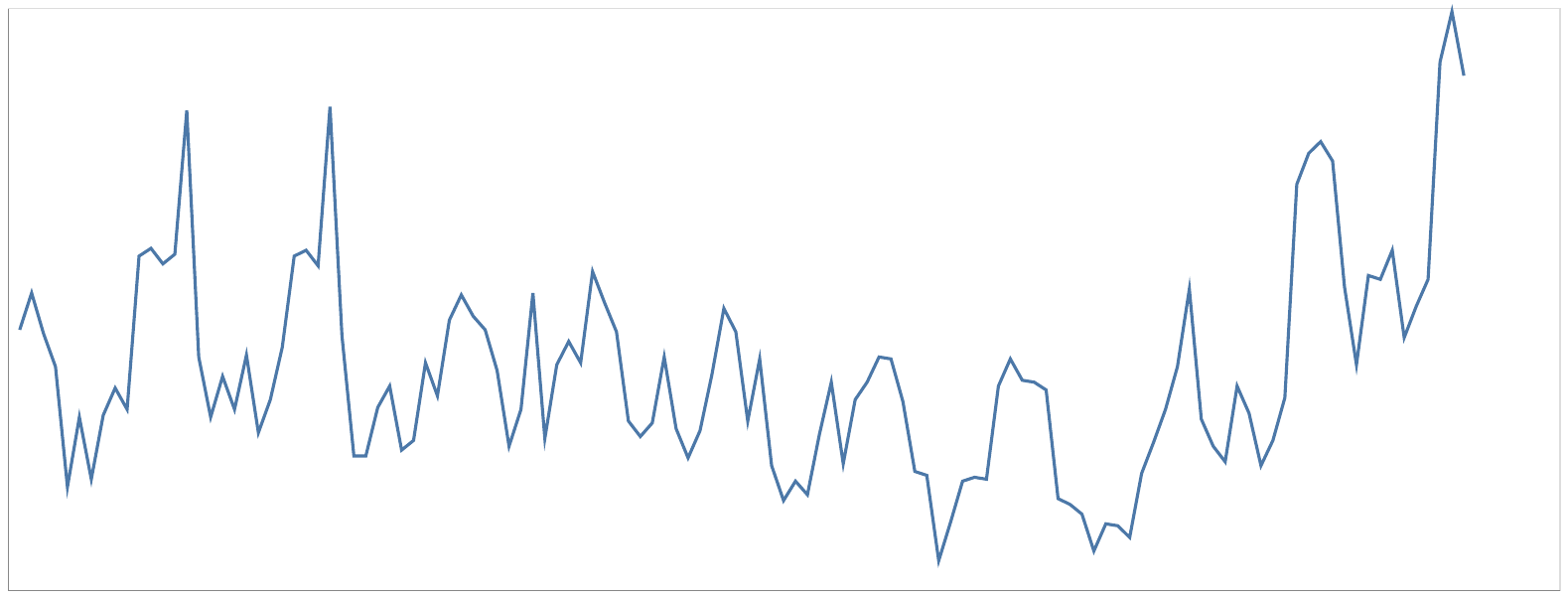}
    }

    \caption{Line charts for all nine datasets used in this study.}
    \label{fig:datasets}
\end{figure}

\section{Primary Study Design}
\label{sec:study_design}

Understanding how individuals attend to and prioritize visual features in noisy line charts offers valuable insights for designing compelling visualizations. To this end, we conducted a laboratory study with 20 participants to examine which features of line charts they would recreate under varying noise levels. Participants first completed a stenography task where they were asked to re-draw (see~\Cref{sec:study_design:exp}) line charts of 
nine real-world datasets (see~\Cref{sec:study_design:datasets}), each with one of five levels of Gaussian noise (see~\Cref{sec:study_design:noise}). We then assessed feature retention in their sketches compared to the stimuli, both quantitatively and qualitatively (see~\Cref{sec:analysis}). Participants were later invited back for a follow-up study to select preferred sketches and provide a verbal rationale for their choices (see~\Cref{sec:followup}).

\subsection{Datasets and Properties}
\label{sec:study_design:datasets}
We selected nine datasets (see \Cref{fig:datasets}), including:
Daily \texttt{Apple} Stock Closing Price \cite{nagadia_2022};
Radio \texttt{Astronomy} Signal \cite{alma};
Monthly Homicide Rate in \texttt{Chicago} \cite{chicago_crime};
Daily High \texttt{Temperature} at JFK Airport \cite{noaa};
Daily \texttt{Doge} Coin Closing Price \cite{dhruvil_dave_2021};
Single Channel of \texttt{EEG} Data \cite{eeg_database}; 
Weekly US Domestic \texttt{Flights} \cite{d3_zoomable};
Monthly \texttt{Tourists} in New Zealand \cite{new_zealand}; and
US Monthly \texttt{Unemployment} in Agriculture \cite{us_bls}. 
Each of the datasets demonstrated properties  (see \Cref{tab:dataset}) often encountered in real-world data visualization scenarios, including long-term stock data with overall directional trends (e.g., \Cref{fig:datasets:apple} and \Cref{fig:datasets:doge}), cyclical or periodic patterns seen in climate data (e.g., \Cref{fig:datasets:cli}), prominent peaks and valleys (e.g., \Cref{fig:datasets:astro} and \Cref{fig:datasets:flights}). Importantly, most datasets demonstrated multiple properties simultaneously.

\begin{table}[!b]
    \centering
    \caption{A description of the datasets used and the properties we observed, including overall trend (\rotatebox{45}{$\rightarrow$}up, \parbox{10pt}{\rotatebox{-45}{$\rightarrow$}}down, or \rotatebox{0}{$\rightarrow$} constant), periodic patterns, prominent peaks and/or valleys. 
    }
    \label{tab:dataset}
    \resizebox{0.975\linewidth}{!}{%
  \begin{tabular}{c@{\hspace{5pt}}p{6.1cm}@{\hspace{5pt}}c@{\hspace{5pt}}c@{\hspace{5pt}}c}
    \toprule
    \multirow{2}{*}{Dataset} & \multirow{2}{*}{Description}                            & Overall     & Periodic           & Peaks and/or   \\
                             &                                        &    Trend        & Pattern           & Valleys    \\
    \midrule
    \texttt{Apple}           & Daily Apple Stock Closing Price \cite{nagadia_2022}    & \rotatebox{45}{$\rightarrow$} &            & \checkmark \\
    \hline
    \texttt{Astronomy}       & Radio Astronomy Signal \cite{alma}                     & \rotatebox{0}{$\rightarrow$}  &            & \checkmark \\
    \hline
    \texttt{Chicago}         & Monthly Homicide Rate in Chicago \cite{chicago_crime}  & \rotatebox{0}{$\rightarrow$}  & \checkmark & \checkmark \\
    \hline
    \texttt{Temperature}     & Daily High Temp.\ at JFK Airport \cite{noaa}      & \rotatebox{0}{$\rightarrow$}  & \checkmark &            \\
    \hline
    \texttt{Doge}            & Daily Doge Coin Closing Price \cite{dhruvil_dave_2021} & \parbox{10pt}{\rotatebox{-45}{$\rightarrow$}} & & \checkmark \\
    \hline
    \texttt{EEG}             & Single Channel of EEG Data \cite{eeg_database}         & \rotatebox{45}{$\rightarrow$} &            & \checkmark \\
    \hline
    \texttt{Flights}         & Weekly US Domestic Flights \cite{d3_zoomable}          & \rotatebox{45}{$\rightarrow$} &            & \checkmark \\
    \hline
    \texttt{Tourists}        & Monthly Tourists in New Zealand \cite{new_zealand}     & \rotatebox{45}{$\rightarrow$} &            &           \\
    \hline
    \texttt{Unemployment}   & US Monthly Unemploy.\ in Agriculture \cite{us_bls}    & \rotatebox{0}{$\rightarrow$}  & \checkmark & \checkmark \\
  \bottomrule
\end{tabular}}
\end{table}

\subsection{Introducing Noise}
\label{sec:study_design:noise}
One aspect of our experimental design was examining how noise impacts visual features in participants' sketches. While each input dataset contained some inherent noise, we initially treated them as noise-free (i.e., signal-only). Afterwards, Gaussian noise, chosen for its resemblance to real‐world noise~\cite{kim2021adaptive}, was introduced across datasets at five distinct signal-to-noise ratio (SNR) levels:

\begin{itemize}[noitemsep,itemsep=2pt]
\item \textbf{None}: The dataset remained unchanged, acting as the control group without additional noise;
\item \textbf{Low}: SNR of $\approx$ 30dB after adding noise;
\item \textbf{Medium}: SNR of $\approx$ 20dB after adding noise;
\item \textbf{High}: SNR of $\approx$ 10dB after adding noise; and
\item \textbf{Max}: SNR of $\approx$ 5dB after adding noise representing a scenario with substantial noise interference.
\end{itemize}

\begin{figure}[!t]
    \captionsetup[subfloat]{labelfont=scriptsize,textfont=scriptsize}
    \centering

    \subfloat[{\scriptsize \texttt{Apple}} \label{fig:datasets_max_noise:apple_max}]{
        \includegraphics[width=0.28\linewidth]{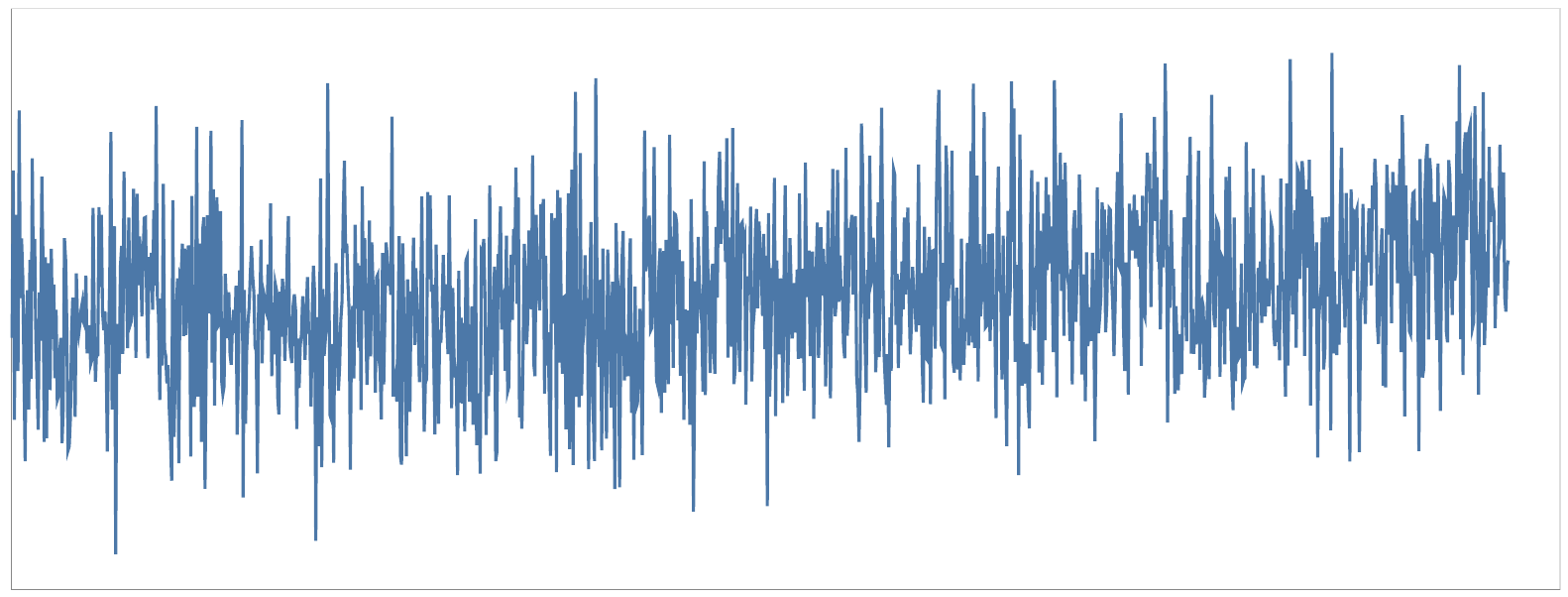}
    }
    \hspace{5pt}
    \subfloat[{\scriptsize \texttt{Astronomy}} \label{fig:datasets_max_noise:astro_max}]{
        \includegraphics[width=0.28\linewidth]{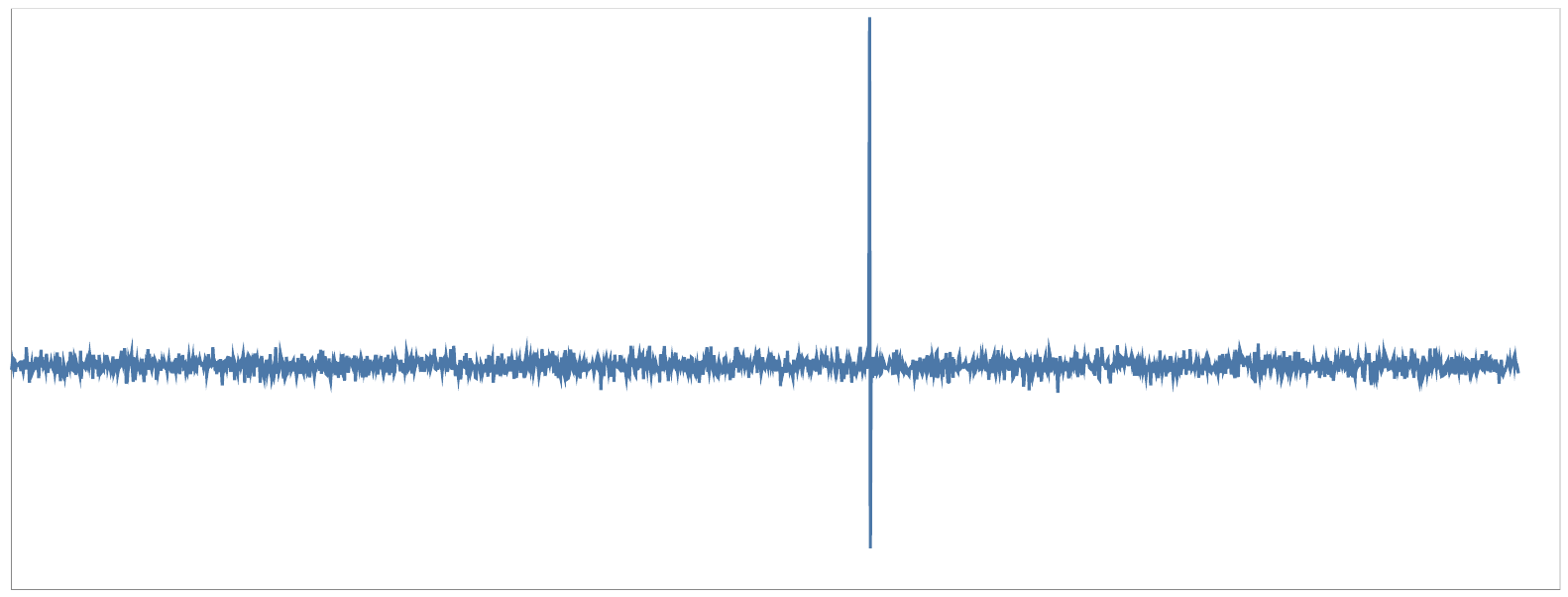}
    }
    \hspace{5pt}
    \subfloat[{\scriptsize \texttt{Chicago}} \label{fig:datasets_max_noise:chi_max}]{
        \includegraphics[width=0.28\linewidth]{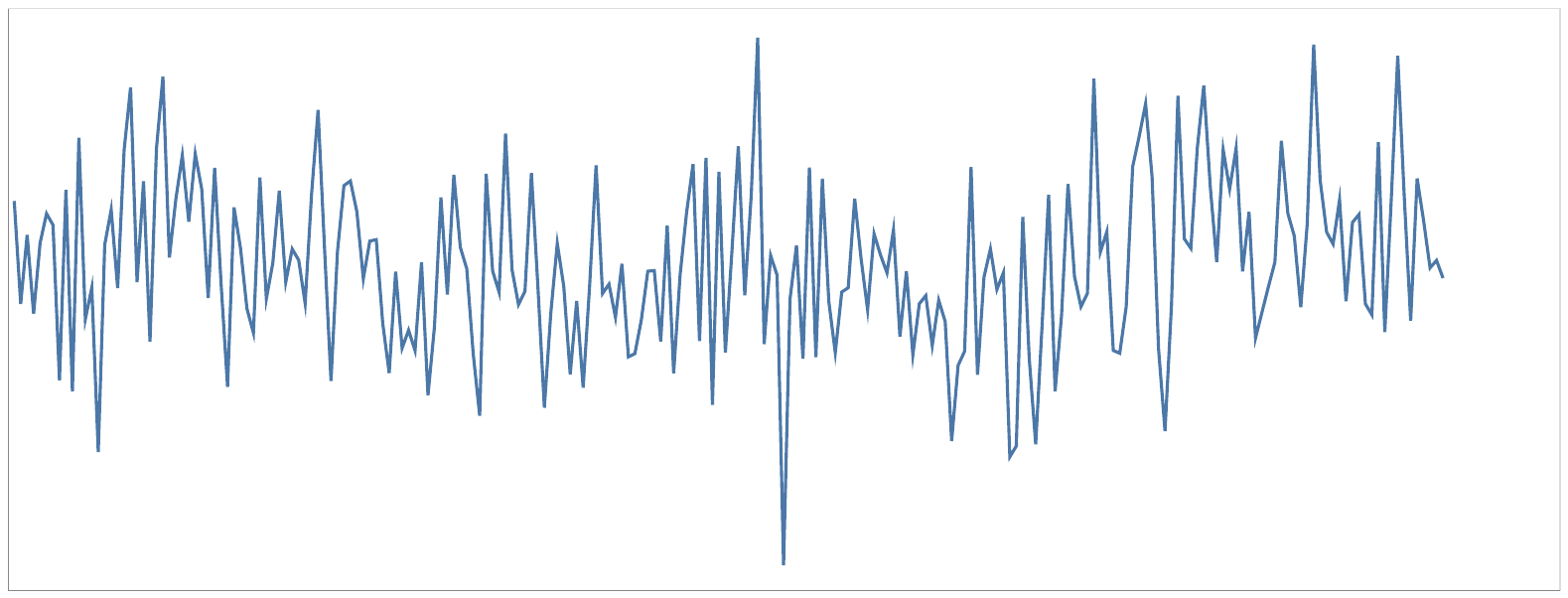}
    }

    \subfloat[{\scriptsize \texttt{Doge}} \label{fig:datasets_max_noise:doge_max}]{
        \includegraphics[width=0.28\linewidth]{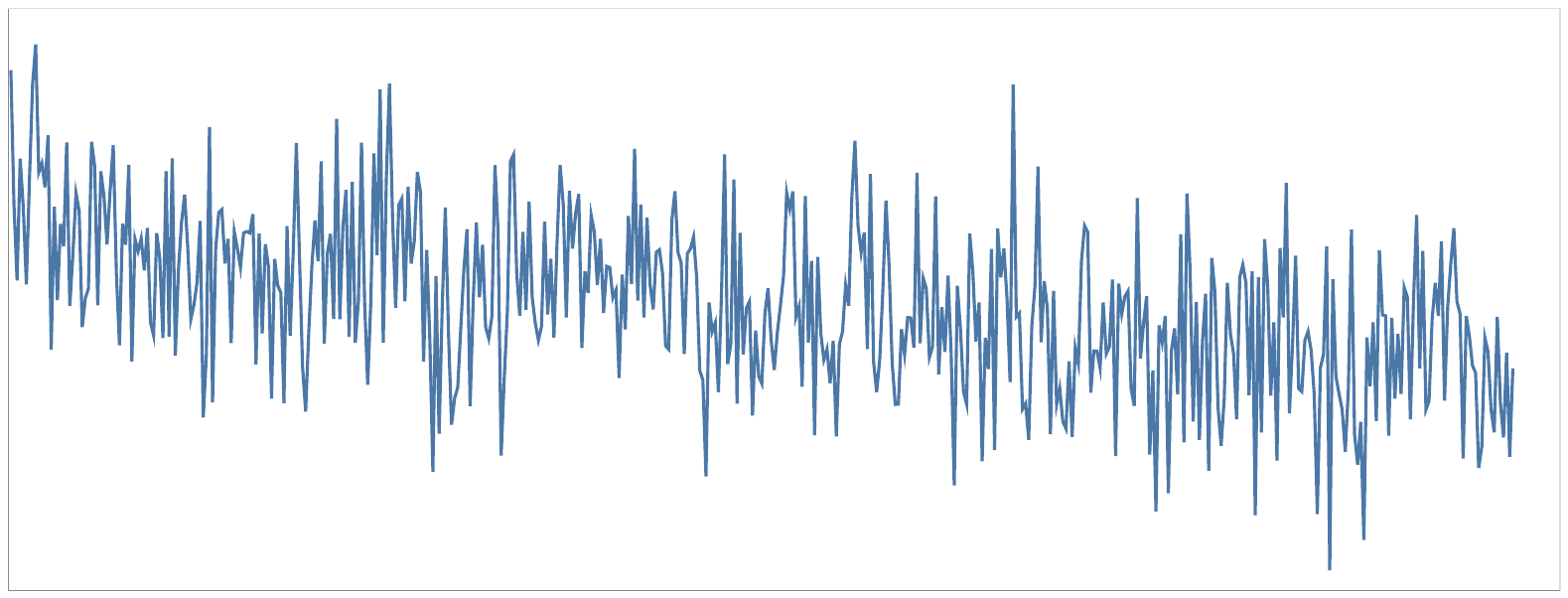}
    }
    \hspace{5pt}
    \subfloat[{\scriptsize \texttt{EEG}} \label{fig:datasets_max_noise:eeg_max}]{
        \includegraphics[width=0.28\linewidth]{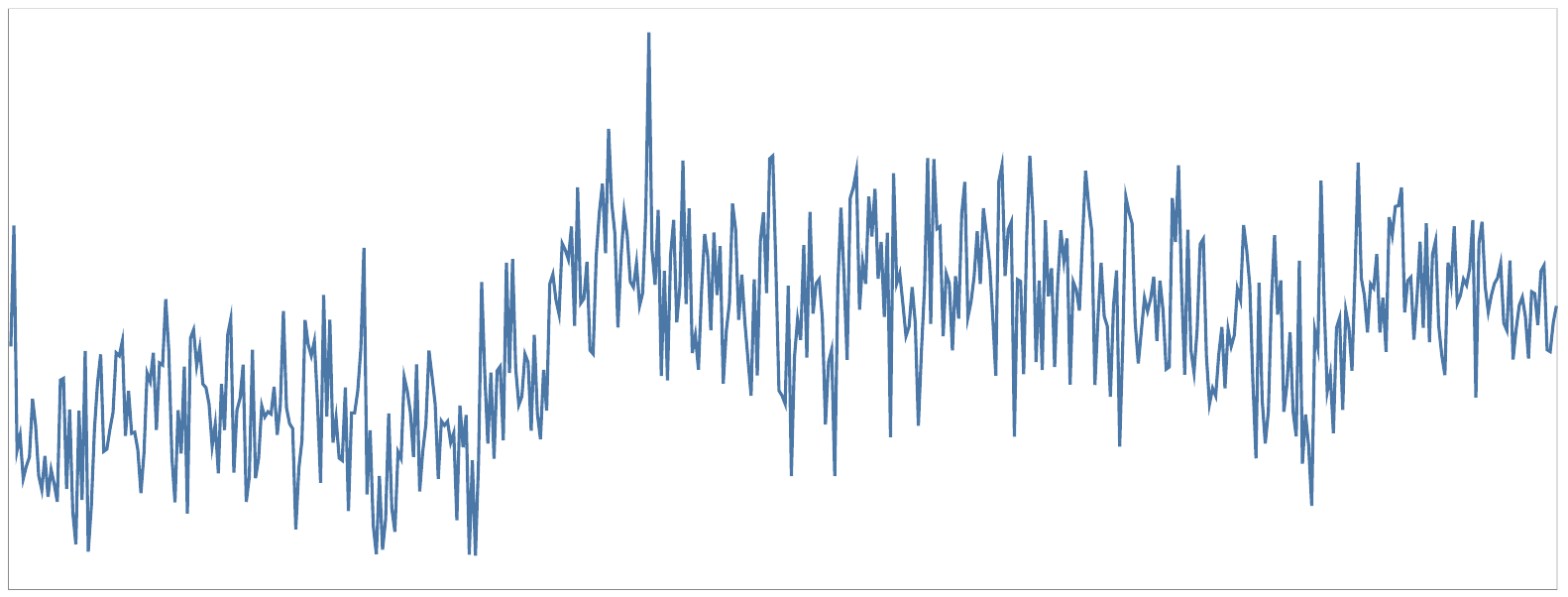}
    }
    \hspace{5pt}
    \subfloat[{\scriptsize \texttt{Flights}} \label{fig:datasets_max_noise:flights_max}]{
        \includegraphics[width=0.28\linewidth]{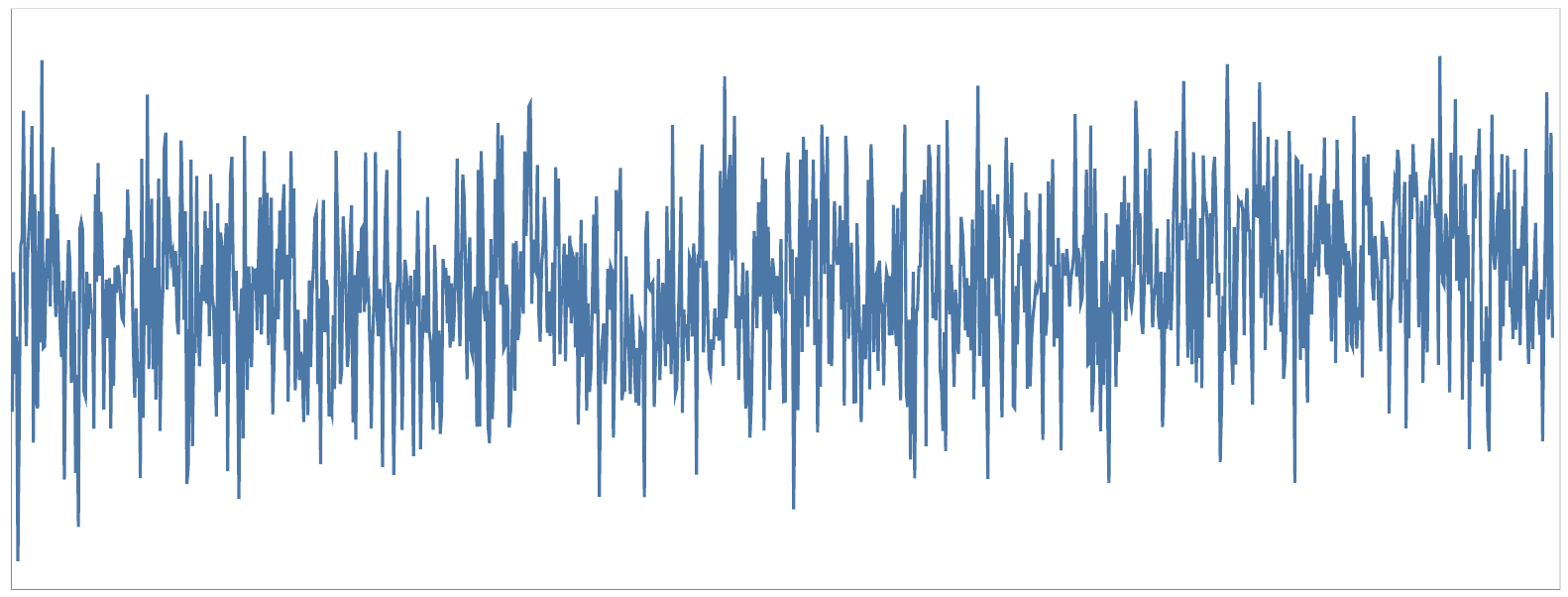}
    }

    \subfloat[{\scriptsize \texttt{Temperature}} \label{fig:datasets_max_noise:cli_max}]{
        \includegraphics[width=0.28\linewidth]{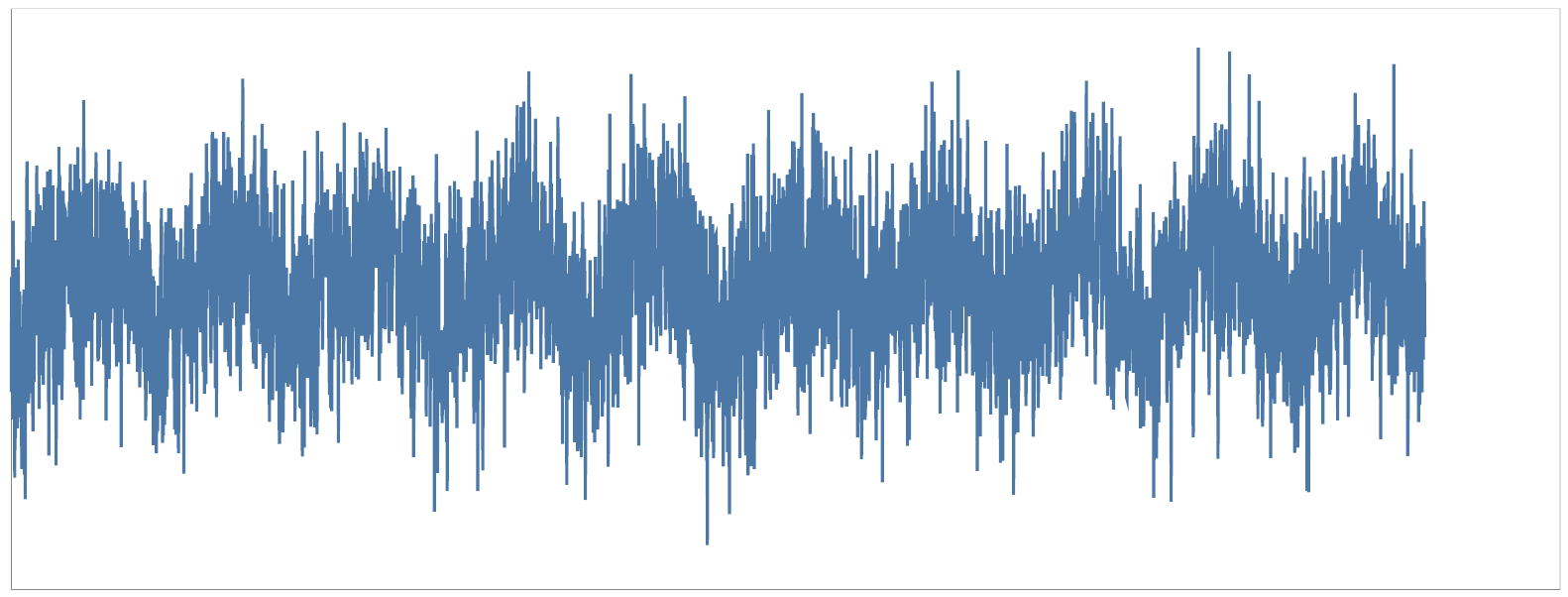}
    }
    \hspace{5pt}
    \subfloat[{\scriptsize \texttt{Tourists}} \label{fig:datasets_max_noise:nz_max}]{
        \includegraphics[width=0.28\linewidth]{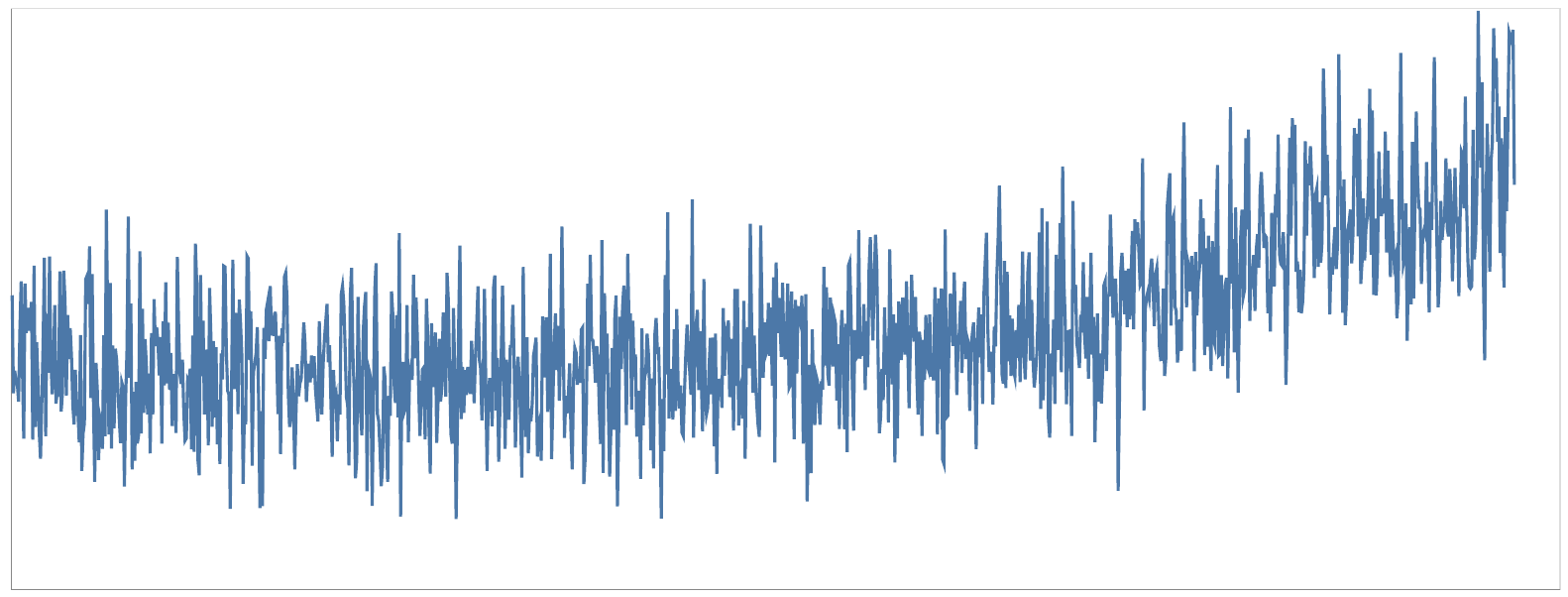}
    }
    \hspace{5pt}
    \subfloat[{\scriptsize \texttt{Unemployment}} \label{fig:datasets_max_noise:unemployment_max}]{
        \includegraphics[width=0.28\linewidth]{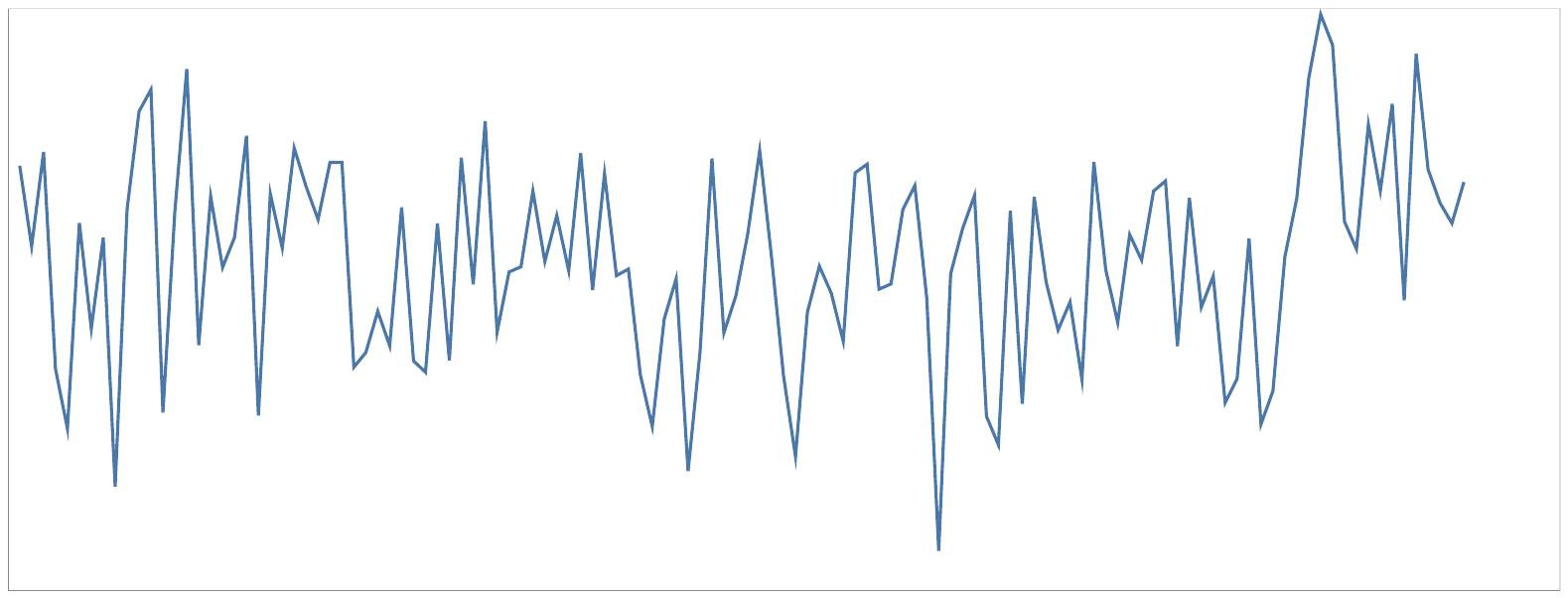}
    }

    \caption{Line charts for all nine datasets with max noise (SNR $\approx$ 5dB).}
    \label{fig:datasets_max_noise}
\end{figure}

One drawback of using real-world series is that each contains some native noise, so it is important to note that the same noise level had different effects on different datasets (see \Cref{fig:datasets_max_noise}) due to data density and structure variations. For example, some were denser (e.g., \Cref{fig:datasets:nz} / \Cref{fig:datasets_max_noise:nz_max}) and others sparser (see \Cref{fig:datasets:unemployment} / \Cref{fig:datasets_max_noise:unemployment_max}), while some contained smaller peaks (e.g., \Cref{fig:datasets:apple} / \Cref{fig:datasets_max_noise:apple_max}) and others contained more prominent peaks (e.g., \Cref{fig:datasets:astro} / \Cref{fig:datasets_max_noise:astro_max}). Figures of each dataset with all noise levels can be found in the supplemental material. While we could have used fully synthetic signals to control noise, working with real data allows preserving the natural trends, periodicity, peaks, and valleys that give the results high ecological validity, while still varying only the noise level as the experimental factor.

\subsection{Experiment}
\label{sec:study_design:exp}

\paragraph{Stimuli}  
We created 45 stimuli with 9 datasets (within-subject) crossed with 5 noise levels (between-subject). Participants saw each dataset once, in the same order, at a randomly selected noise level. The noise levels were shuffled such that all 45 stimuli appeared for every five participants. 
We used Vega-Lite~\cite{satyanarayan2016vega} to generate the line charts. To focus solely on features observed in the chart and to avoid copying the data, none of the generated line charts included axes, gridlines, or ticks. 
Each dataset's minimum and maximum values were used to scale the y-axis to minimize the white space. %

\paragraph{Task}
Participants viewed each of the nine line charts sequentially on a monitor at a resolution of 950 x 375. 
They used an Apple iPad Air (5th Gen) running a web application that captured their sketch via an Apple Pencil (2nd Gen), with both the monitor and iPad canvases set to the same 2.53:1 aspect ratio.
Participants were tasked with re-drawing the line chart without any prior context about the data used to ensure that their sketches reflected purely visual interpretations rather than domain‑driven bias. During the sketching process, the stimulus line chart remained visible at all times. They were encouraged to emphasize elements in their sketches that they deemed significant, with the understanding that an exact replication of the shown stimuli was not required. The web application would record a single stroke only (i.e., the pen always had to stay on the canvas). They were informed that lifting the pen would prompt them to accept or reset their sketch, with unlimited resets allowed. Although no time limit was imposed, the experiment generally took 20 minutes, with one participant taking approximately 40 minutes.

\paragraph{Participants}
   We recruited 20 participants (5 females and 15 males) who were engineering students (1 undergraduate and 19 graduate).  
   Each participant had experience conducting research in the visualization domain and affirmed their familiarity with line charts. Participants did not receive any compensation. The study was IRB-approved.

\section{Analysis}
\label{sec:analysis}
We employ a mixed-methods approach to assess the behaviors observed in the participant-sketched line charts. We performed a qualitative evaluation (see \Cref{sec:qualitative}) to identify which visual features participants retained, while we performed a quantitative analysis (see \Cref{sec:quantitative}) to determine the accuracy of this feature replication in the resulting sketches. This combination captured both numerical similarity and semantically significant representations.

\begin{figure}[!b]
    \centering
    
    \subfloat[{\scriptsize No Noise}\label{fig:coding_figure:chi_none}]{
        \parbox{0.45\linewidth}{
            \centering
            \includegraphics[width=\linewidth]{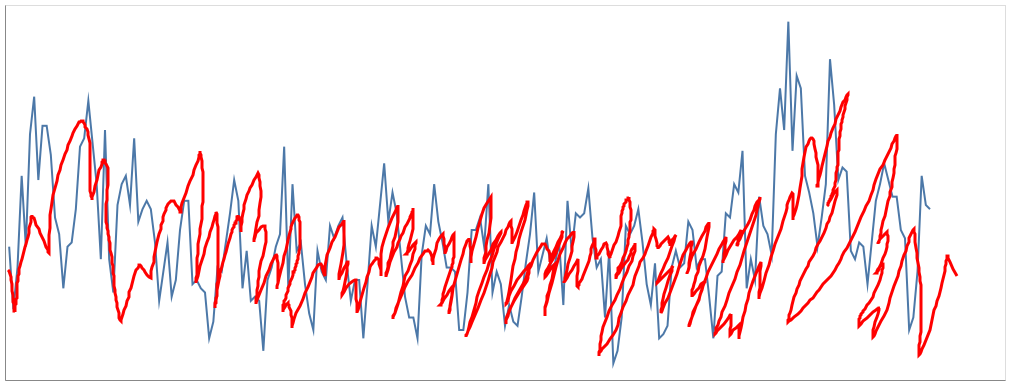}
            \vspace{3pt}
            \resizebox{\linewidth}{!}{%
                \begin{tabular}{ccccccc}
                    \cellcolor{LightGray!10}\textit{Overall} & & \cellcolor{LightGray!10}\textit{Periodicity} & & \cellcolor{LightGray!10}\textit{Peak \& Valley} & & \cellcolor{LightGray!10}\textit{Noise} \\ [2pt]
                    \cellcolor{LightGray!10}\textit{Trend} & & \cellcolor{LightGray!10}\textit{Match} & & \cellcolor{LightGray!10}\textit{Preservation} & & \cellcolor{LightGray!10}\textit{Match} \\ [2pt]
                    \cline{1-1} \cline{3-3} \cline{5-5} \cline{7-7} \\ [-1.1ex]
                    \cellcolor{LightGray!10}\textbf{Very well} & & \cellcolor{LightGray!10}\textbf{Very well} & & \cellcolor{LightGray!10}\textbf{Most of them} & & \cellcolor{LightGray!10}\textbf{Very well} \\ [2pt]
                \end{tabular}
            }
        }
    }
    \hspace{5pt}
    \subfloat[{\scriptsize Low Noise}\label{fig:coding_figure:chi_low}]{
        \parbox{0.45\linewidth}{
            \centering
            \includegraphics[width=\linewidth]{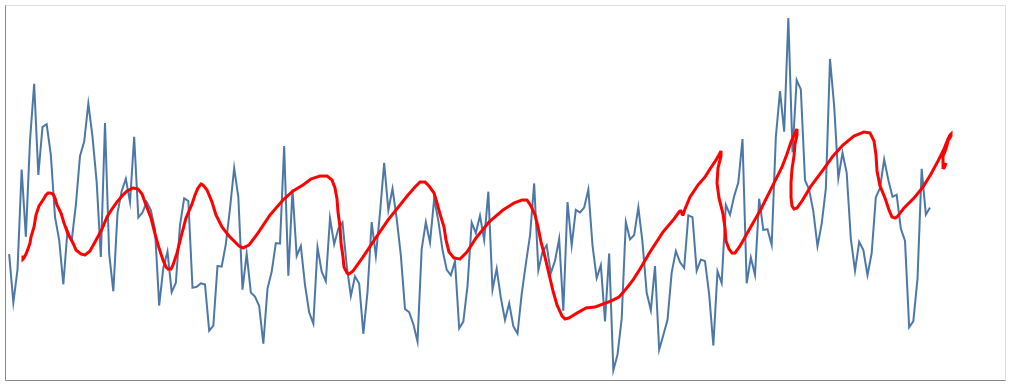}
            \vspace{3pt}
            \resizebox{\linewidth}{!}{%
                \begin{tabular}{ccccccc}
                    \cellcolor{LightGray!10}\textit{Overall} & & \cellcolor{LightGray!10}\textit{Periodicity} & & \cellcolor{LightGray!10}\textit{Peak \& Valley} & & \cellcolor{LightGray!10}\textit{Noise} \\ [2pt]
                    \cellcolor{LightGray!10}\textit{Trend} & & \cellcolor{LightGray!10}\textit{Match} & & \cellcolor{LightGray!10}\textit{Preservation} & & \cellcolor{LightGray!10}\textit{Match} \\ [2pt]
                    \cline{1-1} \cline{3-3} \cline{5-5} \cline{7-7} \\ [-1.1ex]
                    \cellcolor{LightGray!10}\textbf{Somewhat} & & \cellcolor{LightGray!10}\textbf{Somewhat} & & \cellcolor{LightGray!10}\textbf{Some of them} & & \cellcolor{LightGray!10}\textbf{Not at all} \\ [2pt]
                \end{tabular}
            }
        }
    }

    \subfloat[{\scriptsize High Noise}\label{fig:coding_figure:chi_high}]{
        \parbox{0.45\linewidth}{
            \centering
            \includegraphics[width=\linewidth]{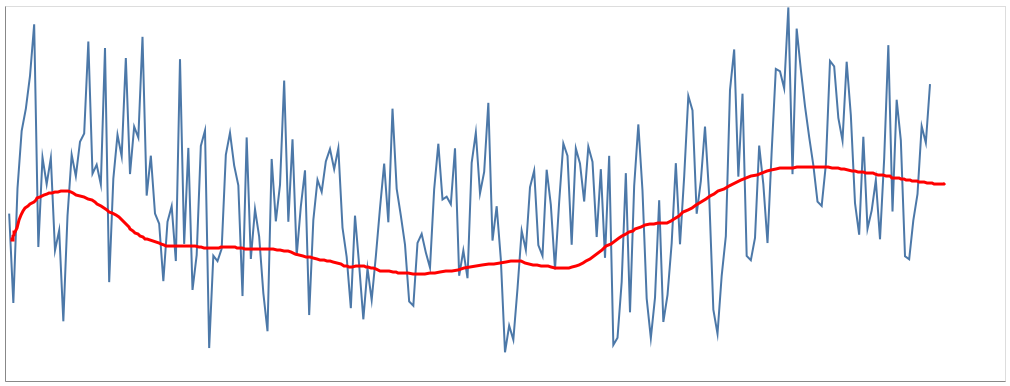}
            \vspace{3pt}
            \resizebox{\linewidth}{!}{%
                \begin{tabular}{ccccccc}
                    \cellcolor{LightGray!10}\textit{Overall} & & \cellcolor{LightGray!10}\textit{Periodicity} & & \cellcolor{LightGray!10}\textit{Peak \& Valley} & & \cellcolor{LightGray!10}\textit{Noise} \\ [2pt]
                    \cellcolor{LightGray!10}\textit{Trend} & & \cellcolor{LightGray!10}\textit{Match} & & \cellcolor{LightGray!10}\textit{Preservation} & & \cellcolor{LightGray!10}\textit{Match} \\ [2pt]
                    \cline{1-1} \cline{3-3} \cline{5-5} \cline{7-7} \\ [-1.1ex]
                    \cellcolor{LightGray!10}\textbf{Very well} & & \cellcolor{LightGray!10}\textbf{Not at all} & & \cellcolor{LightGray!10}\textbf{None of them} & & \cellcolor{LightGray!10}\textbf{Not at all} \\ [2pt]
                \end{tabular}
            }
        }
    }
    \hspace{5pt}
    \subfloat[{\scriptsize Max Noise}\label{fig:coding_figure:chi_max}]{
        \parbox{0.45\linewidth}{
            \centering
            \includegraphics[width=\linewidth]{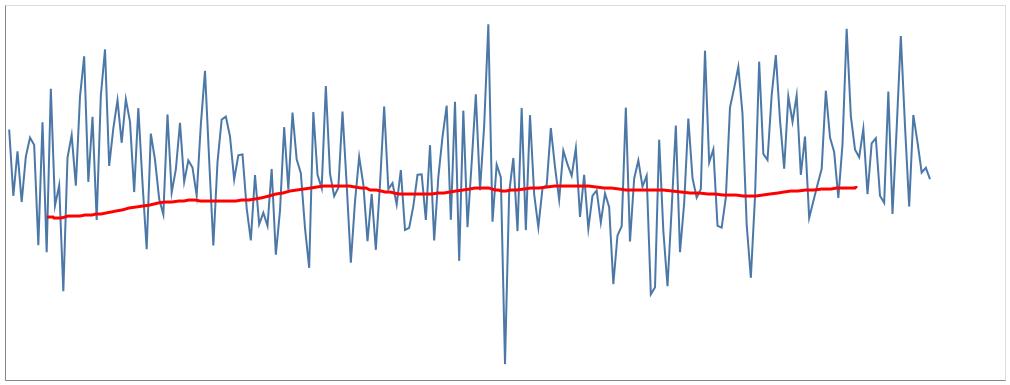}
            \vspace{3pt}
            \resizebox{\linewidth}{!}{%
                \begin{tabular}{ccccccc}
                    \cellcolor{LightGray!10}\textit{Overall} & & \cellcolor{LightGray!10}\textit{Periodicity} & & \cellcolor{LightGray!10}\textit{Peak \& Valley} & & \cellcolor{LightGray!10}\textit{Noise} \\ [2pt]
                    \cellcolor{LightGray!10}\textit{Trend} & & \cellcolor{LightGray!10}\textit{Match} & & \cellcolor{LightGray!10}\textit{Preservation} & & \cellcolor{LightGray!10}\textit{Match} \\ [2pt]
                    \cline{1-1} \cline{3-3} \cline{5-5} \cline{7-7} \\ [-1.1ex]
                    \cellcolor{LightGray!10}\textbf{Not at all} & & \cellcolor{LightGray!10}\textbf{Not at all} & & \cellcolor{LightGray!10}\textbf{None of them} & & \cellcolor{LightGray!10}\textbf{Not at all} \\ [2pt]
                \end{tabular}
            }
        }
    }

    \caption{Participant sketches (red) are overlaid on the stimuli (blue) of the \texttt{Chicago} dataset at four different noise levels. Coders evaluated the sketches and assigned them to categories based on retained features.}
    \label{fig:coding_figure}
\end{figure}

\subsection{Qualitative Analysis}
\label{sec:qualitative}

We adopted a thematic coding process for qualitative analysis to systematically examine key aspects of participant sketches, which involved addressing questions about four factors of the sketches: the overall trend, periodicity, prominent peaks and/or valleys, and the noise level (for datasets where these features were present, see \Cref{tab:dataset}). Each sketch was shown juxtaposed over the stimuli as a reference, as shown in \Cref{fig:coding_figure}. 
The coders were given the following questions with three options each:

\noindent
\begin{center}
\resizebox{0.9\linewidth}{!}{%
\begin{tabular}{p{2.0cm} p{6.2cm} c}
\hline
Feature & Coding Questions & Options \\ 
\hline \\ [-1.5ex]
\textit{Overall Trend} & How well does the trend in the sketch (Red) match the trend in the stimuli (Blue)? & Very well / Somewhat / Not at all \\ [1.5ex]
\textit{Periodicity Match} & How well does the periodicity in the sketch (Red) match the periodicity in the stimuli (Blue)? & Very well / Somewhat / Not at all \\ [1.5ex]
\textit{Peak \& Valley Preservation} & How many of the peaks and/or valleys of the stimuli (Blue) are preserved in the sketch (Red)? & Most / Some / None of them \\ [1.5ex]
\textit{Noise Match} & How well does the noise in the sketch (Red) match the noise in the stimuli (Blue)? & Very well / Somewhat / Not at all \\ [2ex]
\hline
\end{tabular}
}
\end{center}

Two authors of the paper independently coded the sketches. Before coding, they aligned their understanding of trends, periodicity, peaks, valleys, and noise to establish a consensus code. The independent coding yielded a Cohen’s kappa score of 0.733 with 445 matches and 115 mismatches, indicating a ``substantial'' agreement~\cite{mchugh2012interrater}. Moreover, all 115 mismatches indicated only minor interpretive differences (e.g., one coder classified a sketch as matching the original trend \textit{Very well} while another classified it as \textit{Somewhat}) rather than fundamental disagreement. Therefore, results from the first coder were used for all analyses. The complete coding is included in the supplemental material.

\begin{figure*}[!t]
\centering
    \subfloat[{\scriptsize Example Participant Sketch}\label{fig:before_pre_processing}]{\includegraphics[width=0.3\linewidth]{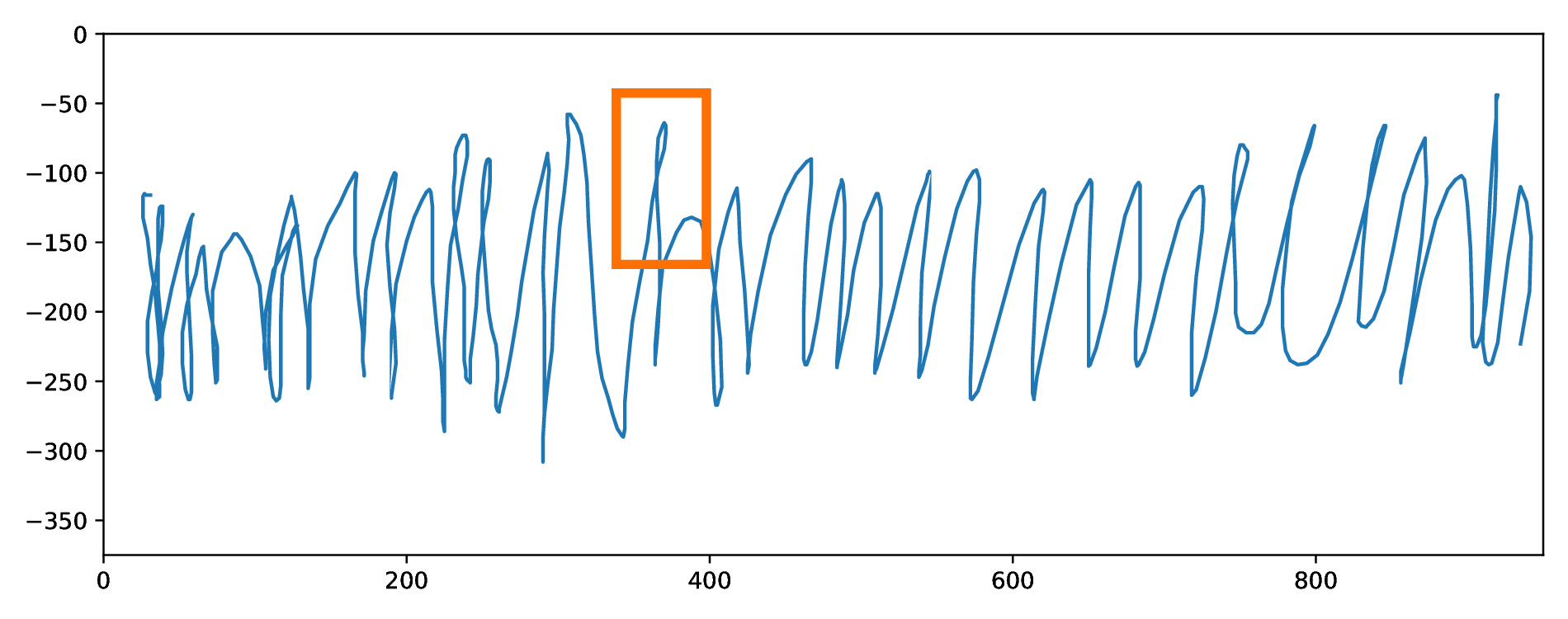}}
    \hspace{5pt}
    \subfloat[{\scriptsize After Removing Artifacts}\label{fig:after_time_series}]{\includegraphics[width=0.3\linewidth]{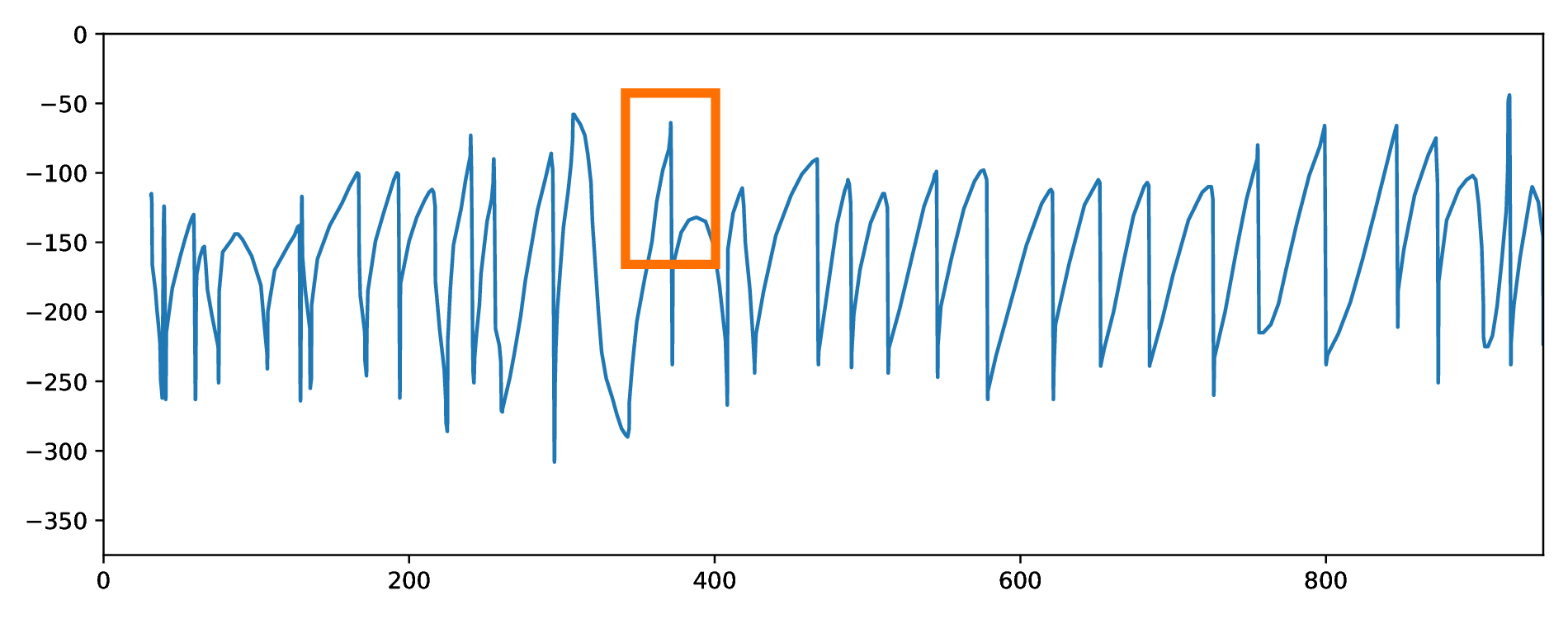}}
     \hspace{5pt}
    \subfloat[{\scriptsize After Normalization}\label{fig:pre_process_stretch_squeeze}]{\includegraphics[width=0.3\linewidth]{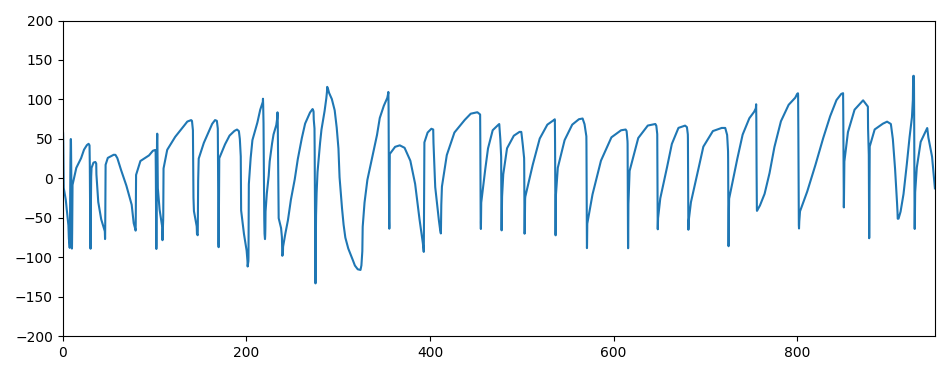}}
     \hspace{5pt}
    \caption{(a)~A participant sketch of the \texttt{Apple} dataset with max noise (SNR $\approx$ 5dB), which is not a valid time series, as pointed out by the orange enclosure. (b)~After artifact removal, the data is now a valid time series. (c)~Data is normalized in order to compare it to the stimuli data.}
    \label{fig:preprocessing}
\end{figure*}

\subsection{Quantitative Analysis}

\label{sec:quantitative}
We conducted a quantitative analysis of participant sketches through first pre-processing to correct artifacts and normalize sketches (see \Cref{pre_processing_participant_drawings}) and then establishing metrics for measuring the visual features (see \Cref{sec:quantitative:metrics}). All the figures from this analysis are included in the supplemental material.

\subsubsection{Pre-processing Participant Sketches}
\label{pre_processing_participant_drawings}
\hfill

\paragraph{Artifact Removal} Due to the free-hand nature of the task, the participant sketches could have artifacts that violated the requirements of time series data, temporal ordering in particular (i.e., data must be in chronological order). This would arise in sketches as strokes that would go backward and/or contain loops (see \Cref{fig:before_pre_processing}). To address this issue, we implemented a correction process. Since the sketch was done in a single stroke, given two temporally consecutive points, $x_{i}$ and $x_{i+1}$, if the x-coordinate of the first data point was larger than the second (i.e., $x_{i}>x_{i+1}$), the x-coordinate of the second data point was set to be 0.1 larger than the first (i.e., $x_{i+1}=x_{i}+0.1$). Here, 0.1 was chosen as a minimal increment on our 950 px–wide canvas, small enough to correct ordering without noticeably altering the sketch’s shape. This approach resolved these issues, such as the loops as shown in~\Cref{fig:after_time_series}.

\paragraph{Normalization} The next step involved normalizing the data, such that the participant sketch and the stimuli can be compared point-by-point. 
First, in the x-direction the time series participant sketches were stretched or compressed to fit the canvas. Next, both the stimuli and participant sketched time series were upsampled to 9,500 samples (10x the original resolution), using linear interpolation (see~\Cref{fig:pre_process_stretch_squeeze}). In the y-direction, both sketches and the stimuli  were aligned on center by subtracting their average value.

\subsubsection{Estimating Features and Metrics for Similarity}
\label{sec:quantitative:metrics}

\begin{figure}[!b]

    \begin{minipage}[t]{0.975\linewidth}
        \includegraphics[width=\linewidth]{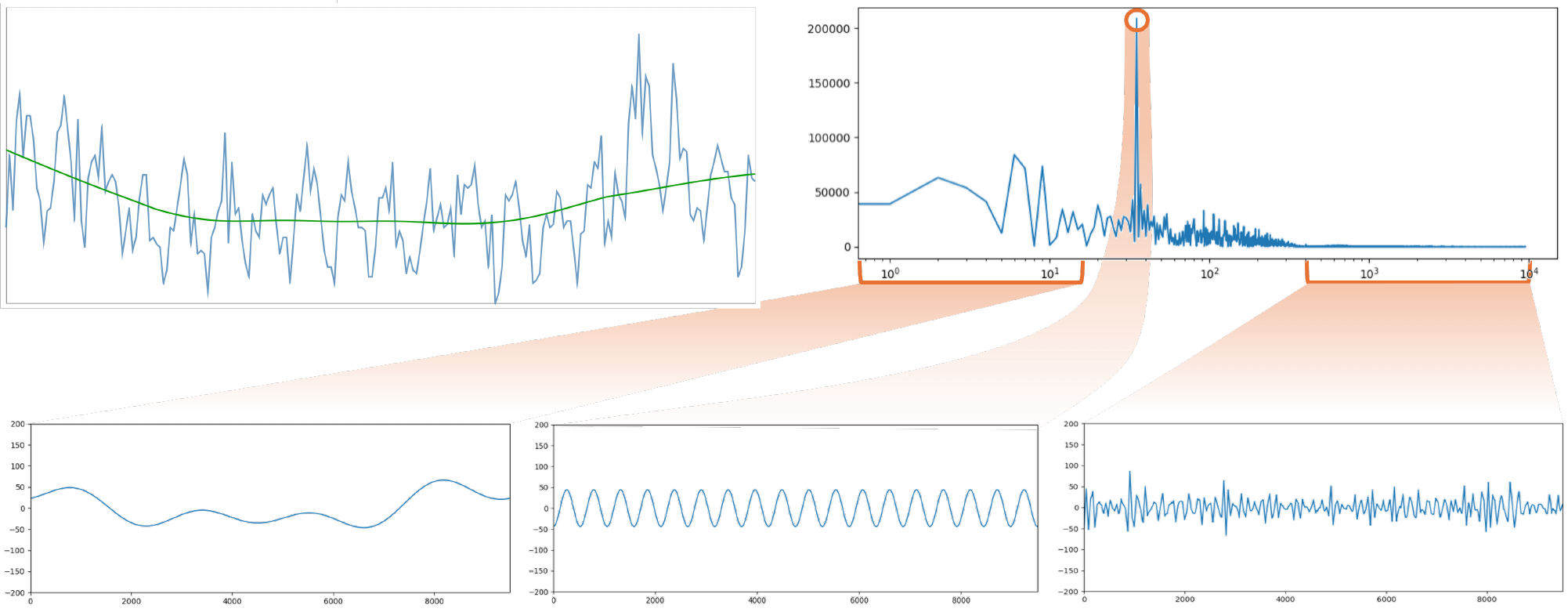}
    \end{minipage}
    \begin{minipage}[t]{1pt}
        \vspace{-62pt}\hspace{-250pt}
        \subfloat[{\scriptsize Estimated trend with LOESS}\label{fig:trend_loess}]{\hspace{110pt}}
        \hspace{25pt}
        \subfloat[{\scriptsize Fast Fourier Transformation}\label{fig:fft_decomposition:fft}]{\hspace{100pt}}

        \vspace{25pt}
        \hspace{-264pt}
        \subfloat[{\scriptsize Estimated trend}\label{fig:fft_decomposition:trend_fft}]{\hspace{83pt}}
        \subfloat[{\scriptsize Estimated periodic pattern}\label{fig:fft_decomposition:periodic_fft}]{\hspace{97pt}}
        \subfloat[{\scriptsize Estimated noise}\label{fig:fft_decomposition:noise_fft}]{\hspace{80pt}}
    \end{minipage}

    \caption{Illustration of feature extraction. (a)~shows the \texttt{Chicago} dataset in a line chart in blue, with its LOESS fit curve trend in green. (b)~The Fast Fourier Transform (FFT) decomposes the data into frequency components. (c)~The inversion of low-frequency components back to the time domain provides the estimated trend. (d)~ The most prominent FFT peak is transformed to the time domain to estimate the periodic pattern. (e)~The high-frequency components are converted to the time domain to estimate noise.}
    \label{fig:fft_decomposition}
\end{figure}

\begin{figure*}[!t]

    \setlength{\fboxrule}{2pt}
    \fcolorbox{green}{white}{
    \begin{minipage}{\linewidth}

        \vspace{5pt}
        \rotatebox{90}{\textcolor{green}{\textbf{Replicator}}}
        \hspace{5pt}
        \subfloat[{\scriptsize SNR  $\approx$ 5dB}\label{fig:astro_replicator}]{\includegraphics[width=0.23\linewidth]{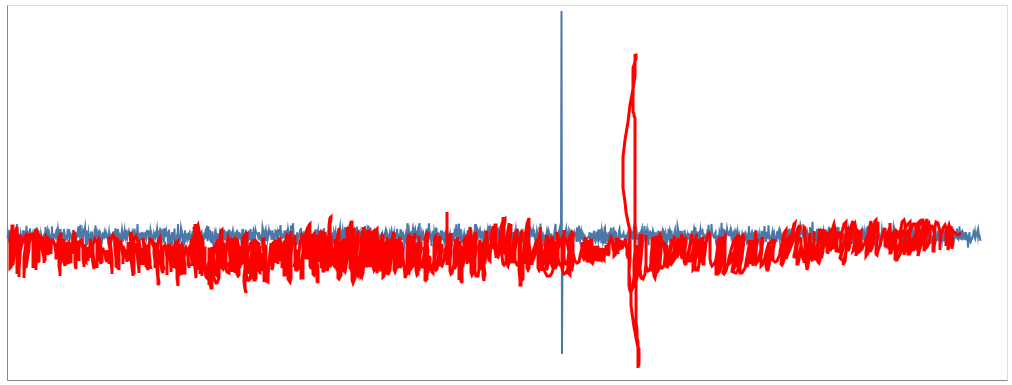}}  
       \hspace{5pt}
        \subfloat[{\scriptsize No added noise}\label{fig:chi_replicator}]{\includegraphics[width=0.23\linewidth]{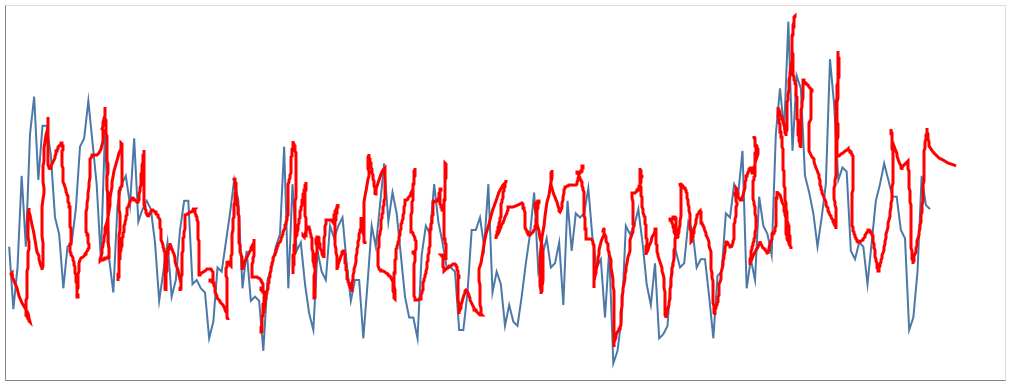}}
        \hspace{5pt}
        \subfloat[{\scriptsize SNR $\approx$ 5dB}\label{fig:doge_replicator}]{\includegraphics[width=0.23\linewidth]{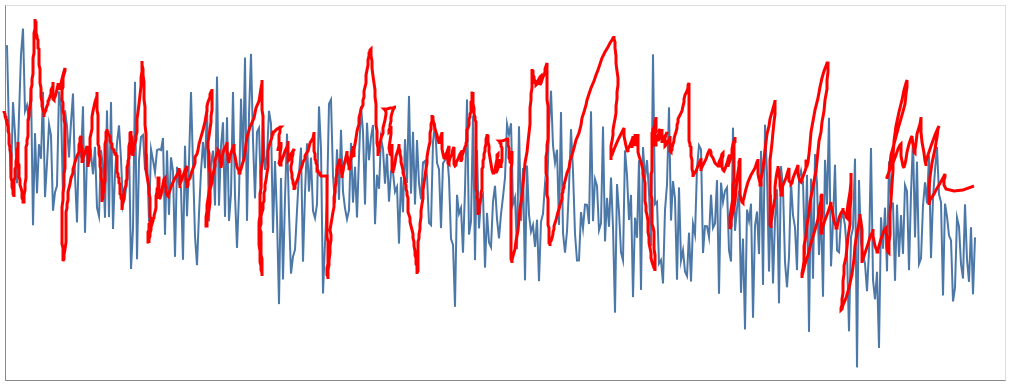}}
        \hspace{5pt}    
        \subfloat[{\scriptsize SNR $\approx$ 30dB}\label{fig:flights_replicator}]{\includegraphics[width=0.23\linewidth]{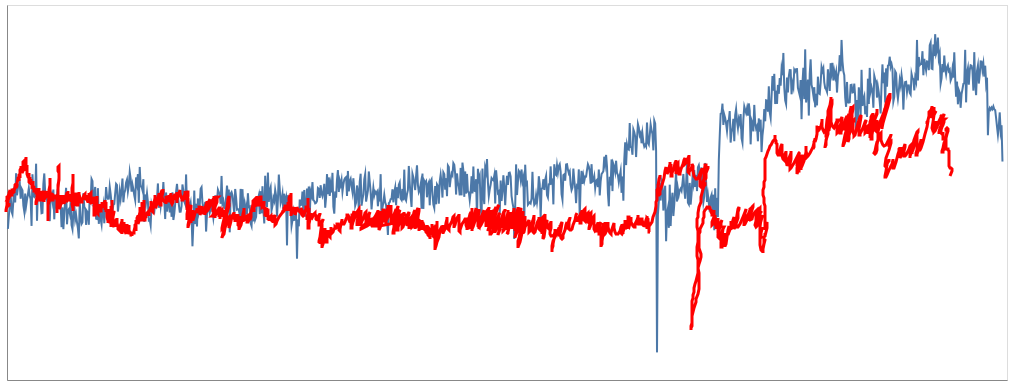}}

    \end{minipage}}

    \setlength{\fboxrule}{2pt}
    \fcolorbox{yellow}{white}{
    \begin{minipage}{\linewidth}

        \vspace{5pt}
        \rotatebox{90}{\textcolor{yellow}{\textbf{Trend Keeper}}}
        \hspace{5pt}
        \subfloat[{\scriptsize SNR $\approx$ 10dB}\label{fig:astro_trend_keeper}]{\includegraphics[width=0.23\linewidth]{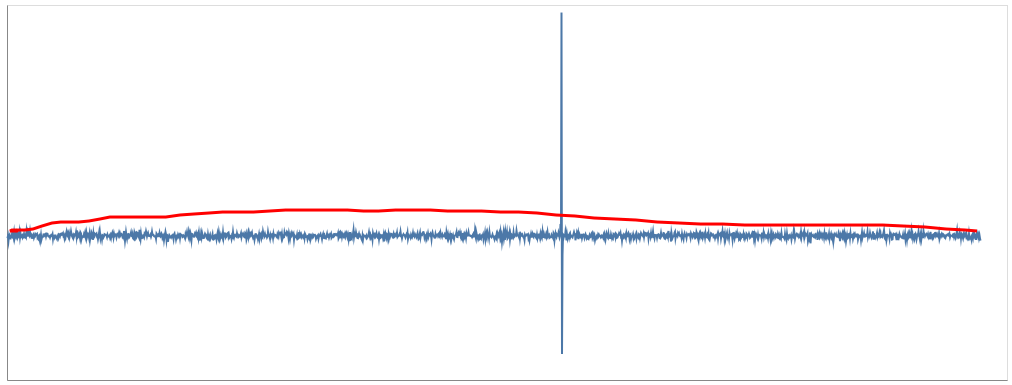}}   
        \hspace{5pt}
        \subfloat[{\scriptsize SNR $\approx$ 10dB}\label{fig:chi_trend_keeper}]{\includegraphics[width=0.23\linewidth]{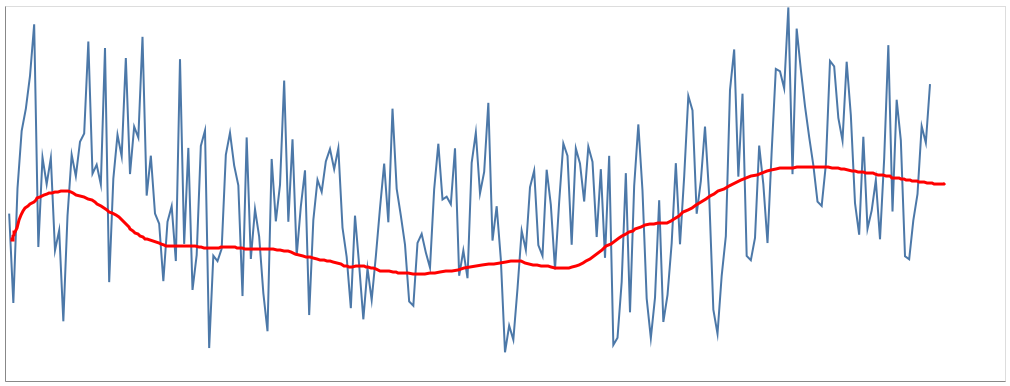}}
        \hspace{5pt}
        \subfloat[{\scriptsize SNR $\approx$ 30dB}\label{fig:doge_trend_keeper}]{\includegraphics[width=0.23\linewidth]{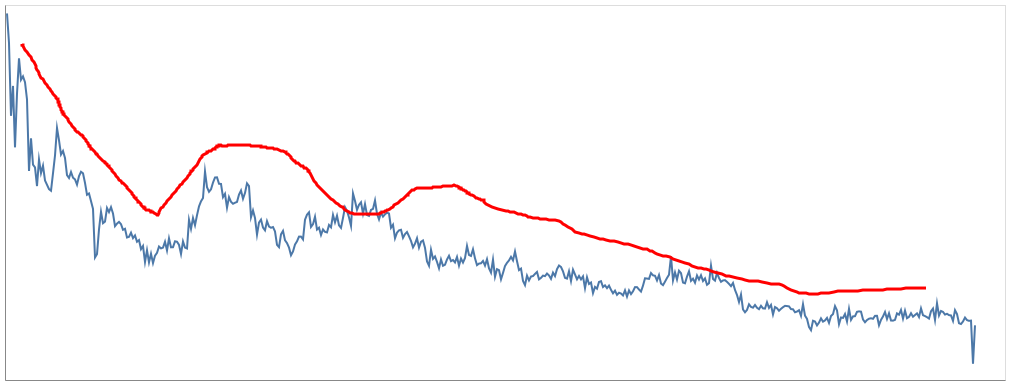}}
        \hspace{5pt}
        \subfloat[{\scriptsize SNR $\approx$ 20dB}\label{fig:flights_trend_keeper}]{\includegraphics[width=0.23\linewidth]{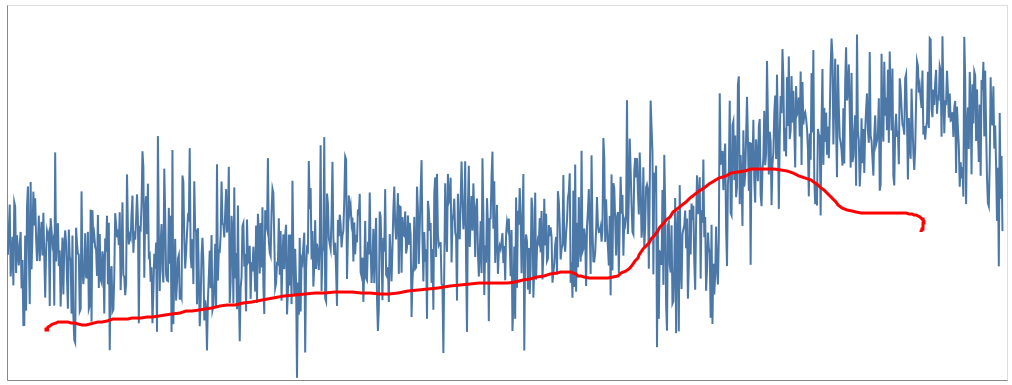}}

    \end{minipage}}

    \setlength{\fboxrule}{2pt} 
    \fcolorbox{pink}{white}{
    \begin{minipage}{\linewidth}
        
        \vspace{5pt}
        \rotatebox{90}{\textcolor{pink}{\textbf{De-noiser}}}
        \hspace{5pt}
        \subfloat[{\scriptsize No added noise}\label{fig:astro_de_noiser}]{\includegraphics[width=0.23\linewidth]{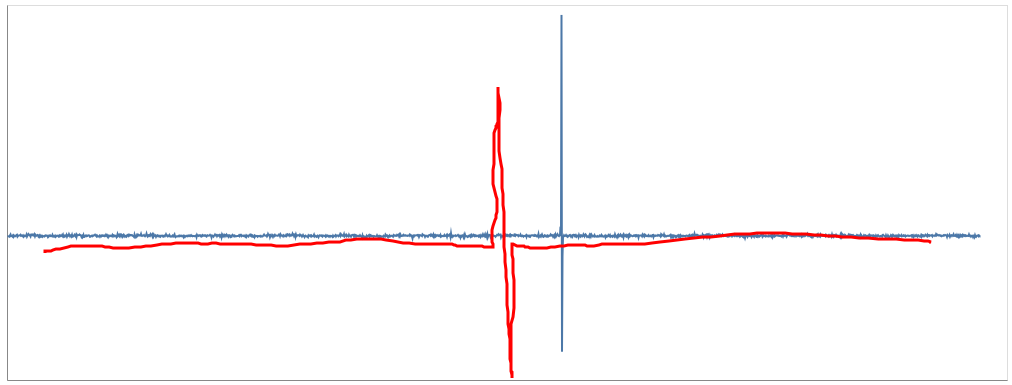}} 
        \hspace{5pt}
        \subfloat[{\scriptsize SNR $\approx$ 30dB}\label{fig:chi_de_noiser}]{\includegraphics[width=0.23\linewidth]{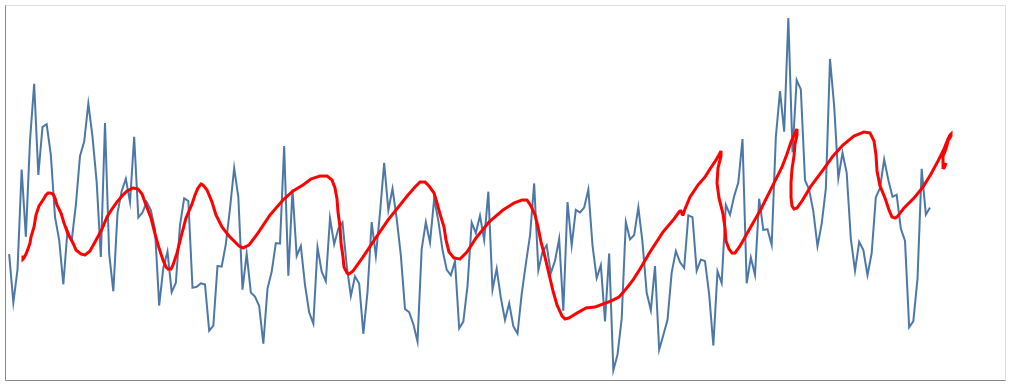}}
        \hspace{5pt}
        \subfloat[{\scriptsize SNR $\approx$ 10dB}\label{fig:doge_de_noiser}]{\includegraphics[width=0.23\linewidth]{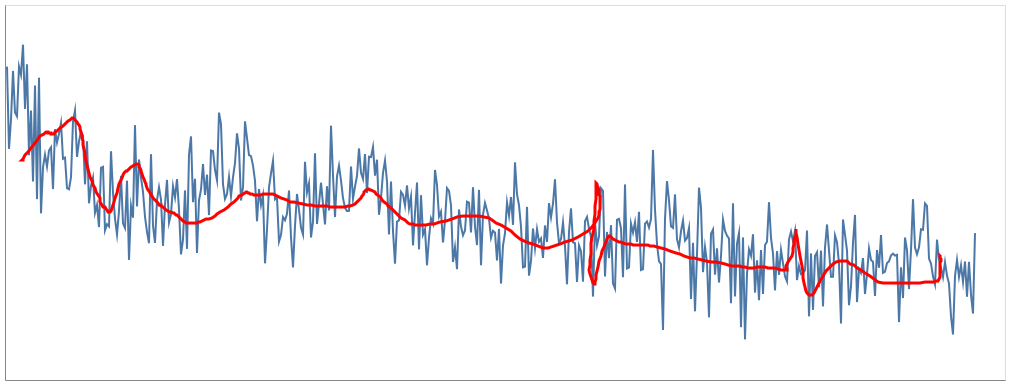}}
        \hspace{5pt}
        \subfloat[{\scriptsize SNR $\approx$ 20dB}\label{fig:flights_de_noiser}]{\includegraphics[width=0.23\linewidth]{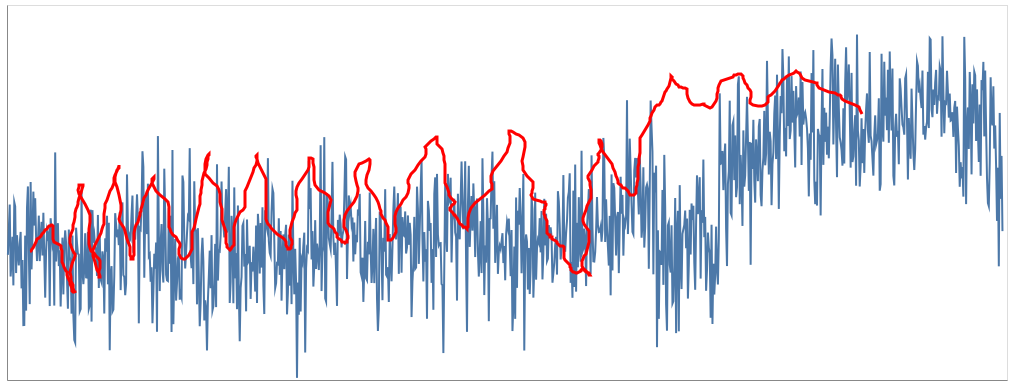}}

    \end{minipage}}

    \vspace{3pt}
    \begin{minipage}{\linewidth}
        \scriptsize

        \hspace{65pt}
        \texttt{Astronomy}
        \hspace{85pt}
        \texttt{Chicago}
        \hspace{95pt}
        \texttt{Doge}
        \hspace{100pt}    
        \texttt{Flights}

    \end{minipage}    
    
    \caption{Participant sketches overlaid on stimuli for four datasets, categorized into clusters based on feature inclusion or exclusion in the sketches.}
    \label{fig:clusters}
\end{figure*}

\begin{figure}[!b]
    \centering
        \vspace{-20pt}
        \subfloat[{\scriptsize Trend Similarity with Total Variation}\label{fig:measures:variation}]{\hspace{10pt}\includegraphics[width=0.4\linewidth]{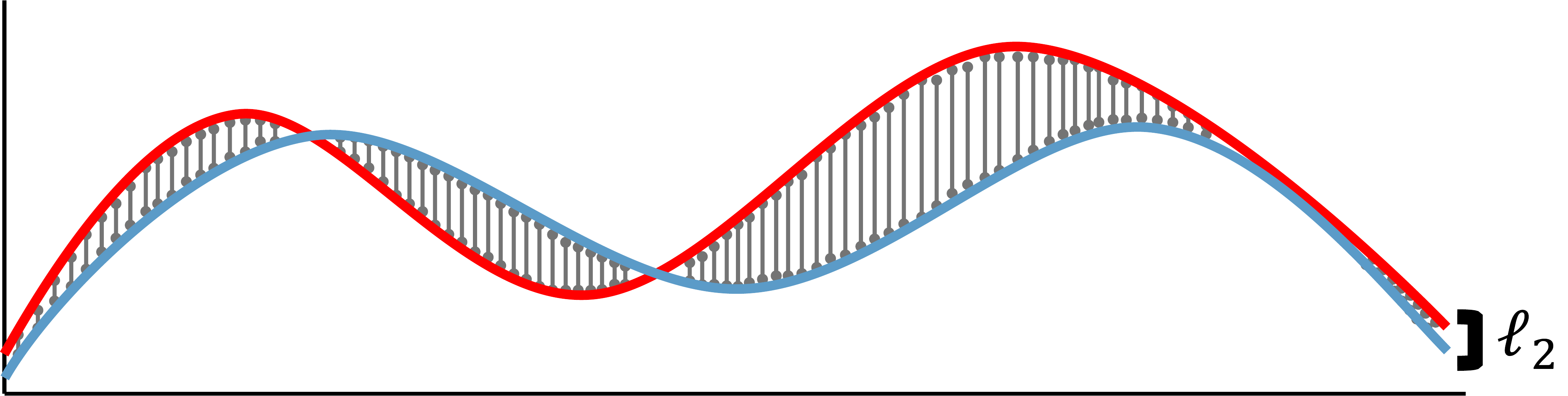}\hspace{10pt}}
        \subfloat[{\scriptsize Comparing Period and Amplitude}\label{fig:measures:period_amplitude}]{\hspace{10pt}\includegraphics[width=0.4\linewidth]{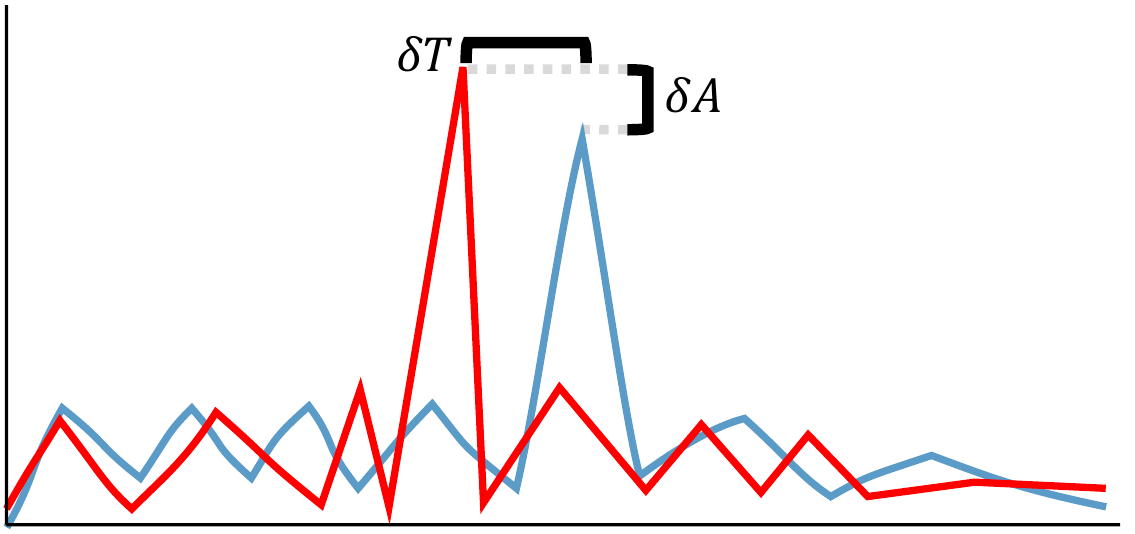}\hspace{10pt}}

        \subfloat[{\scriptsize Matching Peaks and Valleys}\label{fig:measures:peaks}]{\hspace{10pt}\includegraphics[width=0.4\linewidth]{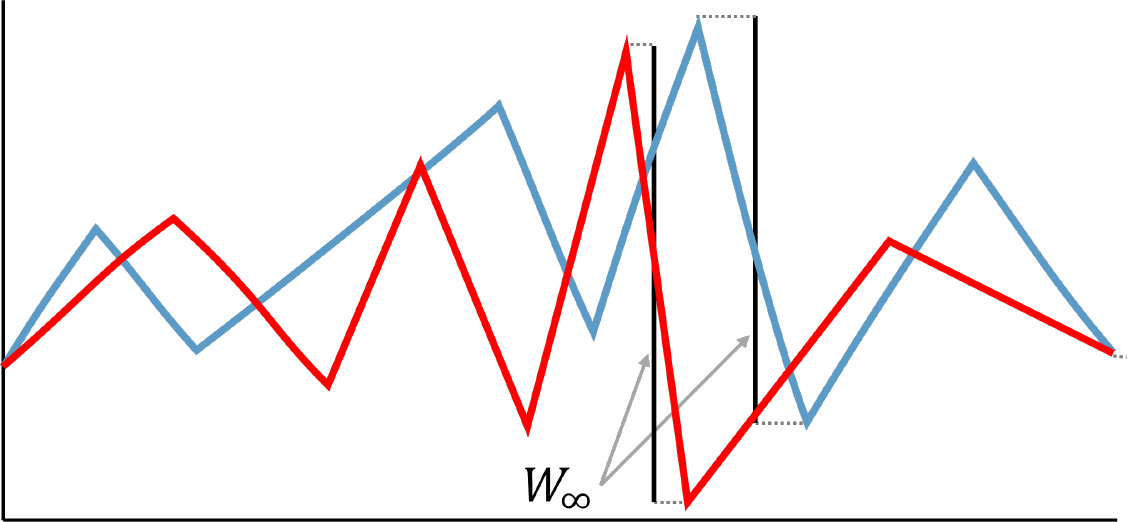}\hspace{10pt}}
        \subfloat[{\scriptsize Measuring Noise Levels}\label{fig:measures:volume}]{\hspace{10pt}\includegraphics[width=0.4\linewidth]{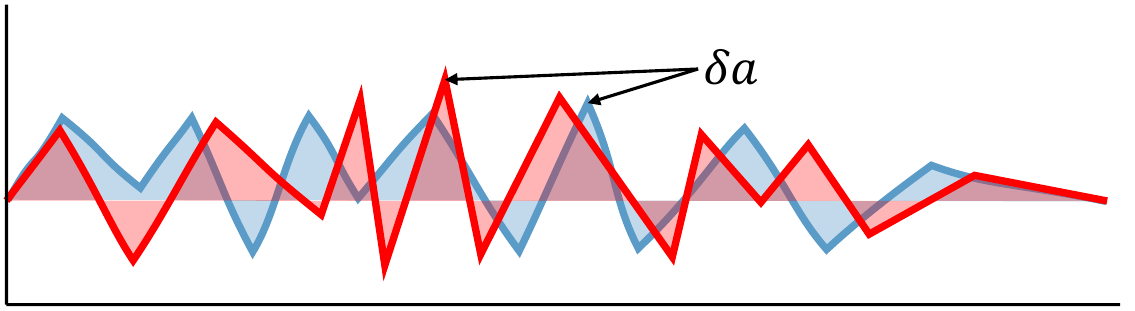}\hspace{10pt}}
    
     \caption{Illustration of similarity measurements used in our approach. (a)~shows measuring trend similarity with a total variation using $\lnorm_{2}$. (b) illustrates comparing the period and amplitude, where $\delta T$ denotes the difference in the number of periods, and $\delta A$ denotes the difference in amplitude between the blue and red signals. (c) illustrates matching the peaks \& valleys in the data, where $W_{\infty}$ measures the largest variation pointed with arrow shows. (d) shows measuring noise levels with area preservation, where $\delta a$ measures the difference between the two shaded areas. }
    \label{fig:measures}
\end{figure}

After converting participant sketches into time series with equal data points to the stimuli, we quantified how well they preserved features discussed in \Cref{sec:background}, including trends, periodic patterns, prominent peaks and valleys, and noise. However, measuring time series similarity is complex, as prior work shows the human perception of similarity is subjective~\cite{gogolou2018comparing, eichmann2015evaluating}. Ding et al.~\cite{ding2008querying} found no universally superior similarity measure, as accuracy is often dataset and domain-dependent. Since selecting metrics tailored to data and domain was beyond our scope, we opted for generalizable metrics.

    \paragraph{Estimating and Measuring Trends} The ground truth trend in real-world datasets is often unavailable~\cite{wang2017line}, which is why it is estimated. We pursued two methods for trend estimation: frequency domain analysis and LOESS curve fitting. We used Fast Fourier Transformation (FFT) for frequency domain analysis on each line chart (see \Cref{fig:fft_decomposition:fft}), applying a low pass filter to approximate the trend. Specifically, we considered only the first one-third of the log scale frequency components, yielding an estimated trend as depicted in \Cref{fig:fft_decomposition:trend_fft}. As an additional validation, we utilized the LOESS curve fitting technique (see \Cref{fig:trend_loess}), inspired by the work of Wang et al.~\cite{wang2017line} to captures the deterministic component of variation in a time series. Consistent with their methodology, we employed a span value of 0.4 to capture large-scale patterns.

    We evaluated several common metrics (\( L^\infty \), \( L^2 \), \( L^1 \)-norms, and Dynamic Time Warping (DTW)), observing consistent results across all (see supplemental material). We report \( L^2 \)-norm due to its widespread use in assessing overall differences in continuous data and its sensitivity to cumulative variation. \( L^2 \)-norm, 
    $\lnorm_{2}(X,Y)=\sqrt{\sum_{i=1}^{n} (x_i - y_i)^2}$ quantifies the overall Euclidean distance between the trends of the stimuli and participant sketches. In \Cref{fig:measures:variation}, it is shown as the cumulative squared differences across all points in the lines.

    \paragraph{Estimating and Measuring Periodicity} In datasets featuring periodic patterns, researchers often utilize frequency domain analysis, as indicated by prior studies~\cite{musbah2020novel, musbah2019identifying}. 
    According to Musbah et al.~\cite{musbah2019identifying}, significant peaks in the FFT represent frequencies corresponding to the periodicity in the signal. Following this methodology, we identified the maximum frequency component in the FFT for line charts with periodic characteristics. We transformed it back to the time domain, as depicted in \Cref{fig:fft_decomposition:periodic_fft}. We calculated the amplitude and period difference between the estimated periodicity from the stimuli and participant sketches as illustrated in \Cref{fig:measures:period_amplitude}. The estimated periodicity in the stimuli is represented by blue line and the participant sketch by a red line. If the number of period in the blue line is ${T_1}$ and ${T_2}$ in the red line, the difference in period is $\delta T = |T_1 - T_2|$.
        Similarly, If the amplitude of the blue line is ${A_1}$ and ${A_2}$ in the red line, the difference in amplitude is $\delta A = |A_1 - A_2|$.

    \paragraph{Estimating and Measuring Peaks and Valleys} To identify peaks and valleys, we utilize techniques from Topological Data Analysis. Peaks and valleys can be defined as the local minima and maxima~\cite{rosen2020linesmooth, palshikar2009simple}. The local minima and maxima from a line chart are paired using a process described in detail in~\cite{suh2019topolines} and placed into an object called a persistence diagram~\cite{EdelsbrunnerHarer2010}. \Cref{fig:measures:peaks} shows examples of such peaks.
    To compare line charts, the persistence diagrams from the two line charts, $\overline{X}$ and $\overline{Y}$ are compared using bottleneck distance, which identifies the peaks with the maximum variance. Let $\eta$ be a bijection between the two sets, 
            The bottleneck distance is $ W_{\infty}(\overline{X},\overline{Y}) = \inf_{\eta: \overline{X} \rightarrow \overline{Y}} \sup_{\overline{x} \in \overline{X}}  \left\| \overline{x}-\eta(\overline{x}) \right\|_\infty$. 

    \paragraph{Estimating and Measuring Noise} Noise can be considered as the high-frequency component in the data. Therefore, we employed FFT to identify and isolate the noise components as suggested by previous research~\cite{musbah2019identifying, musbah2020novel}. To ensure a consistent and generalized interpretation of noise, we applied a high-pass filter to the FFT. Specifically, we focused on the last one-third of the frequency components and transformed them back to the time domain, as illustrated in~\Cref{fig:fft_decomposition:noise_fft}.

    The area preserved was calculated for the estimated noise (see \Cref{fig:fft_decomposition:noise_fft}). The change in area, $\delta a(X,Y)=\left|\sum x_{i}- \sum y_{i}\right|$, was computed by taking the difference between the stimuli~($X$) and the participant sketch ($Y$). The process is shown in \Cref{fig:measures:volume}. The sum of the blue bars represents the area covered by the estimated noise in the stimuli, and the sum of the red bars indicates the participant sketches. 
   In addition, a second metric was used to estimate the noise, namely Pixel Approximate Entropy (PAE). PAE is a measure of the perceptual complexity in line charts~\cite{2018entropy}, and served as an indirect measure of noise level, with higher entropy values indicating greater noise.

\begin{figure*}[ht]
     \centering
   \begin{minipage}{0.65\linewidth}
        \subfloat[Preserved visual features per cluster\label{fig:cluster_feature_upset_plot}]{%
        \includegraphics[trim=295pt 115pt 280pt 155pt, clip, width=0.6\linewidth]{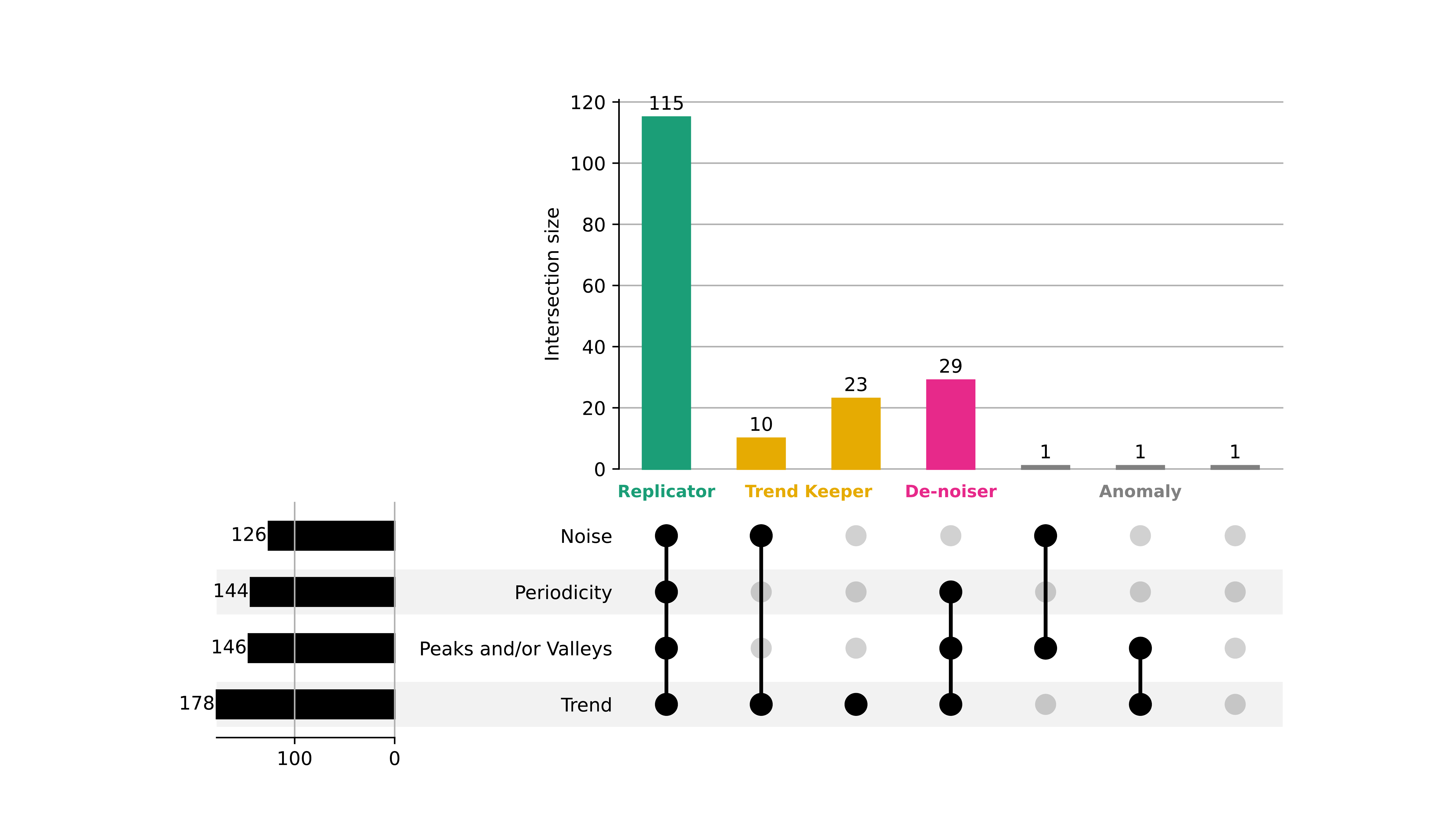}}%
    \hfill
    \subfloat[Clusters across participant sketches\label{fig:participant_cluster_distribution}]{%
        \includegraphics[width=0.35\linewidth]{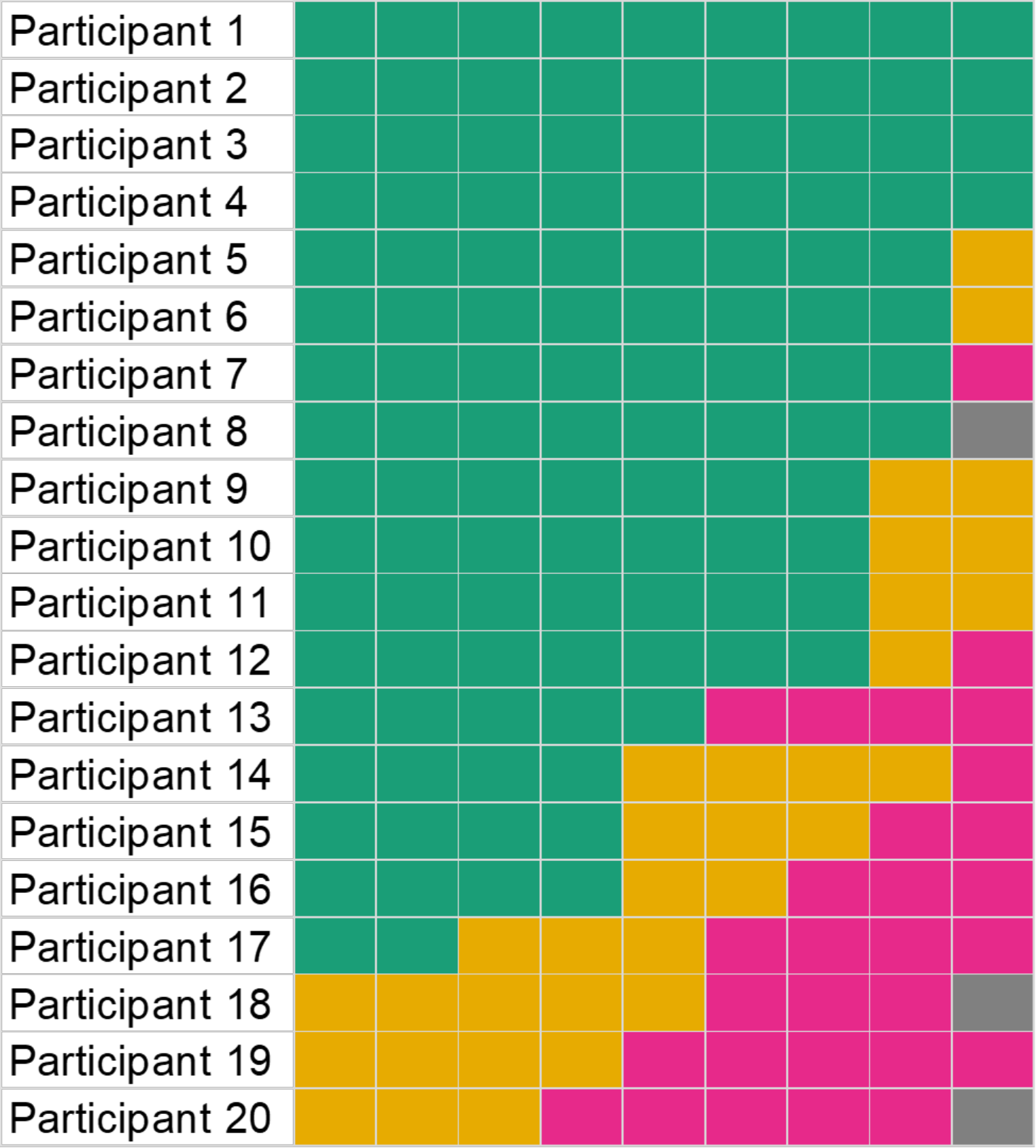}}%
   \end{minipage}
    \hfill
    \begin{minipage}{0.28\linewidth}
        \subfloat[Clusters across different noise levels\label{fig:cluster_across_noise}]{%
            \includegraphics[trim=100pt 20pt 25pt 15pt,width=\linewidth]{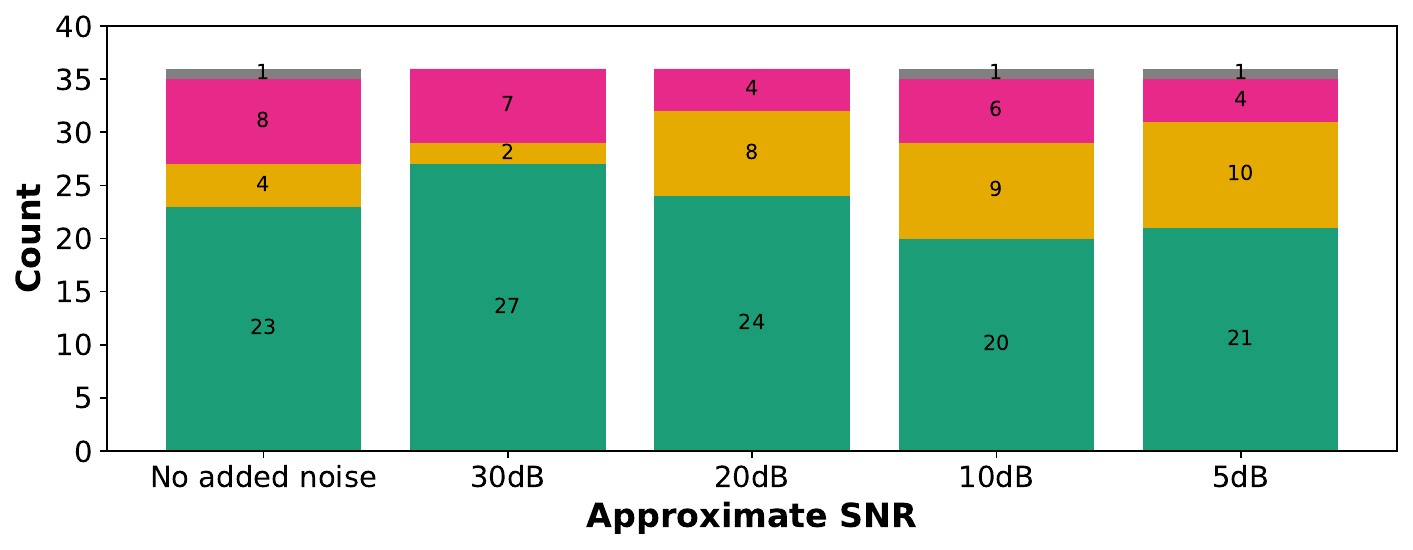}}%
        \vspace{1pt} %
        \subfloat[Clusters across different datasets\label{fig:cluster_across_datasets}]{%
            \includegraphics[trim=140pt 60pt 40pt 15pt,width=\linewidth]{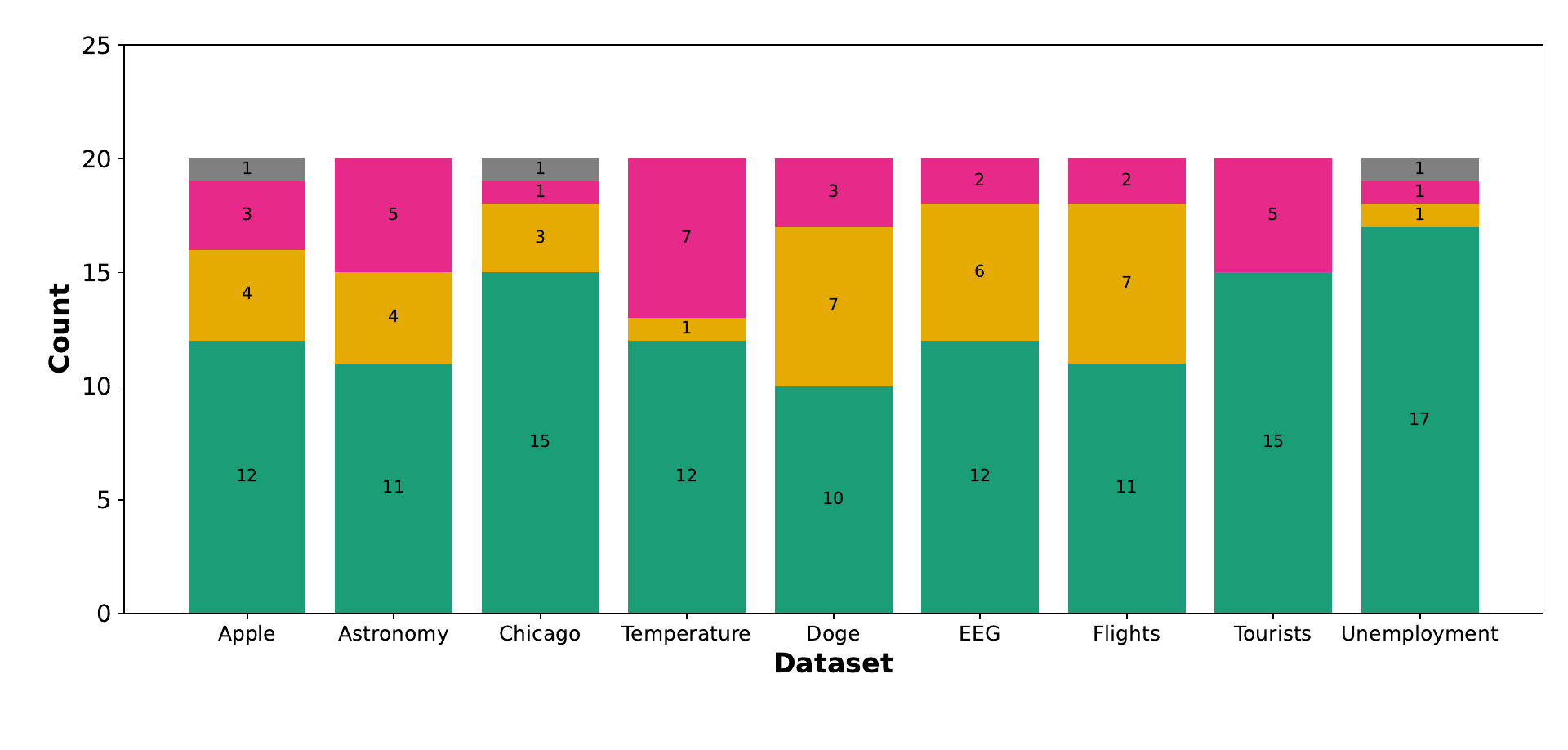}}%
    \end{minipage}
    
    \caption{The figure illustrates the distribution of different behavioral patterns—\textcolor{green}{Replicator}, \textcolor{yellow}{Trend Keeper}, \textcolor{pink}{De-noiser}, and \textcolor{gray}{Anomaly}—across features, participants, noise levels, and datasets. (a) An Upset plot~\cite{lex2014upset} of the visual features preserved by cluster type is shown. (b) The color-coded 2D grid depicts each participant's behaviors in their 9 sketches. It was organized in decreasing order to highlight the prevalence of \textcolor{green}{Replicator} behavior and subsequently \textcolor{yellow}{Trend Keeper}, \textcolor{pink}{De-noiser}, and \textcolor{gray}{Anomaly}. (c)~The distribution of these clusters at different noise levels is shown, where each bar signifies a distinct noise level. (d)~The cluster distribution across the 9 datasets.}
    \label{fig:heatmap_clusters}
\end{figure*}

\section{Results}
\label{results}
We present three main findings. First, the qualitative analysis revealed that individual \textbf{participants did not consistently exhibit the same behavior across all stimuli} but instead clustered into three behavior types: \textcolor{green}{Replicator}, when they replicated all features of the stimuli including noise in sketches; \textcolor{yellow}{Trend Keeper}, when they retained only the general trend in their sketches with or without noise; and the \textcolor{pink}{De-noiser}, when they removed the noise from the shown stimuli and retained all the other features in sketches.

Second, we found that viewers \textbf{retained most features in line charts across varying noise levels}. However, the same noise level introduced different amounts of visual clutter depending on the dataset. Despite this, most participants effectively retained features, showing robustness to noise.

Finally, while participants retained features, they did so either \textbf{\textit{faithfully}} (with low quantitative error) or \textbf{\textit{semantically}} (with higher quantitative error but capturing the presence of features qualitatively). For example, in some sketches, the \textit{amplitude} and \textit{period} of the periodic pattern differed from the stimuli, yet they still conveyed the \textit{presence} of these features.

\subsection{Clustering Participant Behavior}
\label{clusters}

Due to the free‑form nature of participant sketches, there is no reliable numeric measure to group them by shared drawing strategies, and the artifact removal and normalization required for any quantitative analysis would erase the nuances of sketches we aimed to capture. Consequently, we applied qualitative thematic analysis in \Cref{sec:qualitative} to identify behavior patterns directly from the sketches. Based on the analysis, we formed three clusters of participant behavior based on the inclusion or exclusion of visual features in sketches (see \Cref{fig:cluster_feature_upset_plot}):
\begin{itemize}[noitemsep,itemsep=4pt]
    \item \textcolor{green}{\textbf{Replicator}}: These sketches (e.g.,~\Cref{fig:astro_replicator}-\Cref{fig:flights_replicator}) generally maintained all features in the stimuli. Trends and periodicity (if present) were preserved either \textit{very well} or \textit{somewhat}, along with \textit{most} or \textit{some} peaks and valleys (if present). Noise was also replicated \textit{very well} or \textit{somewhat}. 

    \item \textcolor{yellow}{\textbf{Trend Keeper}}: These sketches
    (e.g.,~\Cref{fig:astro_trend_keeper}-\Cref{fig:flights_trend_keeper}) predominantly maintained trends \textit{very well} or \textit{somewhat}, but did not preserve periodicity, peaks, or valleys, even when these features were present in the stimuli. Noise was optionally preserved in these sketches.
    
    \item \textcolor{pink}{\textbf{De-noiser}}: These sketches (e.g.,~\Cref{fig:astro_de_noiser}-\Cref{fig:flights_de_noiser}) maintained all the features of the stimuli like trends and periodicity (if present in the stimuli) \textit{very well} or \textit{somewhat} and \textit{most} or \textit{some} of the peaks and valleys (if present in the stimuli) but removed the noise present in the stimuli. The only difference between a Replicator and a De-noiser is the presence of noise. 
\end{itemize}

Three \textcolor{gray}{\textbf{Anomaly}} sketches did not fit any cluster and were excluded from further analysis. Feature inclusion/exclusion for these sketches is shown in \Cref{fig:cluster_feature_upset_plot}, with additional figures in the supplemental materials.

Observations of these behavior patterns varied both across and within participants. \Cref{fig:participant_cluster_distribution} shows the clusters assigned to all participant sketches, sorted by the proportion of \textcolor{green}{Replicator}, \textcolor{yellow}{Trend Keeper}, and \textcolor{pink}{De-noiser} behavior. The majority of sketches, 64\% (115/180), were labeled \textcolor{green}{Replicator}, 18\% (33/180) \textcolor{yellow}{Trend Keeper}, and 16\% (29/180) \textcolor{pink}{De-noiser}. Four participants [P01-P04] consistently exhibited \textcolor{green}{Replicator} behavior across all stimuli, while others showed a mix of different behaviors in reaction to different stimuli.

\paragraph{Cluster Relationship to Noise and Dataset}
To understand variation in participant behaviors, we examined cluster distribution across noise levels (see \Cref{fig:cluster_across_noise}) and datasets (see \Cref{fig:cluster_across_datasets}). The analysis revealed a dependency of behaviors on both the dataset type and the noise level. As noise increases, \textcolor{green}{Replicator} and \textcolor{pink}{De-noiser} behaviors decrease, while \textcolor{yellow}{Trend Keeper} behavior increases (\Cref{fig:cluster_across_noise}).

\begin{figure}[!tb]
    \centering

    \subfloat[{\scriptsize \texttt{Astronomy}} \label{fig:astro_max}]{
        \includegraphics[width=0.45\linewidth]{figs/astro/astro_noise_max.pdf}
    }
    \hspace{5pt}
    \subfloat[{\scriptsize \texttt{Temperature}} \label{fig:climate_max}]{
        \includegraphics[width=0.45\linewidth]{figs/climate/climate_noise_max.pdf}
    }\\[10pt]  %

    \subfloat[{\scriptsize \texttt{Chicago}} \label{fig:chi_max}]{
        \includegraphics[width=0.45\linewidth]{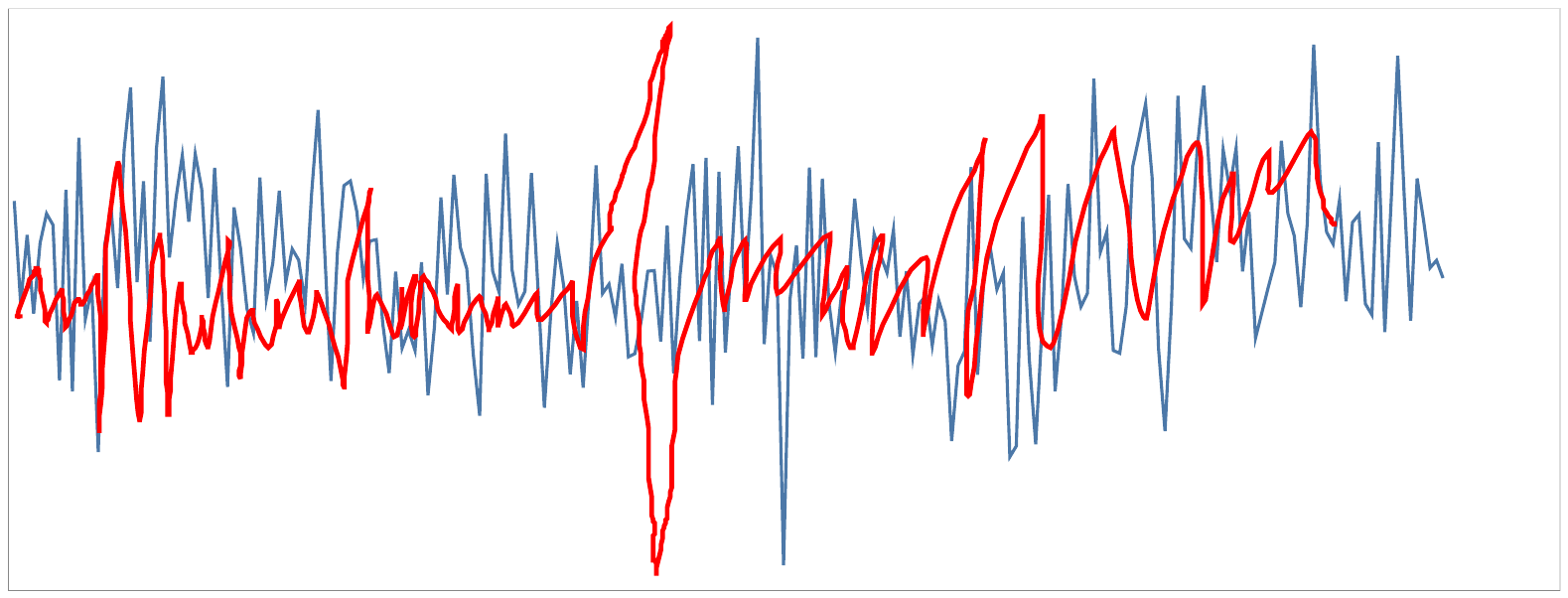}
    }
    \hspace{5pt}
    \subfloat[{\scriptsize \texttt{Unemployment}} \label{fig:unemployment_max}]{
        \includegraphics[width=0.45\linewidth]{figs/unemployment/unemployment_noise_max.pdf}
    }

    \caption{Participant sketches superimposed on their corresponding stimuli, with an SNR $\approx$ 5dB. In these datasets, participants demonstrated robustness to noise by capturing features, such as trends, periodicity, peaks, and valleys, in their sketches despite the presence of a significant level of noise. These sketches were classified as \textcolor{green}{Replicator}}
    \label{fig:dataset_with_overwhelmed_behavior}
\end{figure}

Among all the datasets,\texttt{Temperature} had the most number of \textcolor{pink}{De-noiser} sketches, while \texttt{Unemployment} had the most \textcolor{green}{Replicator} sketches (see \Cref{fig:cluster_across_datasets}). One possible reason behind this could be the nature of the datasets. \texttt{Temperature} dataset’s periodic pattern (see \Cref{fig:climate_max}) remained visible even at higher noise levels, making it less susceptible to noise. In contrast, the \texttt{Unemployment} (see \Cref{fig:unemployment_max}) dataset remained less cluttered under high noise, making replication easier. The \texttt{Tourists} dataset had no \textcolor{yellow}{Trend Keeper} sketches, as it lacked periodicity or prominent peaks and valleys (\Cref{tab:dataset}). As such, sketches preserving trends without noise fell into \textcolor{green}{Replicator} or \textcolor{pink}{De-noiser} clusters.

\begin{figure*}[!b]

    \begin{minipage}[b]{\linewidth}
        \centering
        \texttt{Tourists}

        \vspace{-5pt}
        \subfloat[{\scriptsize Trend estimated using FFT}\label{fig:tourists_fft_trend_l_2}]{%
            \includegraphics[width=0.325\linewidth]{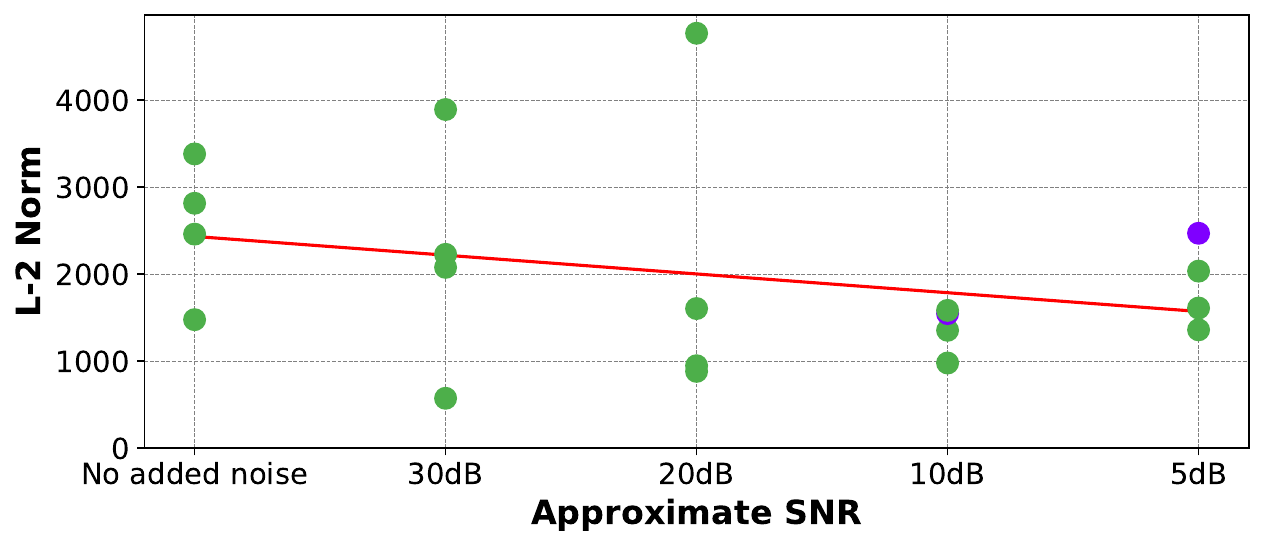}
        }
        \subfloat[{\scriptsize Trend estimated using LOESS}\label{fig:tourists_loess_trend_l_2}]{%
            \includegraphics[width=0.325\linewidth]{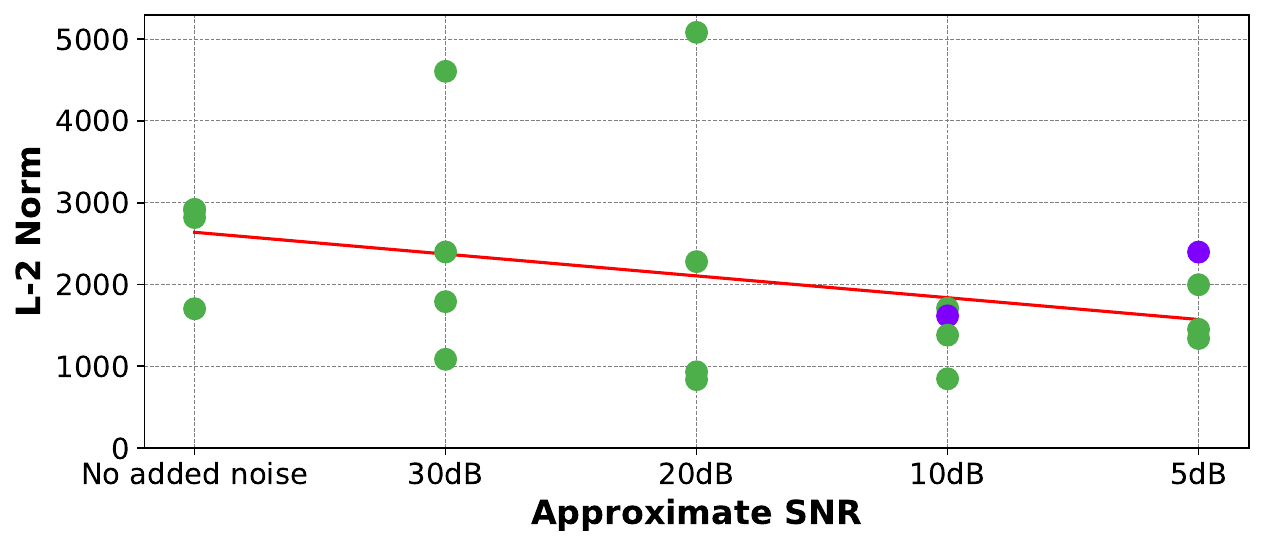}
        }
    \end{minipage}
    \vspace{10pt}

    \begin{minipage}[b]{\linewidth}
        \centering
        \texttt{Unemployment}

        \vspace{-5pt}
        \subfloat[{\scriptsize Trend estimated using FFT}\label{fig:unemployment_fft_trend_l_2}]{%
            \includegraphics[width=0.325\linewidth]{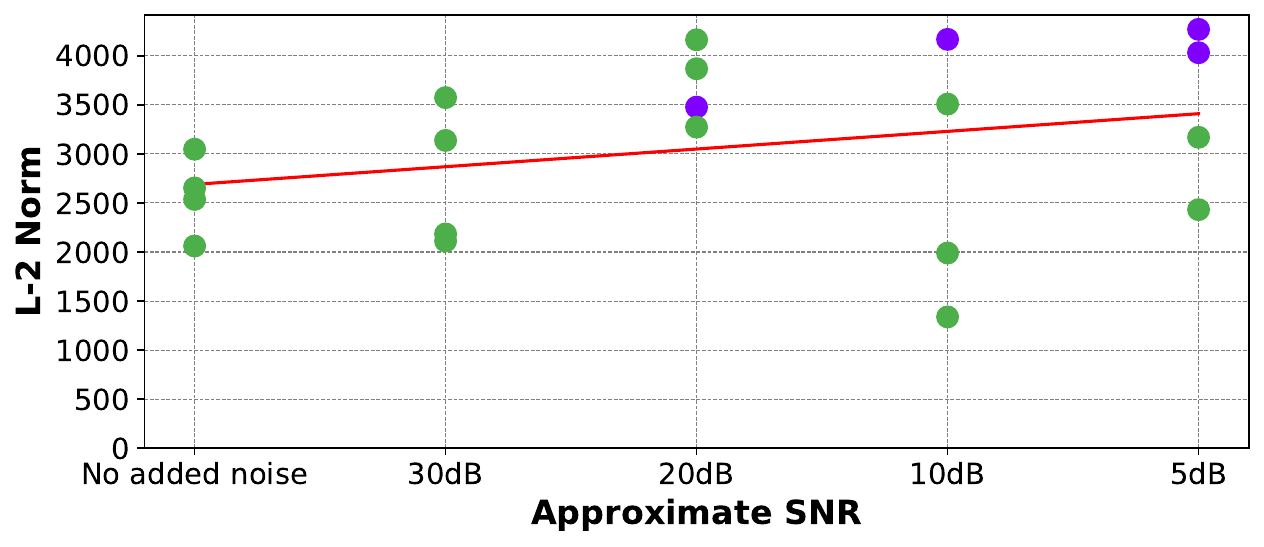}
        }
        \subfloat[{\scriptsize Trend estimated using LOESS}\label{fig:unemployment_loess_trend_l_2}]{%
            \includegraphics[width=0.325\linewidth]{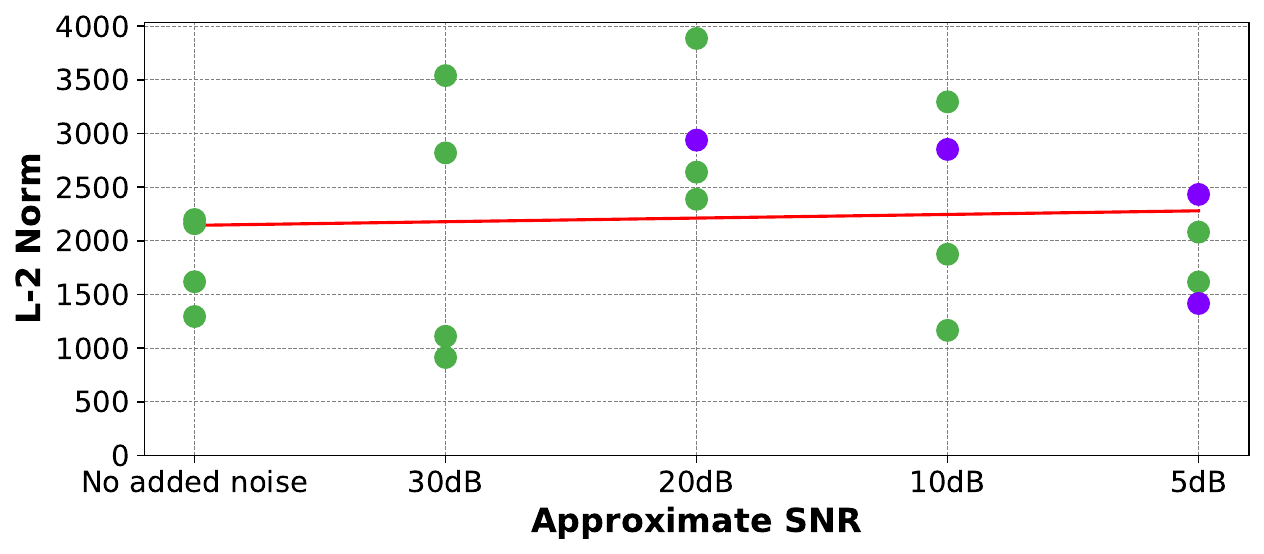}
        }
        
    \end{minipage}
    \begin{minipage}[b]{\linewidth}
        \centering
        \raisebox{0.8\height}{\includegraphics[width=0.2\linewidth]{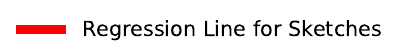}}
        \includegraphics[width=0.15\linewidth]{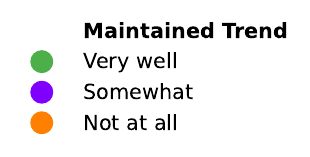}
    \end{minipage}
    
    \caption{Scatterplot of the $L^2$-norm of estimated trends via FFT and LOESS across \texttt{Tourists} and \texttt{Unemployment} dataset. Each point represents the $L^2$-norm between the participant sketch and the stimuli. The point colors correspond to the qualitative coding for the associated participant sketch.}
    \label{fig:trend_l_2_plot}
\end{figure*}

\subsection{Robustness to Noise}
\label{robustnes_to_noise}

We observed varying participant robustness to noise, with feature preservation differing across datasets. We calculated a regression line per dataset and quantitative metric to measure error increases in participant sketches as noise increases (e.g., red lines in \Cref{fig:trend_l_2_plot}). The slope between no noise and max noise (SNR $\approx$ 5dB) represents the percentage error increase. Results for all datasets, features, and metrics are in \Cref{tab:error_increment}. To illustrate distinct slope patterns (upward, downward, or constant), we included plots from specific datasets in the paper, with additional plots available in the supplemental materials.

\paragraph{Trend Robustness to Noise}
\label{trend_robustness_to_noise}
\Cref{fig:tourists_fft_trend_l_2} and \Cref{fig:tourists_loess_trend_l_2} show downward regression lines, indicating reduced error (difference between the estimated trend in stimuli and participant sketches) as noise increases in the \texttt{Tourists} dataset. Conversely, \Cref{fig:unemployment_fft_trend_l_2} and \Cref{fig:unemployment_loess_trend_l_2} show upward trends, indicating increased error for the \texttt{Unemployment} as noise increased.

As seen in \Cref{tab:error_increment}, apart from the \texttt{Temperature} and \texttt{Unemployment} datasets, most datasets show flat or even lower trend preservation errors as noise increases.
This suggests that trend preservation depends on both dataset characteristics and noise levels. However, given the general error reduction across most datasets for substantial increases in noise, we conclude that participants' trend preservation was generally robust to noise.

\begin{table}[!hb]
    \centering
    \caption{The percent change in error for participant sketches for trends, periodicity, and peaks and valleys from noise-free to max noise stimuli for all datasets. Negative values (i.e., decreased error with increasing noise) are highlighted.}
    \label{tab:error_increment}
    \resizebox{0.975\linewidth}{!}{%
    \begin{tabular}{c|rr|rr|r}
        \multirow{2}{*}{Dataset}   & \multicolumn{2}{c|}{Trend}          & \multicolumn{2}{c|}{Periodicity}        & Peaks/Valleys   \\ 
        \cline{2-6} 
                                   & \multicolumn{1}{c}{FFT}    & LOESS   & Amplitude & Period & \multicolumn{1}{c}{Bottleneck} \\ \hline
        \texttt{Apple}             & \cellcolor{SkyBlue!30}-41.7\%  & \cellcolor{SkyBlue!30}-31.6\%  &                    &        & \cellcolor{White!30}61.3\%      \\ \hline
        \texttt{Astronomy}         & \cellcolor{SkyBlue!30}-20.8\%  & \cellcolor{White!30}26.4\%  &                    &        & \cellcolor{SkyBlue!30}-3.0\%     \\ \hline
        \texttt{Chicago}           & \cellcolor{White!30}0.93\%    & \cellcolor{SkyBlue!30}-5.4\%  & \cellcolor{SkyBlue!30}-41.7\%  & \cellcolor{SkyBlue!30}-89.5\%  & \cellcolor{White!30}44.0\%     \\ \hline
        \texttt{Temperature}       & \cellcolor{White!30}9.7\%     & \cellcolor{White!30}37.4\%   & \cellcolor{SkyBlue!30}-47.7\%  & 0\%                                &            \\ \hline
        \texttt{Doge}              & \cellcolor{SkyBlue!30}-8.8\%   & \cellcolor{White!30}3.2\%    &                    &        & \cellcolor{SkyBlue!30}-23.7\%     \\ \hline
        \texttt{EEG}               & \cellcolor{SkyBlue!30}-11.0\%  & \cellcolor{SkyBlue!30}-9.1\%  &                    &        & \cellcolor{SkyBlue!30}-18.6\%     \\ \hline
        \texttt{Flights}           & \cellcolor{SkyBlue!30}-27.4\%  & \cellcolor{SkyBlue!30}-15.7\% &                    &        & \cellcolor{White!30}85.0\%       \\ \hline
        \texttt{Tourists}          & \cellcolor{SkyBlue!30}-35.4\%  & \cellcolor{SkyBlue!30}-40.4\% &                    &        &                                     \\ \hline
        \texttt{Unemployment}      & \cellcolor{White!30}26.8\%    & \cellcolor{White!30}6.3\%    & \cellcolor{SkyBlue!30}-46.6\%  & \cellcolor{SkyBlue!30}-56.7\%  & \cellcolor{White!30}49.22\%     \\ \hline
    \end{tabular}}
\end{table}

\begin{figure}[!b]

     \begin{minipage}[b]{\linewidth}
        \centering
        \texttt{Temperature}

        \vspace{-5pt}
        \subfloat[{\scriptsize Amplitude difference}\label{fig:amplitude_plot_climate}]{%
            \includegraphics[width=0.75\linewidth]{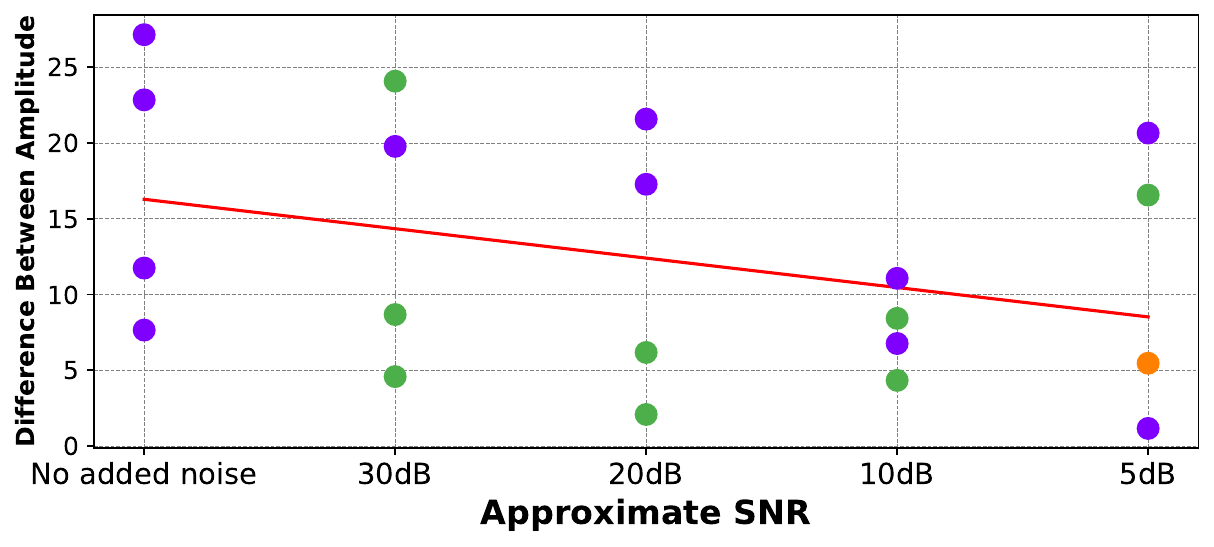}
        }
        \vspace{10pt}
        \subfloat[{\scriptsize Period difference}\label{fig:period_plot_climate}]{%
            \includegraphics[width=0.75\linewidth]{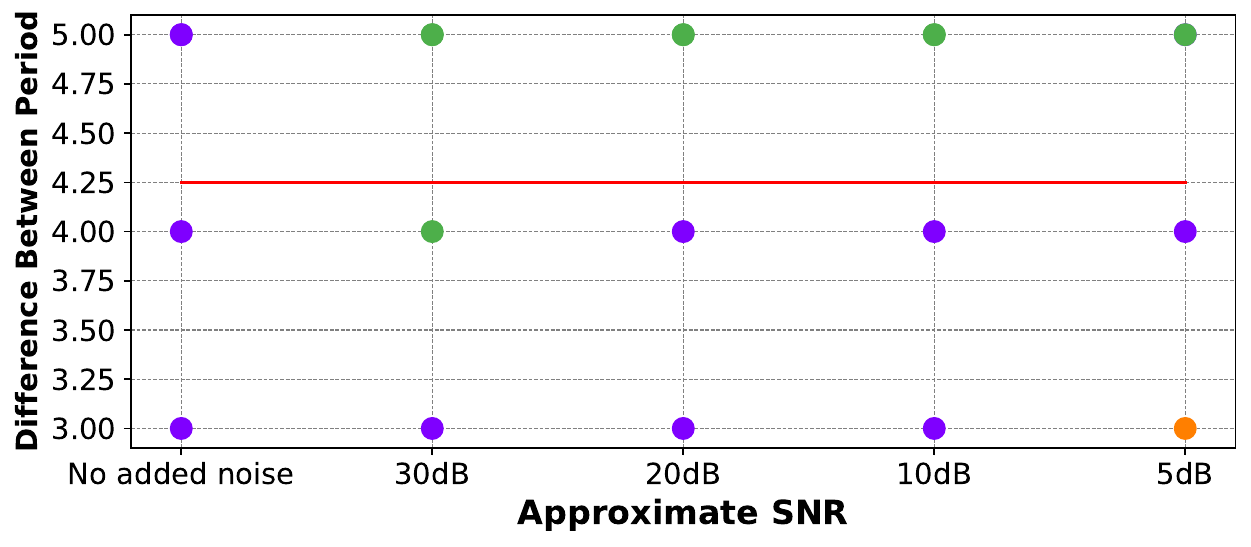}
        }
        
    \end{minipage}
    \begin{minipage}[b]{\linewidth}
        \centering
        \raisebox{0.8\height}{\includegraphics[width=0.45\linewidth]{figs/trend/regression_legend.pdf}}
        \includegraphics[width=0.35\linewidth]{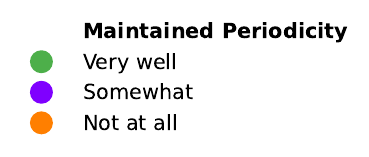}
    \end{minipage}

    \caption{Scatterplot displaying (a)~amplitude difference and (b)~period difference between all participant sketches and the stimuli across \texttt{Temperature} dataset. The data points are color-coded based on their assigned code in the qualitative analysis.}
    
    \label{fig:climate_period_amplitude_plot}
\end{figure}

\paragraph{Periodicity Robustness to Noise}
Participants identified and preserved periodicity in their sketches despite varying noise levels. \Cref{fig:climate_period_amplitude_plot} shows the amplitude and period difference between the stimuli and all the participant sketches for the \texttt{Temperature} dataset. The regression line in the amplitude plot (\Cref{fig:amplitude_plot_climate}) indicates a decrease in error with higher noise levels, while the period plot (\Cref{fig:period_plot_climate}) shows a consistent error rate. \Cref{tab:error_increment} also shows that for both \texttt{Chicago} and \texttt{Unemployment} datasets, error decreased for amplitude and period differences, suggesting a general insensitivity to noise levels when estimating amplitude and periodicity for these datasets.

Quantitative analysis confirms participants identified periodicity across noise levels, though the accuracy of period count or amplitude varied. As shown in \Cref{fig:cluster_across_datasets}, 53 of 60 sketches from periodic datasets (\texttt{Chicago}, \texttt{Temperature}, \texttt{Unemployment}) were classified as \textcolor{green}{Replicator} or \textcolor{pink}{De-noiser}, indicating participants’ robustness in identifying and reproducing periodicity despite noise.

\begin{figure}[!t]
    \centering
    \begin{minipage}[b]{0.75\linewidth} %
        \centering
        \texttt{Chicago}
        \vspace{-5pt}
        \subfloat{%
            \includegraphics[width=\linewidth]{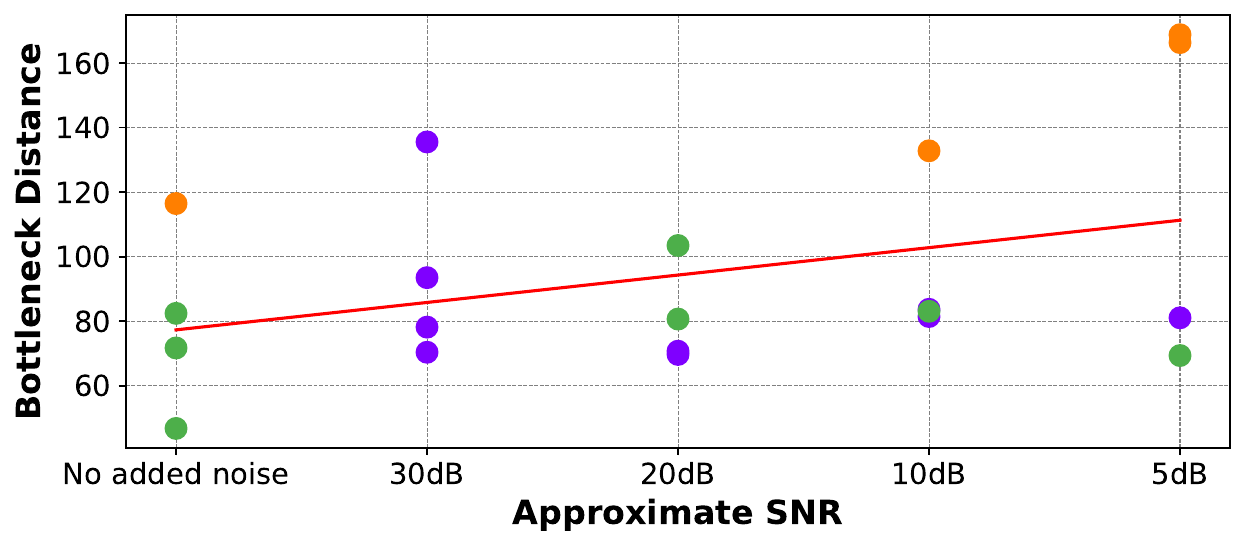}
        }
    
        \vspace{15pt}
        \texttt{EEG}
        \vspace{-5pt}
        \subfloat{%
            \includegraphics[width=\linewidth]{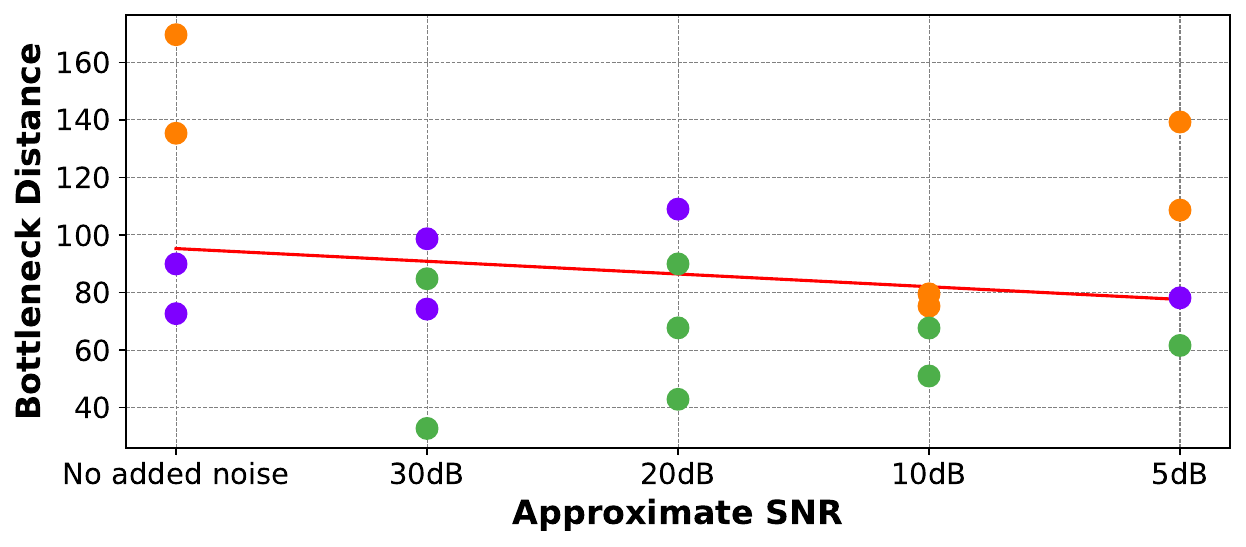}
        }
    \end{minipage}

      \begin{minipage}[b]{\linewidth}
        \centering
        \vspace{10pt}
        \raisebox{0.8\height}{\includegraphics[width=0.45\linewidth]{figs/trend/regression_legend.pdf}}
        \includegraphics[width=0.45\linewidth]{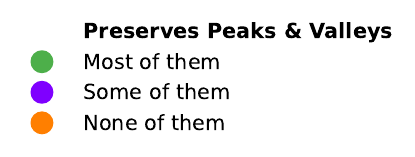}
    \end{minipage}

    \caption{Scatterplot displaying the Bottleneck distance between prominent peaks and valleys for \texttt{Chicago} and \texttt{EEG} dataset. Data point colors correspond to the qualitative coding for the associated participant sketch. The regression line in red is derived from participant sketches.}
    \label{fig:bottleneck_distance_plot}
\end{figure}

\paragraph{Peak and Valley Robustness to Noise}
Preserving peaks and valleys showed different patterns than trends and periodicity as noise increased. \Cref{fig:bottleneck_distance_plot} shows the bottleneck distance between the stimuli and participant sketches for two datasets. \Cref{fig:bottleneck_distance_plot}(\texttt{Chicago}) shows an increasing error with higher noise levels in the \texttt{Chicago} dataset, a tendency also seen in \texttt{Apple}, \texttt{Flights}, and \texttt{Unemployment} (\Cref{tab:error_increment}). In contrast, \Cref{fig:bottleneck_distance_plot}(\texttt{EEG}) shows decreasing error in the \texttt{EEG} dataset, a pattern also seen in \texttt{Doge} and \texttt{Astronomy} (\Cref{tab:error_increment}). As noise can create spurious peaks and valleys, only very prominent peaks remained visible with increased noise, making these features harder to preserve in some datasets.

In conclusion, feature preservation amidst noise varies by dataset. Datasets like \texttt{Astronomy}, \texttt{Chicago}, \texttt{Doge}, \texttt{EEG}, and \texttt{Tourists} show robustness, enabling accurate feature replication even under high noise. This may result from fewer data points or more prominent features. In contrast, \texttt{Apple}, \texttt{Temperature}, \texttt{Flights}, and \texttt{Unemployment} are more susceptible to noise, leading to greater errors. Peaks and valleys were harder to preserve in \texttt{Apple} and \texttt{Unemployment}, while trend preservation was more challenging in \texttt{Temperature} and \texttt{Unemployment}. The higher errors in \texttt{Apple} and \texttt{Flights} may be due to more data points, increasing visual clutter at lower SNR levels and blurring peaks and valleys (\Cref{fig:datasets_max_noise:apple_max}, \Cref{fig:datasets_max_noise:flights_max}).

\begin{figure*}[!t]
 \centering
    \begin{minipage}[b]{0.75\linewidth}
        \centering
        \subfloat[{\scriptsize Pixel Approximate Entropy}\label{fig:nz_approx_entropy}]{%
            \includegraphics[width=0.475\linewidth]{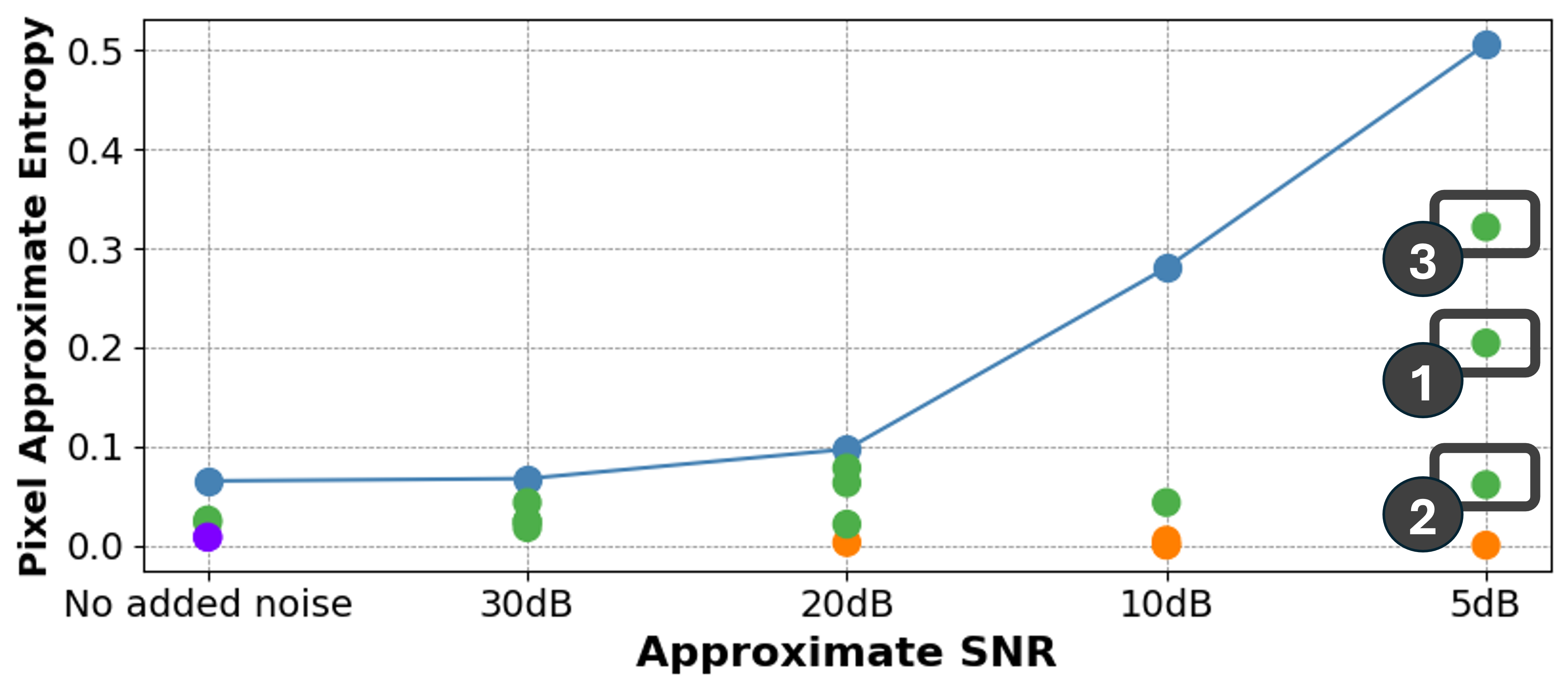}
        }
        \hfill
        \subfloat[{\scriptsize Difference of Area Preserved}\label{fig:nz_area_plot}]{%
            \includegraphics[width=0.475\linewidth]{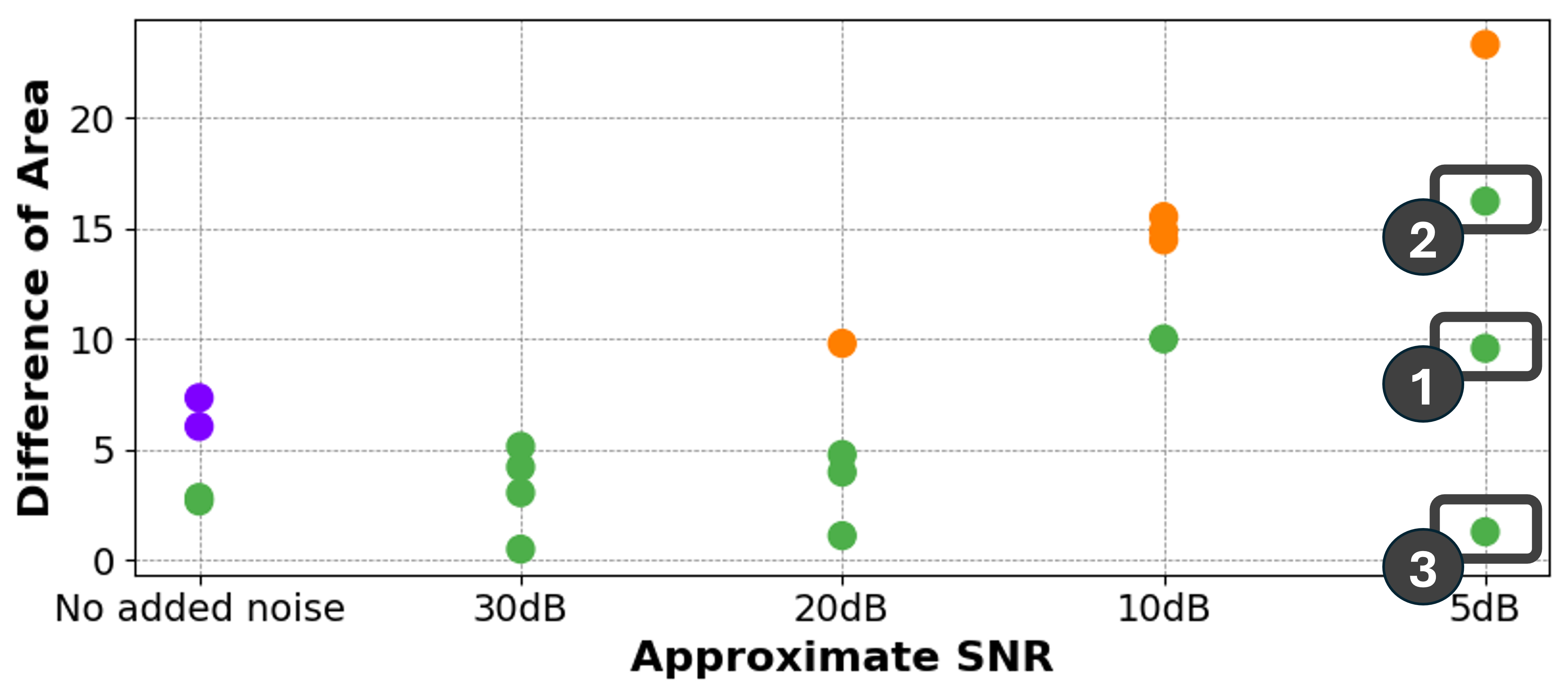}
        }
    \end{minipage}%
    \hfill
    \begin{minipage}[b]{0.25\linewidth}
        \centering
        \includegraphics[trim=40pt 0 0 0,width=0.25\linewidth]{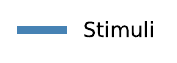}
        \includegraphics[trim=0 0 5pt 5pt,width=0.8\linewidth]{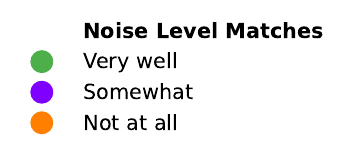}
    \end{minipage}

    \vspace{15pt} %

    \begin{minipage}[b]{0.3\linewidth}
        \centering
        \subfloat[{\scriptsize Semantic representation of noise in sketch 1}\label{fig:nz_semantic_1}]{%
            \includegraphics[width=\linewidth]{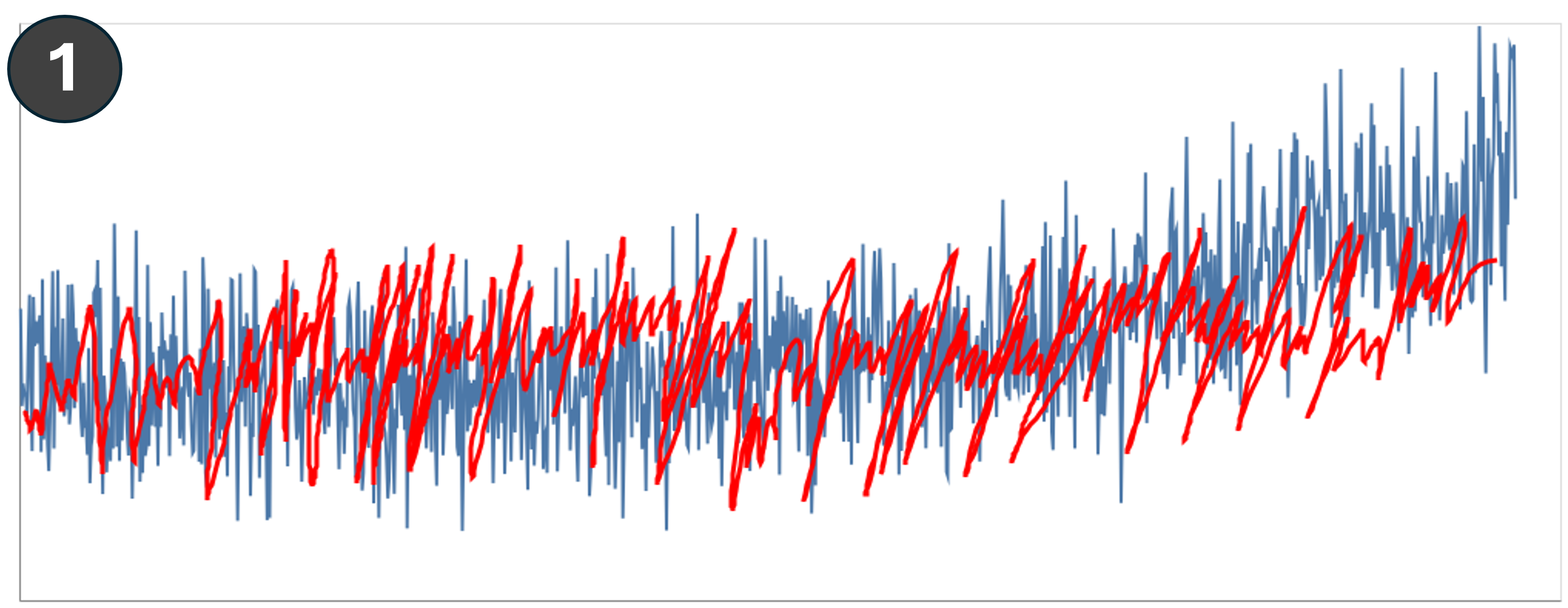}
        }
    \end{minipage}%
    \hfill
    \begin{minipage}[b]{0.3\linewidth}
        \centering
        \subfloat[{\scriptsize Semantic representation of noise in sketch 2}\label{fig:nz_semantic_2}]{%
            \includegraphics[width=\linewidth]{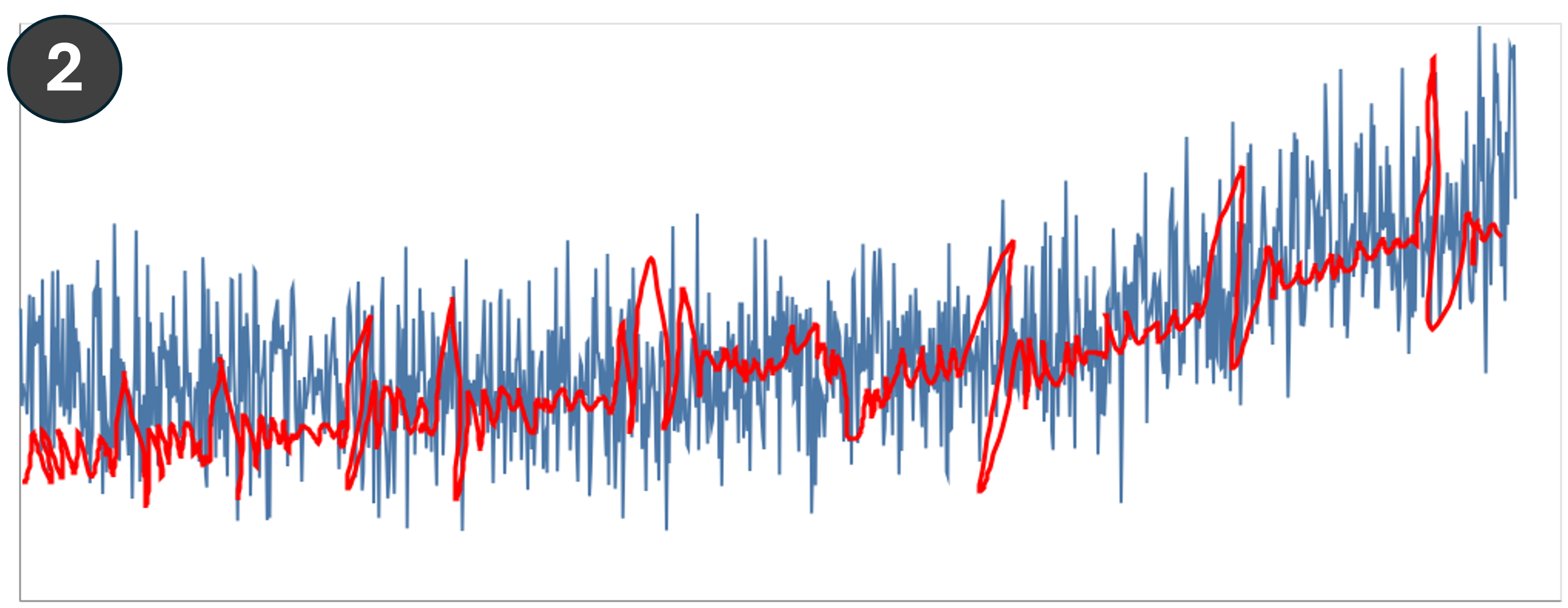}
        }
    \end{minipage}%
    \hfill
    \begin{minipage}[b]{0.3\linewidth}
        \centering
        \subfloat[{\scriptsize Faithful representation of noise in sketch 3}\label{fig:nz_faithful_3}]{%
            \includegraphics[width=\linewidth]{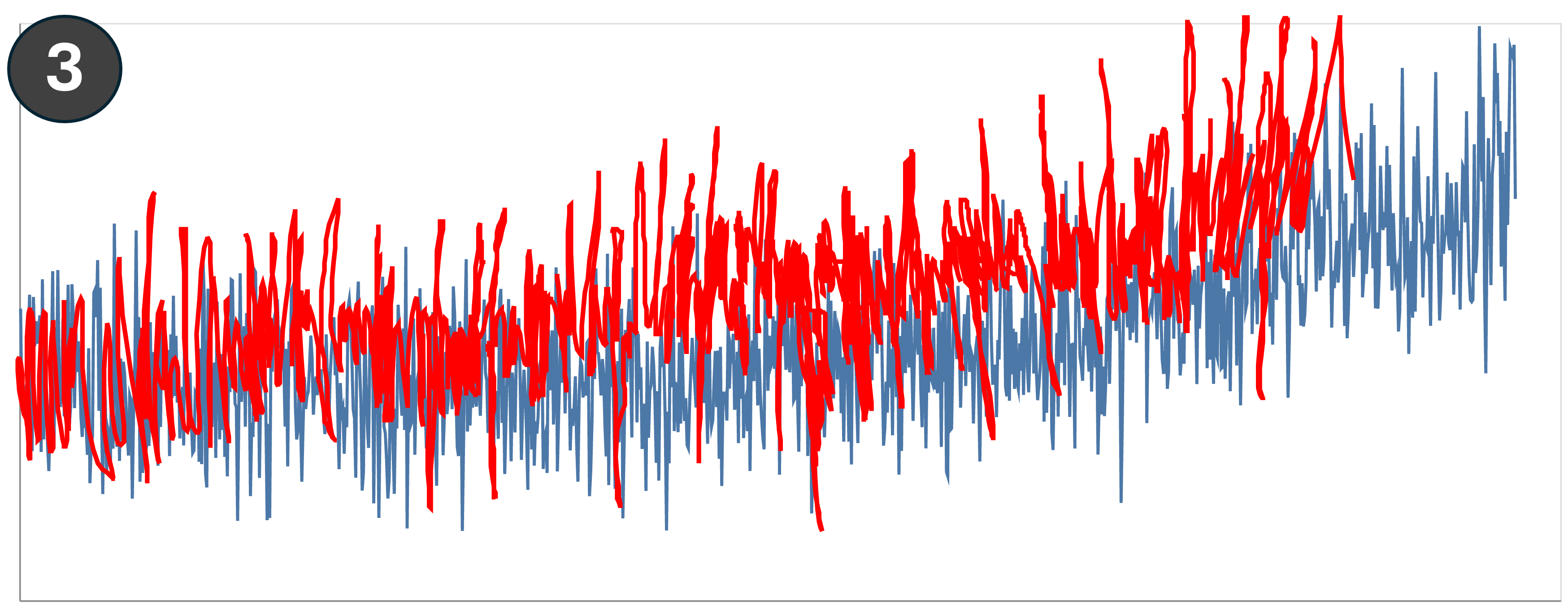}
        }
    \end{minipage}

    \caption{The (a)~Pixel Approximate Entropy (PAE) and (b)~Area Preserved by the estimated noise of all participant drawn sketches for the \texttt{Tourists} dataset. Data point colors denote assigned qualitative code. The blue line in (a) indicates PAE for stimuli at different noise levels. The annotated data points as {\tiny\textcircled{1}} and {\tiny\textcircled{2}} highlight two sketches (c-d), coded as closely matching the noise level in the stimuli. However, (a)~reveals these sketches have lower noise levels than their stimuli. In (b), the annotated data points confirm a discrepancy in noise preservation, suggesting a semantic interpretation of noise in these sketches. Whereas the annotated datapoint as {\tiny\textcircled{3}} highlights a sketch (e) which shows more faithful representation of noise.}
    \label{fig:nz_noise_rep_semantically}
\end{figure*}

\subsection{Semantic and Faithful Representation of Features}
The analyses in \Cref{clusters} show that participants preserved many salient features in their sketches. Where features were preserved accurately in sketches, we deem this representation \textit{faithful}.
Where features were represented visually, but in ways that diverge from their quantitative properties, we deem this representation \textit{semantic}.  Quantitative analysis indicated that periodicity and noise were mainly represented semantically, while trends, peaks, and valleys were more faithfully preserved. However, we acknowledge the ambiguity and subjectiveness between the \textit{semantic} and \textit{faithful} distinction.

\subsubsection{More Faithful Features}
Trends, Peaks and Valleys

\paragraph{Trends}
The qualitative analysis showed that participants effectively preserved the trends, with all clusters coded as either \textit{Very well} or \textit{Somewhat} preserving trends. \Cref{fig:trend_l_2_plot} shows the error in the estimated trends across noise levels for \texttt{Tourists} and \texttt{Unemployment}. As discussed in \Cref{trend_robustness_to_noise}, participants exhibited robustness to noise, consistently preserving trends across all datasets and noise levels.

\paragraph{Peaks and Valleys}
 As noted in \Cref{robustnes_to_noise}, participants were not consistently robust to noise when preserving peaks and valleys. However, \Cref{fig:bottleneck_distance_plot} shows that sketches retaining most peaks and valleys exhibited a lower Bottleneck distance, indicating a closer resemblance to the stimuli with lower error. Despite phase shifts in sketches where peak and valley locations differed, prominent peaks and valleys were faithfully represented, as seen in \Cref{fig:astro_max} and \Cref{fig:chi_max}. This tendency was also observed in other datasets. This indicates that the qualitative code of participant sketches regarding preserving prominent peaks and valleys matched the quantitative measurement, further supporting a faithful representation.

\subsubsection{More Semantic Features}
Periodicity and Noise

\paragraph{Periodicity}
Analysis in \Cref{robustnes_to_noise} showed that participants' sketches did not precisely replicate the amplitude and period of the stimuli. Yet, they consistently conveyed its periodic characteristics. \Cref{fig:amplitude_plot_climate} and \Cref{fig:period_plot_climate} show the difference in amplitude and number of periods preserved, respectively, for the \texttt{Temperature} dataset. For this dataset, the number of periods remained the same across all the stimuli despite varying noise levels. This shows an apparent discrepancy: despite the uniformity in the number of periods across all stimuli, the sketches predominantly did not capture the precise period of the stimuli. However, most sketches were coded as preserving periodicity \textit{Very well} or \textit{Somewhat}, with only one exception marked as \textit{Not at all}.  This suggests participants grasped the underlying periodicity of the data but did not reproduce it precisely, as demonstrated in \Cref{fig:climate_max}. This semantic representation pattern without exact replication was also observed in the \texttt{Chicago} and \texttt{Unemployment} datasets, which also exhibit periodic trends. %

\paragraph{Noise}
We observed a similar pattern in the participants' sketches for preserving noise. Using Pixel Approximate Entropy (PAE), we compared noise levels in the stimuli to those in sketches. \Cref{fig:nz_approx_entropy} shows PAE values for stimuli with various noise levels and corresponding participant sketches, with each data point for sketch color-coded based on qualitative coding. The figure reveals that captured noise in sketches was generally lower than in the stimuli, even when qualitatively coded as matching noise \textit{Very well}. This indicates participants tended to underrepresent the noise in their sketches, but semantically, they captured the noisiness as shown in \Cref{fig:nz_semantic_1} and \Cref{fig:nz_semantic_2}. However, noise levels in sketches were closer to stimuli noise at SNR $\approx$ 5dB (\Cref{fig:nz_area_plot}). One sketch \Cref{fig:nz_faithful_3} captured the noise in the shown stimuli both qualitatively and quantitatively, indicating a more faithful representation. However, semantic representation was more common across all datasets. Additional PAE values and area difference plots are available in the supplemental material.

\section{Validation Study Design}
\label{sec:followup}
Our primary visual stenography experiment had some limitations, such as a lack of verbal feedback from participants regarding their thought processes and imposed sketching constraints, such as using a single continuous stroke. Moreover, most people are not faithful artists~\cite{eitz2012humans} and struggle to depict what they want to convey accurately~\cite{cohen1997can}. To address these limitations, we designed a follow-up evaluation to validate the findings from the visual stenography experiment and gain a deeper insight into feature prioritization and reasoning. Participants were presented with reference stimuli and pre-generated sketches from the primary experiment. This approach allowed participants to select a well-drawn sketch that they believed best represented the important aspects of the reference stimuli, regardless of their sketching abilities.
\subsection{Experiment}
\label{sec:follow_up_evaluation:exp}

\paragraph{Stimuli}
\label{para:follow_up_stimuli}
Participants viewed nine reference stimuli (one per dataset) in a fixed order, followed by corresponding sketches we selected from the initial experiment. For each dataset, we selected sketches from the noise level with the most diversity in behavior cluster, excluding \textcolor{gray}{Anomaly}, which we skipped due to their rarity and lack of consistent patterns. We prioritized higher noise levels to explore noise perception. An exception was the \texttt{Flights} dataset: we selected SNR $\approx$ 20dB instead of SNR $\approx$ 5dB, as it uniquely produced \textcolor{yellow}{Trend Keeper} and \textcolor{pink}{De-noiser} behavior, with no \textcolor{green}{Replicator} behavior. 
For each task, one sketch per cluster type was shown.
When multiple sketches were available for a cluster and noise level, sketches were selected at random to reduce bias.
A figure illustrating the selected noise level for each dataset is provided in the supplemental material.

\paragraph{Task}
For this task, each reference stimulus was displayed for 10 seconds before disappearing, after which participants were presented with multiple sketches. Participants were asked the following questions:

\begin{enumerate}[label=Q\arabic*.]
    \item \textit{Based on your understanding of the reference image shown earlier, which one would you prefer to sketch to capture its important aspects? If none of them capture what you find important, feel free to say so.}
    \item \textit{What is the reason behind your choice?}
\end{enumerate}

Like the visual stenography task experiment described in \Cref{sec:study_design:exp}, participants were encouraged to select a sketch that preserved the significant features. It was emphasized that this was not a matching task but a choice based on their interpretation of important aspects. No specific criteria or guidelines for assessing similarity were provided to avoid influencing their judgment. Instead, we aimed to capture their intuitive perceptions of visual features, an approach aligned with methods used in prior studies~\cite{gogolou2018comparing, correll2016semantics}.

After completing the sketch selection for all nine reference stimuli, participants were directed to a follow-up questions page. This page had a draggable list of features, including trend, periodicity, peaks and valleys, and noise. They were asked the following additional questions:

\begin{enumerate}[label=Q\arabic*., start=3]
    \item \textit{When seeing a line chart, how do you prioritize its features? Drag and drop the following features to rank them from highest to lowest priority.}
    \item \textit{What other features do you think are important when viewing a line chart?}
\end{enumerate}

We asked participants for a general ranking of features, without providing dataset examples or task context, to assess their intuitive, context‐free prioritization of visual features.

The study was conducted online via Zoom. Consent was obtained before each session, and interviews were recorded. The experiment took around 20 minutes.
     
\paragraph{Participants}
12 of the 20 participants from the first study (2 females and 10 males) participated, with no compensation. All were graduate students. To minimize recall bias, we ran this validation study six months after the initial study. For clarity, participants in the validation study were re-anonymized using alphabetical labels.

\begin{figure}[!b]
        \centering
    \begin{minipage}{0.6\linewidth}
        \centering
        \subfloat[{\scriptsize Cluster across sketches chosen by participants}\label{fig:follow_up_participant_cluster_distribution}]{\includegraphics[width=\linewidth]{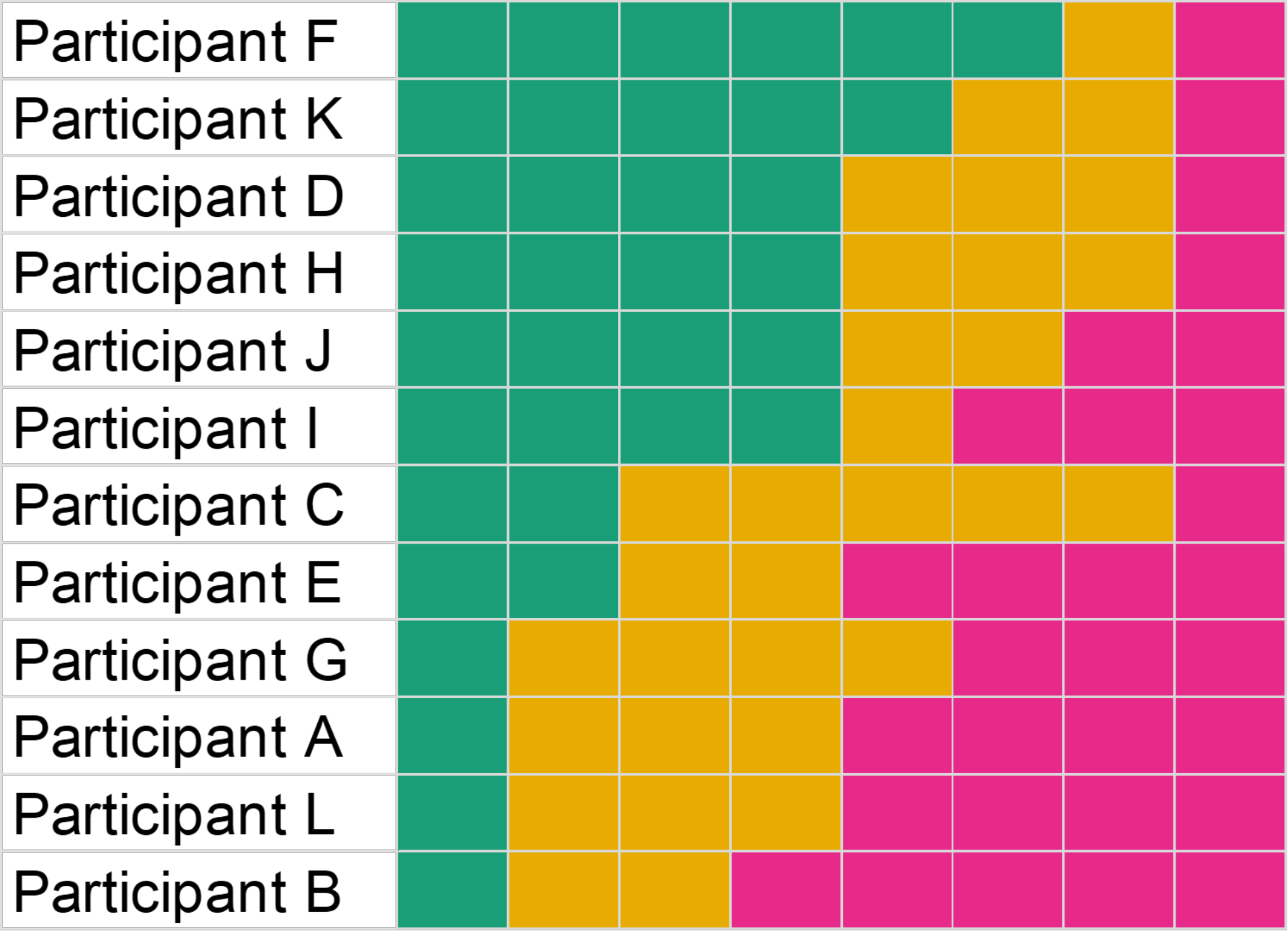}}
    \end{minipage}
    
    \begin{minipage}{0.8\linewidth}
         \vspace{10pt}
        \subfloat[{\scriptsize  Clusters across noise levels}\label{fig:follow_up_cluster_across_noise}]{\includegraphics[width=\linewidth]{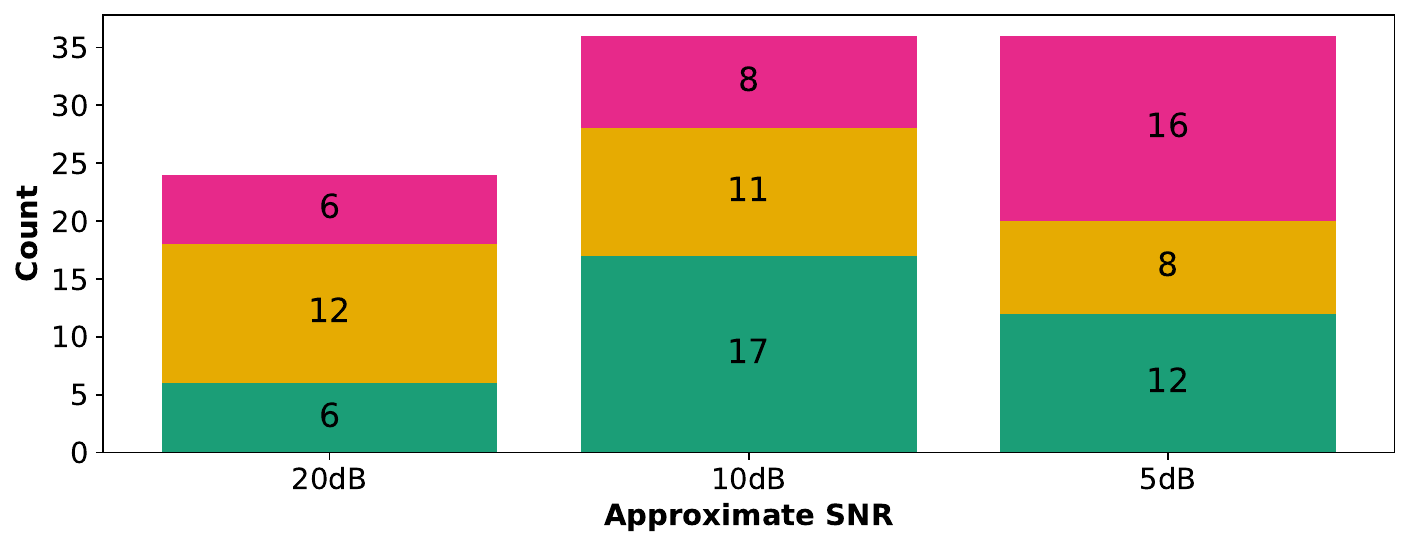}}

        \subfloat[{\scriptsize  Clusters across datasets}\label{fig:follow_up_cluster_across_datasets}]{\includegraphics[trim=0pt 0pt 0pt 80pt,width=\linewidth]{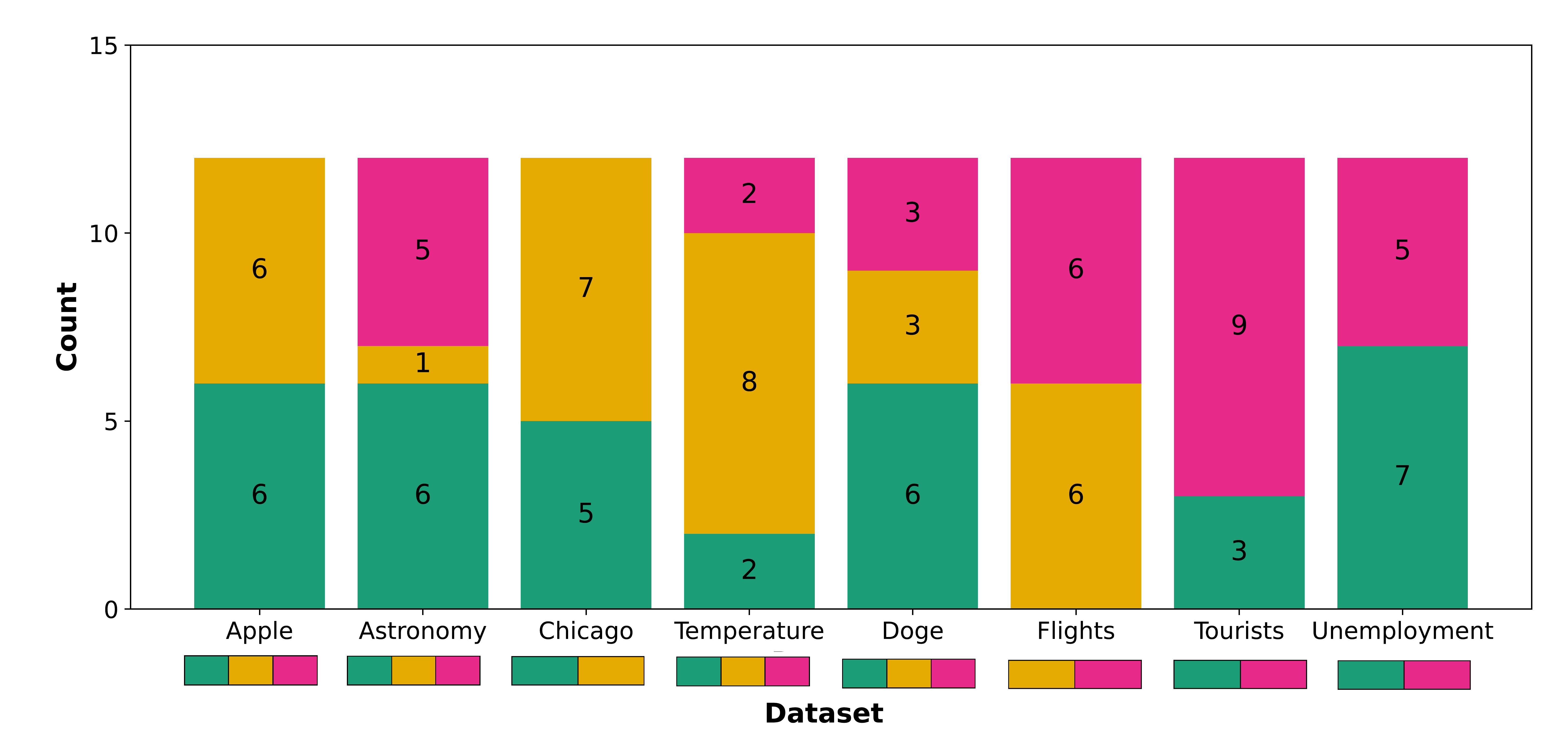}}        
    \end{minipage}

    \caption{ Chosen sketches from different behavioral patterns—\textcolor{green}{Replicator}, \textcolor{yellow}{Trend Keeper}, \textcolor{pink}{De-noiser}—across participants, noise levels, and datasets. (a)~A color-coded 2D grid depicting the assigned behavior pattern of the sketch chosen by each participant for all 8 stimuli. The data was arranged in descending order \textcolor{green}{Replicator}, \textcolor{yellow}{Trend Keeper}, and \textcolor{pink}{De-noiser} behaviors. (b)~The distribution of these clusters at different noise levels is shown, where each bar signifies a distinct noise level, with colors indicating the behavioral clusters. (c)~The cluster distribution across the 8 datasets. The color boxes below each dataset label represent stimuli sketch types associated with each dataset.}
    \label{fig:follow_up_heatmap_clusters}
\end{figure}

\subsection{Analysis}
\label{sec:follow_up_analysis}

We used a mixed-methods approach to analyze participant responses. Sketch choices were categorized into \textcolor{green}{Replicator}, \textcolor{yellow}{Trend Keeper}, and \textcolor{pink}{De-noiser} across participants, datasets, and noise levels. The feature priority rankings provided quantitative data, while participants' explanations required qualitative analysis to understand their thought processes.

\paragraph{Data Preprocessing}\label{sec:follow_up_evaluation:preprocessing} 
Responses were captured via Zoom transcripts, with one missing transcript generated using Descript. Transcripts were anonymized and organized by stimulus. 
An error in the \texttt{EEG} dataset selection (SNR $\approx$ 30dB instead of SNR $\approx$ 5dB) was discovered. As a result, it was excluded from the analysis, reducing it to eight datasets. All plots, including \texttt{EEG} data, are in the supplemental material.

\paragraph{Thematic Coding}
\label{sec:follow_up_evaluation:thematic_coding}
Participants' explanations for sketch choices were coded by one of the authors to identify alignment with behavior patterns (i.e., \textcolor{green}{Replicator}, \textcolor{yellow}{Trend Keeper}, or \textcolor{pink}{De-noiser}). Mixed behaviors or deviations from predefined clusters were noted. Feature ranking responses were analyzed to find common patterns. The anonymized transcripts and coding are in the supplemental material.

\subsection{Results}
\label{sec:follow_up_evaluation:res}

The follow-up study confirmed that participants' sketch choices varied across clusters, consistent with the initial visual stenography experiment. \Cref{fig:follow_up_participant_cluster_distribution} shows that \textcolor{green}{Replicator} sketches were chosen most frequently (35/96), followed by \textcolor{yellow}{Trend Keeper} (31/96) and \textcolor{pink}{De-noiser} (30/96). We did not ask participants whether they were prioritizing \textit{faithful} versus \textit{semantic} representations of features. We avoided asking this directly to prevent framing effects that might bias their selections, and so we cannot definitively know their underlying intentions. Nonetheless, the selections indicate that for periodicity and noise, participants favored sketches that conveyed those features \textit{semantically} (i.e., clearly signaled their presence), whereas for trends, peaks, and valleys, they leaned more toward \textit{faithful} representations.
Sketches from all clusters were selected across noise levels (\Cref{fig:follow_up_cluster_across_noise}) and datasets (\Cref{fig:follow_up_cluster_across_datasets}). However, the \texttt{Apple} dataset differed, with participants consistently choosing \textcolor{green}{Replicator} or \textcolor{yellow}{Trend Keeper} sketches despite being shown sketches from all clusters.

\begin{figure}[!b]
     \centering
        \begin{minipage}[t]{0.5\linewidth}
            \centering
            \includegraphics[trim=10pt 20pt 10pt 350pt,width=0.9\linewidth]{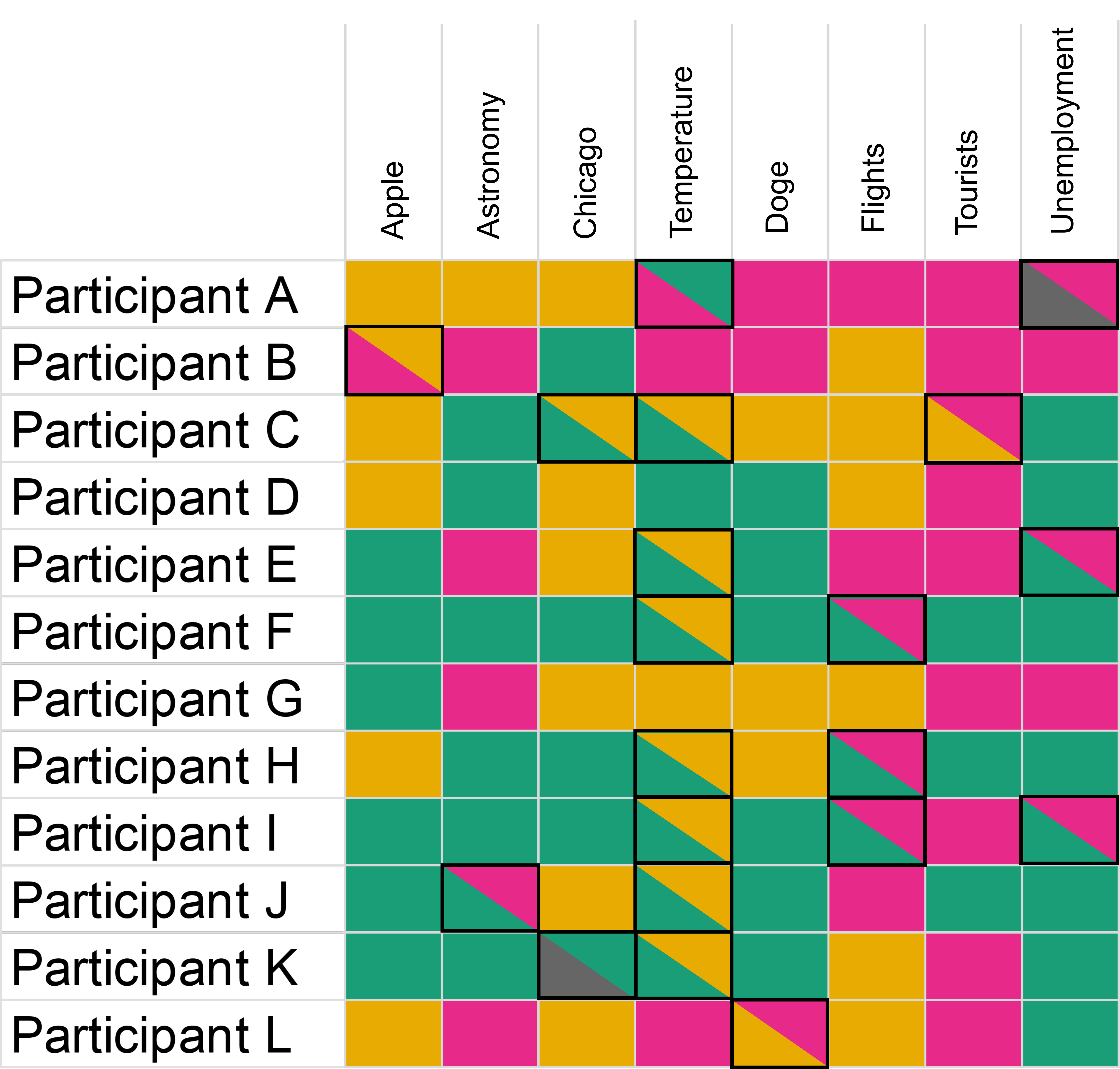}
        \end{minipage}%
        \hspace{5pt}
        \begin{minipage}[t]{0.4\linewidth}
            \centering
            \vspace{-30mm} 
            \includegraphics[trim=100pt 100pt 10pt 10pt,width=\linewidth]{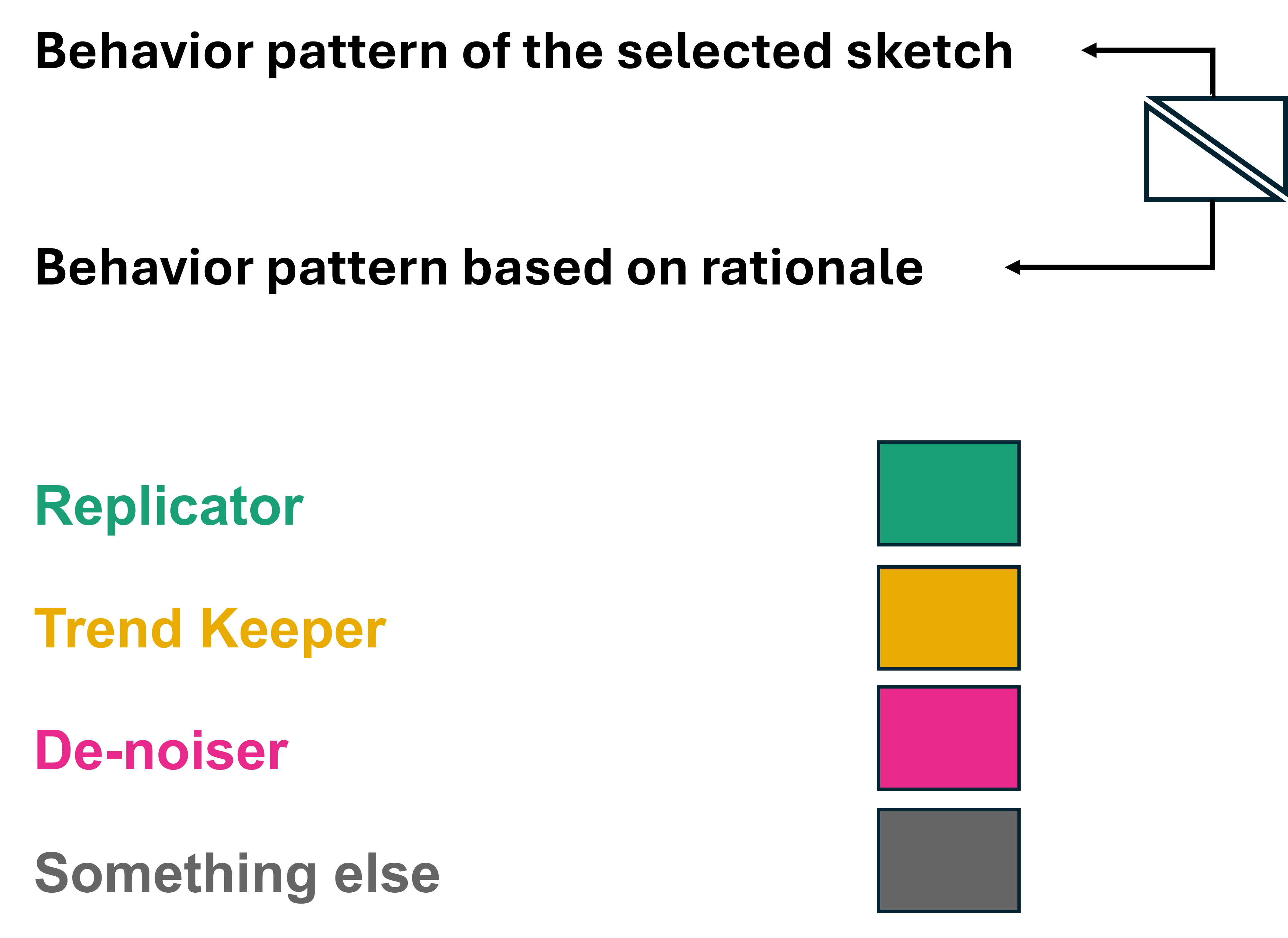}
        \end{minipage}
       
    \caption{The colored matrix compares the selected sketch's behavior pattern and the rationale for choosing them. Single-colored cells indicate a match between the behavior pattern between the selected sketch and rationale, while two-colored cells indicate a mismatch. The bottom triangle represents the behavior pattern demonstrated through the rationale, and the top triangle represents the assigned behavior pattern of the selected sketch during the primary study. Two \textcolor{darkgrey}{Gray} triangles indicate instances where the participant rationale did not align with predefined behavior patterns.}
    \label{fig:cluster_match}
\end{figure}

The alignment between chosen sketches and participants' rationale was analyzed qualitatively. \Cref{fig:cluster_match} shows that most sketch selections (76/96) matched participants’ rationale, while 20 instances showed mismatches, particularly in datasets like \texttt{Chicago}, \texttt{Temperature}, \texttt{Flights}, and \texttt{Unemployment}.  
The \texttt{Temperature} dataset had the most mismatches (8/12). In these cases, participants often provided rationale aligned with \textcolor{green}{Replicator} behavior but chose \textcolor{yellow}{Trend Keeper} sketches. Most participants selected sketches through elimination, rejecting those they found too noisy or insufficiently detailed. 
A similar elimination pattern was seen in \texttt{Chicago}, \texttt{Flights}, and \texttt{Unemployment} datasets. For \texttt{Chicago}, Participant C mentioned that their preference fluctuated between the simpler \textcolor{yellow}{Trend Keeper} and more detailed \textcolor{green}{Replicator} sketches. Regarding \texttt{Flights}, Participant H noted that both sketches captured key features, but the \textcolor{pink}{De-noiser} sketch was clearer.
Two instances did not align with predefined behavior patterns (\Cref{fig:cluster_match}). For \texttt{Unemployment}, Participant A focused on valley position over trends or noise, leading to a \textcolor{pink}{De-noiser} selection. For \texttt{Chicago}, Participant K saw no clear pattern in the reference image and chose the \textcolor{green}{Replicator} sketch's recognizable peaks and valleys.

These findings reveal two key insights: first, the alignment between sketch clusters and rationale clusters supports the validity of the visual stenography task, indicating that the sketches generally preserved the key features participants aimed to capture. Second, line chart interpretation is more nuanced than initially assumed. Participants' intentions often diverged from their final choices, suggesting that what they \textit{intended} to preserve did not always align with what they \textit{perceived} in the available sketches.

\paragraph{Visual Features in Line charts}

\begin{figure}[!h]
        \centering
        \begin{minipage}[t]{0.75\linewidth}
            \centering
            \includegraphics[trim= 50pt 10pt 50pt 30pt,width=0.9\linewidth]{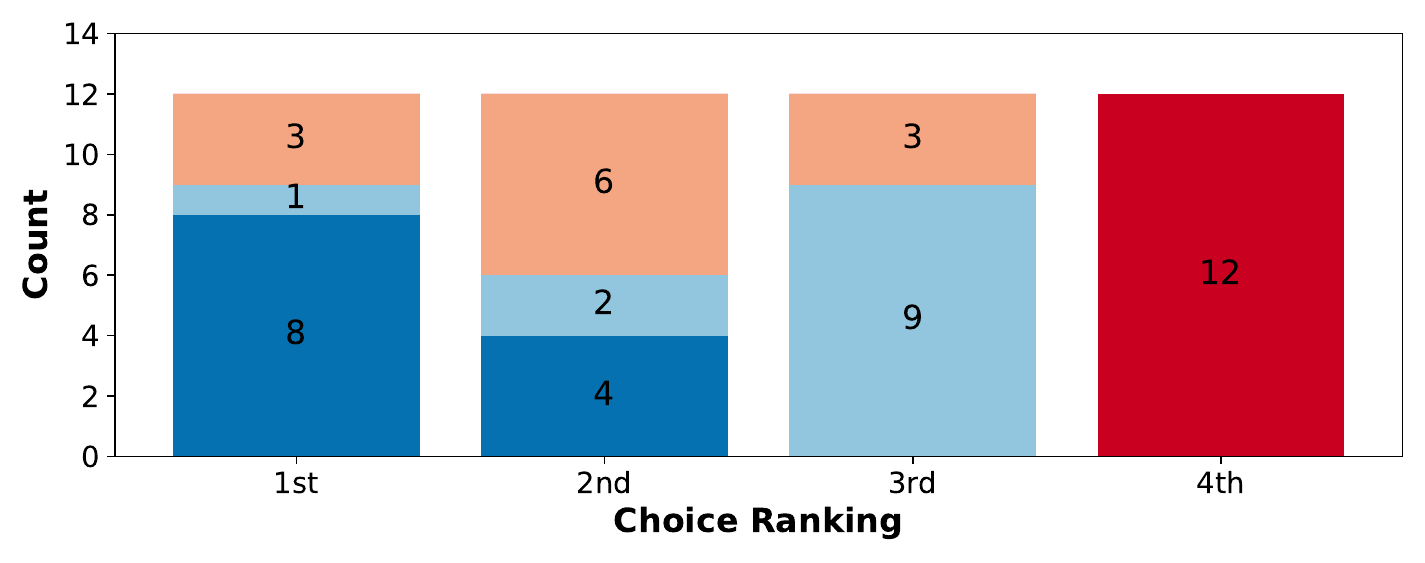}
        \end{minipage}%
        \hfill
        \begin{minipage}[t]{0.2\linewidth}
            \centering
            \vspace{-20mm} %
            \includegraphics[width=\linewidth, trim=35mm 10mm 20mm 10mm, clip]{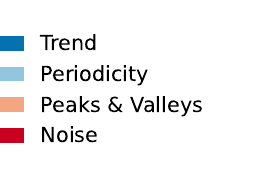}
        \end{minipage}
        \caption{The ranking distribution of features—trend, periodicity, peaks and valleys, and noise by participants. A rank of 1 indicates the highest priority, while 4 represents the lowest priority. The figure shows that the trend was ranked first most of the time, while noise was ranked last consistently.}
        \label{fig:feature_rank}
    \end{figure}

Participants ranked four features—trend, periodicity, peaks and valleys, and noise—by priority when viewing line charts. Most participants ranked trend as the highest priority, followed by peaks and valleys, periodicity, and noise (\Cref{fig:feature_rank}). This aligns with findings from the primary experiment, where sketches consistently preserved trends across clusters. Noise was consistently ranked as the lowest priority feature.

While trends were prioritized without context, participants emphasized that priorities shift based on the data's purpose. 
In specific contexts, noise may become a top priority. Participant C shared a personal example involving microscopy data, where analyzing noise was essential to evaluate the performance and sensitivity of the microscope.
In other situations, noisy line charts can obscure critical features like peaks and valleys, underscoring the importance of filtering noise or adding annotations to enhance visibility.
Participant F highlighted the difficulty in distinguishing between peaks and valleys and noise without data context. Participant C commented: \textit{``I think these may depend a little bit on the task and the data you are dealing with. For instance, if you have a line chart showing a patient's heart rate over time, extreme values are actually very, very important."}

Some participants suggested additional relevant features such as contextual markers, axes, change points, average lines, and uncertainty boundaries. Although opinions varied on whether noise added value or was distracting, those who mentioned additional features preferred context-driven enhancements for transparent and accurate visualizations.

\section{Discussion}
Given the findings of our primary study (\autoref{results}) and follow-up validation study (\autoref{sec:follow_up_evaluation:res}), we identified three behavior groups while re-creating line charts. Further analysis showed that some people treated the same stimuli differently regarding feature inclusion or exclusion. Also, a single individual would follow different patterns for line charts of different datasets or noise levels. This shows that what people see and re-create in line charts is subjective and depends on specific data properties, noise levels, and the person. Therefore, predicting what people will notice in a line chart is tough. However, generally speaking, people showed robustness to different noise levels by capturing trends and periodicity in their sketches. However, they were not quite as robust to noise in re-creating the peaks and valleys. 

Another significant finding of our research highlights the tendency of participants to interpret and depict periodicity and noise in line charts through a semantic lens rather than with exact replication. However, the representation of trends, peaks, and valleys is more faithful to the data. These findings suggest the following implications for design:

\paragraph{Smoothing May Not Always Be Necessary to Show Trend and Periodicity} Both studies indicate that individuals can effectively recreate trends and periodic patterns within datasets. Therefore, smoothing or de-noising may be unnecessary if the noise level is still reasonable and the primary objective is highlighting trends or periodicity. However, what counts as \emph{reasonable} noise is often data and context dependent, as reflected in our analysis (\Cref{robustnes_to_noise}). Since noise can also adversely affect judgments~\cite{harvey1997effects} or trust~\cite{elhamdadi2022we} in data, determining reasonable thresholds and their effects remains an open question that requires further investigation. %

\paragraph{Highlight Peaks and Valleys in Noisy Line Charts} Findings from the primary study reveal that noise significantly impacts the re-creation of peaks and valleys compared to periodicity and trends in line chart sketches. The validation study also reveals that people struggle to distinguish between true extrema in the data and fluctuations caused by noise. Some ways to ensure viewers notice peaks and valleys in noisy line charts include smoothing or using annotations to highlight significant peaks and/or valleys~\cite{fan2022annotating, kong2009perceptual}. Depending on the dataset, task, and features to be preserved, selecting an appropriate level of smoothing can make peaks and valleys more prominent. Such levels can be explored using tools like LineSmooth~\cite{rosen2020linesmooth}, which allow designers to adjust smoothing levels and compare multiple algorithms, including TopoLines~\cite{suh2019topolines}, designed to reduce noise while preserving extrema.

\paragraph{Annotate Important Features} Both studies highlight the difficulty in predicting which aspects of a line chart will capture attention due to variability in behavior. Participants in the validation study noted the importance of annotations~\cite{Rahman2025Annotate}, making it a reliable approach to emphasize key features and ensure effective communication.
    
\paragraph{Make Visual Query Systems More Adaptable}
 Participants often abstracted periodic patterns, sketching varying periods and amplitudes that did not match the stimuli. Instead, they aimed to semantically convey the cycles in the data. Similarly, participants represented noise semantically rather than precisely replicating it.
 This observation sheds light on the potential challenges and considerations in visual querying, particularly in sketch-based query interfaces where diverse algorithms may be necessary to interpret ambiguous or vague human input.
 These findings resonate with Correll et al.'s discussion on the potential irrelevance of specific data attributes, such as amplitude and noise, in sketch-based queries, calling such attributes invariants~\cite{correll2016semantics}. Our study suggests that periodicity could also be one such invariant.

 \paragraph{Representation of Line Charts May Require Less Detail Than You Think}
Individuals often express periodicity and noisiness in data more semantically, suggesting that detailed information on these aspects might not significantly enhance line chart representation. Unless the quantity of periods or the amplitude holds particular importance, incorporating such details into visual representation like sparklines~\cite{tufte2006beautiful} could be considered unnecessary. Conversely, trends, peaks, and valleys are typically more precisely or faithfully represented. Therefore, emphasizing these elements in representation could provide more value, offering a clearer understanding of the data's key features.

\section{Limitations}
Our studies provide insights into user strategies for preserving visual features in line charts. Nevertheless, these findings must be considered within the scope of the study design choices we made. Notably, our primary study did not explicitly consider participants' sketching abilities, domain knowledge, dataset familiarity, or task framing, factors that could influence the interpretation and representation of visual features~\cite{lee2019you, xiong2019curse}. To partially mitigate these concerns, we conducted a second validation study designed to assess user priorities and intentions independently of their sketching abilities. However, since this validation relied on sketches produced by participants from the first study, the influence of individual differences remains partially embedded within the evaluated materials.

We utilized participant-generated sketches, drawn in a single continuous stroke, to systematically capture user strategies for preserving visual features in line charts. This design simplified qualitative analysis by reducing variability and allowed direct comparisons across different conditions. However, the single-stroke limitation potentially restricted participants' ability to depict complex visual features accurately. The alternative approach of enabling multi-stroke sketches could offer greater representational flexibility, but at the expense of introducing analytical complexity, losing the clarity of comparison achieved by our chosen constraint.

Another limitation of our study is occasional mismatch between the qualitative codes and the quantitative metrics for sketches. For example, a sketch that we coded qualitatively as “Very well” at preserving a feature may show a relatively large $L^2$-norm value, while a sketch coded as “Not at all” may score low as shown in \Cref{fig:trend_l_2_plot}. Such deviations can stem from multiple factors: the pre-processing steps (i.e., artifact removal, normalization, interpolation) that reshape sketches, the methods used to estimate features, or the choice of similarity metric itself. These factors can produce artificially high or low quantitative fidelity that does not fully reflect participants’ original sketching choices. 

In our second study, we intentionally limited the sketches presented per noise level (four per condition) to reduce participant fatigue and enhance focus, thus ensuring high-quality responses. This facilitated clear insights into user feature preferences and priorities. However, the small set of sketches might have restricted the representational breadth, thereby affecting the generalizability and robustness of our validation findings. Although a larger sketch pool would provide broader insights, it would also necessitate careful design to avoid cognitive overload and reduced participant attention.

\section{Conclusion}
In conclusion, we explored how individuals recreate features of line charts, focusing on trends, periodicity, peaks, and valleys. Our findings reveal three significant insights. First, participants exhibited diverse strategies in recreating visual features, balancing fidelity with abstraction based on perceived importance. Second, individuals demonstrated resilience in retaining underlying data patterns despite varying noise levels. Finally, our research indicates that people may represent periodicity and noise semantically or conceptually rather than precisely. In contrast, their depictions of trends, peaks, and valleys are more faithful to the actual data. Based on these findings, we propose guidelines for designing line charts that better align with human perceptual priorities, providing valuable direction for the visualization community to create more effective line chart designs. All the supplemental materials are available at: \url{https://osf.io/4p3ze/?view_only=1dbd44ff26c24728b21a65f2907ccabb}

\setstretch{0.96}

\bibliographystyle{IEEEtran}

\bibliography{main}

\begin{thebibliography}{10}
\providecommand{\url}[1]{#1}
\csname url@samestyle\endcsname
\providecommand{\newblock}{\relax}
\providecommand{\bibinfo}[2]{#2}
\providecommand{\BIBentrySTDinterwordspacing}{\spaceskip=0pt\relax}
\providecommand{\BIBentryALTinterwordstretchfactor}{4}
\providecommand{\BIBentryALTinterwordspacing}{\spaceskip=\fontdimen2\font plus
\BIBentryALTinterwordstretchfactor\fontdimen3\font minus
  \fontdimen4\font\relax}
\providecommand{\BIBforeignlanguage}[2]{{%
\expandafter\ifx\csname l@#1\endcsname\relax
\typeout{** WARNING: IEEEtran.bst: No hyphenation pattern has been}%
\typeout{** loaded for the language `#1'. Using the pattern for}%
\typeout{** the default language instead.}%
\else
\language=\csname l@#1\endcsname
\fi
#2}}
\providecommand{\BIBdecl}{\relax}
\BIBdecl

\bibitem{mussweiler2003goes}
T.~Mussweiler and K.~Schneller, ``What goes up must come down-how charts
  influence decisions to buy and sell stocks,'' \emph{Journal of Behavioral
  Finance}, vol.~4, no.~3, pp. 121--130, 2003.

\bibitem{assfalg2009periodic}
J.~Assfalg, T.~Bernecker, H.-P. Kriegel, P.~Kr{\"o}ger, and M.~Renz, ``Periodic
  pattern analysis in time series databases,'' in \emph{Database Systems for
  Advanced Applications (DASFAA)}, 2009, pp. 354--368.

\bibitem{stein2023effect}
O.~Stein, A.~Jacobson, and F.~Chevalier, ``The effect of smoothing on the
  interpretation of time series data: A covid-19 case study,'' \emph{arXiv
  preprint}, 2023.

\bibitem{correll2012comparing}
M.~Correll, D.~Albers, S.~Franconeri, and M.~Gleicher, ``Comparing averages in
  time series data,'' in \emph{ACM SIGCHI Conference on Human Factors in
  Computing Systems}, 2012, pp. 1095--1104.

\bibitem{correll2017regression}
M.~Correll and J.~Heer, ``Regression by eye: Estimating trends in bivariate
  visualizations,'' in \emph{ACM SIGCHI Conference on Human Factors in
  Computing Systems}, 2017, pp. 1387--1396.

\bibitem{batista2014cid}
G.~E. Batista, E.~J. Keogh, O.~M. Tataw, and V.~M. De~Souza, ``Cid: an
  efficient complexity-invariant distance for time series,'' \emph{Data Mining
  and Knowledge Discovery}, vol.~28, pp. 634--669, 2014.

\bibitem{kim2021towards}
D.~H. Kim, V.~Setlur, and M.~Agrawala, ``Towards understanding how readers
  integrate charts and captions: A case study with line charts,'' in \emph{ACM
  SIGCHI Conference on Human Factors in Computing Systems}, 2021, pp. 1--11.

\bibitem{xiong2019curse}
C.~Xiong, L.~Van~Weelden, and S.~Franconeri, ``The curse of knowledge in visual
  data communication,'' \emph{IEEE Trans.\ on Visualization and Computer
  Graphics}, vol.~26, no.~10, pp. 3051--3062, 2019.

\bibitem{vredeveldt2011eyeclosure}
A.~Vredeveldt, G.~J. Hitch, and A.~D. Baddeley, ``Eyeclosure helps memory by
  reducing cognitive load and enhancing visualisation,'' \emph{Memory \&
  cognition}, vol.~39, pp. 1253--1263, 2011.

\bibitem{hu2023application}
J.~Hu, F.~Jia, and W.~Liu, ``Application of fast fourier transform,''
  \emph{Highlights in Science, Engineering and Technology}, vol.~38, pp.
  590--597, 2023.

\bibitem{lee2013sketchstory}
B.~Lee, R.~H. Kazi, and G.~Smith, ``Sketchstory: Telling more engaging stories
  with data through freeform sketching,'' \emph{IEEE Trans.\ on Visualization
  and Computer Graphics}, vol.~19, no.~12, pp. 2416--2425, 2013.

\bibitem{xia2018dataink}
H.~Xia, N.~Henry~Riche, F.~Chevalier, B.~De~Araujo, and D.~Wigdor, ``Dataink:
  Direct and creative data-oriented drawing,'' in \emph{ACM SIGCHI Conference
  on Human Factors in Computing Systems}, 2018, pp. 1--13.

\bibitem{muthumanickam2016shape}
P.~K. Muthumanickam, K.~Vrotsou, M.~Cooper, and J.~Johansson, ``Shape grammar
  extraction for efficient query-by-sketch pattern matching in long time
  series,'' in \emph{2016 IEEE Conference on Visual Analytics Science and
  Technology (VAST)}.\hskip 1em plus 0.5em minus 0.4em\relax IEEE, 2016, pp.
  121--130.

\bibitem{cohen1997can}
D.~J. Cohen and S.~Bennett, ``Why can't most people draw what they see?''
  \emph{Journal of Experimental Psychology: Human Perception and Performance},
  vol.~23, no.~3, p. 609, 1997.

\bibitem{lee2019you}
D.~J.-L. Lee, J.~Lee, T.~Siddiqui, J.~Kim, K.~Karahalios, and A.~Parameswaran,
  ``You can't always sketch what you want: Understanding sensemaking in visual
  query systems,'' \emph{IEEE Trans.\ on Visualization and Computer Graphics},
  vol.~26, no.~1, pp. 1267--1277, 2019.

\bibitem{rosen2020linesmooth}
P.~Rosen and G.~J. Quadri, ``Linesmooth: An analytical framework for evaluating
  the effectiveness of smoothing techniques on line charts,'' \emph{IEEE
  Trans.\ on Visualization and Computer Graphics}, vol.~27, no.~2, pp.
  1536--1546, 2020.

\bibitem{steinarsson2013downsampling}
S.~Steinarsson, ``Downsampling time series for visual representation,'' Ph.D.
  dissertation, University of Iceland, 2013.

\bibitem{correll2016semantics}
M.~Correll and M.~Gleicher, ``The semantics of sketch: Flexibility in visual
  query systems for time series data,'' in \emph{IEEE Visual Analytics Science
  and Technology (VAST)}, 2016, pp. 131--140.

\bibitem{tufte2006beautiful}
E.~R. Tufte, \emph{Beautiful evidence}.\hskip 1em plus 0.5em minus 0.4em\relax
  Graphics Press Cheshire, CT, 2006.

\bibitem{cleveland1984graphical}
W.~S. Cleveland and R.~McGill, ``Graphical perception: Theory, experimentation,
  and application to the development of graphical methods,'' \emph{Journal of
  the American Statistical Association}, vol.~79, no. 387, pp. 531--554, 1984.

\bibitem{franconeri2021science}
S.~L. Franconeri, L.~M. Padilla, P.~Shah, J.~M. Zacks, and J.~Hullman, ``The
  science of visual data communication: What works,'' \emph{Psychological
  Science in the Public Interest}, vol.~22, no.~3, pp. 110--161, 2021.

\bibitem{szafir2016four}
D.~A. Szafir, S.~Haroz, M.~Gleicher, and S.~Franconeri, ``Four types of
  ensemble coding in data visualizations,'' \emph{Journal of Vision}, vol.~16,
  no.~5, pp. 11--11, 2016.

\bibitem{gleicher2013perception}
M.~Gleicher, M.~Correll, C.~Nothelfer, and S.~Franconeri, ``Perception of
  average value in multiclass scatterplots,'' \emph{IEEE Trans.\ on
  Visualization and Computer Graphics}, vol.~19, no.~12, pp. 2316--2325, 2013.

\bibitem{harvey1997effects}
N.~Harvey Teresa Ewart Robert~West, ``Effects of data noise on statistical
  judgement,'' \emph{Thinking \& Reasoning}, vol.~3, no.~2, pp. 111--132, 1997.

\bibitem{ciccione2021can}
L.~Ciccione and S.~Dehaene, ``Can humans perform mental regression on a graph?
  accuracy and bias in the perception of scatterplots,'' \emph{Cognitive
  Psychology}, vol. 128, p. 101406, 2021.

\bibitem{hong2021weighted}
M.-H. Hong, J.~K. Witt, and D.~A. Szafir, ``The weighted average illusion:
  Biases in perceived mean position in scatterplots,'' \emph{IEEE Trans.\ on
  Visualization and Computer Graphics}, vol.~28, no.~1, pp. 987--997, 2021.

\bibitem{ciccione2022analyzing}
L.~Ciccione, M.~Sabl{\'e}-Meyer, and S.~Dehaene, ``Analyzing the misperception
  of exponential growth in graphs,'' \emph{Cognition}, vol. 225, p. 105112,
  2022.

\bibitem{reimann2021visual}
D.~Reimann, C.~Blech, N.~Ram, and R.~Gaschler, ``Visual model fit estimation in
  scatterplots: Influence of amount and decentering of noise,'' \emph{IEEE
  Trans.\ on Visualization and Computer Graphics}, vol.~27, no.~9, pp.
  3834--3838, 2021.

\bibitem{kim2018assessing}
Y.~Kim and J.~Heer, ``Assessing effects of task and data distribution on the
  effectiveness of visual encodings,'' \emph{Computer Graphics Forum}, vol.~37,
  no.~3, pp. 157--167, 2018.

\bibitem{ciccione2023outlier}
L.~Ciccione, G.~Dehaene, and S.~Dehaene, ``Outlier detection and rejection in
  scatterplots: Do outliers influence intuitive statistical judgments?''
  \emph{Journal of Experimental Psychology: Human Perception and Performance},
  vol.~49, no.~1, p. 129, 2023.

\bibitem{gogolou2018comparing}
A.~Gogolou, T.~Tsandilas, T.~Palpanas, and A.~Bezerianos, ``Comparing
  similarity perception in time series visualizations,'' \emph{IEEE Trans.\ on
  Visualization and Computer Graphics}, vol.~25, no.~1, pp. 523--533, 2019.

\bibitem{eitz2012humans}
M.~Eitz, J.~Hays, and M.~Alexa, ``How do humans sketch objects?'' \emph{ACM
  Trans.\ on Graphics (TOG)}, vol.~31, no.~4, pp. 1--10, 2012.

\bibitem{mannino2018expressive}
M.~Mannino and A.~Abouzied, ``Expressive time series querying with hand-drawn
  scale-free sketches,'' in \emph{ACM SIGCHI Conference on Human Factors in
  Computing Systems}, 2018, pp. 1--13.

\bibitem{lekschas2020peax}
F.~Lekschas, B.~Peterson, D.~Haehn, E.~Ma, N.~Gehlenborg, and H.~Pfister,
  ``Peax: Interactive visual pattern search in sequential data using
  unsupervised deep representation learning,'' in \emph{Computer Graphics
  Forum}, vol.~39, no.~3.\hskip 1em plus 0.5em minus 0.4em\relax Wiley Online
  Library, 2020, pp. 167--179.

\bibitem{wilkinson2005graph}
L.~Wilkinson, A.~Anand, and R.~Grossman, ``Graph-theoretic scagnostics,'' in
  \emph{IEEE Symposium on Information Visualization}, 2005, pp. 21--21.

\bibitem{friedman1974projection}
J.~H. Friedman and J.~W. Tukey, ``A projection pursuit algorithm for
  exploratory data analysis,'' \emph{IEEE Trans.\ on computers}, vol. 100,
  no.~9, pp. 881--890, 1974.

\bibitem{wang2017line}
Y.~Wang, F.~Han, L.~Zhu, O.~Deussen, and B.~Chen, ``Line graph or scatter plot?
  automatic selection of methods for visualizing trends in time series,''
  \emph{IEEE Trans.\ on Visualization and Computer Graphics}, vol.~24, no.~2,
  pp. 1141--1154, 2017.

\bibitem{lawless2012testing}
J.~Lawless, C.~{\c{C}}i{\u{g}}{\c{s}}ar, and R.~Cook, ``Testing for monotone
  trend in recurrent event processes,'' \emph{Technometrics}, vol.~54, no.~2,
  pp. 147--158, 2012.

\bibitem{alexander2006global}
L.~V. Alexander, X.~Zhang, T.~C. Peterson, J.~Caesar, B.~Gleason,
  A.~Klein~Tank, M.~Haylock, D.~Collins, B.~Trewin, F.~Rahimzadeh
  \emph{et~al.}, ``Global observed changes in daily climate extremes of
  temperature and precipitation,'' \emph{Journal of Geophysical Research:
  Atmospheres}, vol. 111, no.~D5, 2006.

\bibitem{seregina2019new}
L.~S. Seregina, A.~H. Fink, R.~van~der Linden, N.~A. Elagib, and J.~G. Pinto,
  ``A new and flexible rainy season definition: Validation for the greater horn
  of africa and application to rainfall trends,'' \emph{International Journal
  of Climatology}, vol.~39, no.~2, pp. 989--1012, 2019.

\bibitem{wang2009temporal}
T.~D. Wang, C.~Plaisant, B.~Shneiderman, N.~Spring, D.~Roseman, G.~Marchand,
  V.~Mukherjee, and M.~Smith, ``Temporal summaries: Supporting temporal
  categorical searching, aggregation and comparison,'' \emph{IEEE Trans.\ on
  Visualization and Computer Graphics}, vol.~15, no.~6, pp. 1049--1056, 2009.

\bibitem{robertson2008effectiveness}
G.~Robertson, R.~Fernandez, D.~Fisher, B.~Lee, and J.~Stasko, ``Effectiveness
  of animation in trend visualization,'' \emph{IEEE Trans.\ on Visualization
  and Computer Graphics}, vol.~14, no.~6, pp. 1325--1332, 2008.

\bibitem{fan2022annotating}
A.~Fan, Y.~Ma, M.~Mancenido, and R.~Maciejewski, ``Annotating line charts for
  addressing deception,'' in \emph{ACM SIGCHI Conference on Human Factors in
  Computing Systems}, 2022, pp. 1--12.

\bibitem{ahmed2022semantics}
S.~Ahmed, M.~J. Islam, and H.~Rajan, ``Semantics and anomaly preserving
  sampling strategy for large-scale time series data,'' \emph{ACM/IMS Trans.\
  on Data Science (TDS)}, vol.~2, no.~4, pp. 1--25, 2022.

\bibitem{rahman2017ve}
S.~Rahman, M.~Aliakbarpour, H.~K. Kong, E.~Blais, K.~Karahalios,
  A.~Parameswaran, and R.~Rubinfield, ``I've seen" enough" incrementally
  improving visualizations to support rapid decision making,''
  \emph{Proceedings of the VLDB Endowment}, vol.~10, no.~11, pp. 1262--1273,
  2017.

\bibitem{filipowicz2023visual}
A.~Filipowicz, S.~Carter, N.~Bravo, R.~Iliev, S.~Hakimi, D.~A. Shamma,
  K.~Lyons, C.~Hogan, and C.~Wu, ``Visual elements and cognitive biases
  influence interpretations of trends in scatter plots,'' \emph{arXiv
  preprint}, 2023.

\bibitem{johnson1991trend}
M.~Johnson and K.~Ottenbacher, ``Trend line influence on visual analysis of
  single-subject data in rehabilitation research,'' \emph{International
  Disability Studies}, vol.~13, no.~2, pp. 55--59, 1991.

\bibitem{cleveland1979robust}
W.~S. Cleveland, ``Robust locally weighted regression and smoothing
  scatterplots,'' \emph{Journal of the American Statistical Association},
  vol.~74, no. 368, pp. 829--836, 1979.

\bibitem{popivanov2002similarity}
I.~Popivanov and R.~J. Miller, ``Similarity search over time-series data using
  wavelets,'' in \emph{International Conference on Data Engineering}, 2002, pp.
  212--221.

\bibitem{musbah2020novel}
H.~Musbah, H.~H. Aly, and T.~A. Little, ``A novel approach for seasonality and
  trend detection using fast fourier transform in box-jenkins algorithm,'' in
  \emph{IEEE Canadian Conference on Electrical and Computer Engineering
  (CCECE)}, 2020, pp. 1--5.

\bibitem{duda2018detection}
J.~T. Duda and T.~Pe{\l}ech-Pilichowski, ``Detection of periodic components
  from seasonal time series with moving trend method and low pass filtering,''
  in \emph{Advanced Solutions in Diagnostics and Fault Tolerant Control}, 2018,
  pp. 192--202.

\bibitem{denholm1998practical}
J.~Denholm-Price and J.~Rees, ``A practical example of low-frequency trend
  removal,'' \emph{Boundary-layer Meteorology}, vol.~86, pp. 181--187, 1998.

\bibitem{elfeky2005periodicity}
M.~G. Elfeky, W.~G. Aref, and A.~K. Elmagarmid, ``Periodicity detection in time
  series databases,'' \emph{IEEE Trans.\ on Knowledge and Data Engineering},
  vol.~17, no.~7, pp. 875--887, 2005.

\bibitem{lazcano2022improved}
S.~Lazcano and C.~D. Johnson, ``Improved data interpretation through
  identification of time series periodicity changes,'' Pacific Northwest
  National Laboratory, Richland, WA, USA, Tech. Rep., 2022.

\bibitem{shehu2023efficient}
Y.~Shehu and R.~Harper, ``Efficient periodicity analysis for real-time anomaly
  detection,'' in \emph{IEEE/IFIP Network Operations and Management Symposium},
  2023, pp. 1--6.

\bibitem{wen2021robustperiod}
Q.~Wen, K.~He, L.~Sun, Y.~Zhang, M.~Ke, and H.~Xu, ``Robustperiod: Robust
  time-frequency mining for multiple periodicity detection,'' in
  \emph{International Conference on Management of Data}, 2021, pp. 2328--2337.

\bibitem{musbah2019identifying}
H.~Musbah, M.~El-Hawary, and H.~Aly, ``Identifying seasonality in time series
  by applying fast fourier transform,'' in \emph{IEEE Electrical Power and
  Energy Conference (EPEC)}, 2019, pp. 1--4.

\bibitem{wen2023robust}
Q.~Wen, L.~Yang, and L.~Sun, ``Robust dominant periodicity detection for time
  series with missing data,'' in \emph{IEEE International Conference on
  Acoustics, Speech and Signal Processing (ICASSP)}, 2023, pp. 1--5.

\bibitem{mohammed2019developing}
M.~Y.~Y. Mohammed and M.~Celik, ``Developing fast techniques for periodicity
  analysis of time series,'' in \emph{International Symposium on
  Multidisciplinary Studies and Innovative Technologies (ISMSIT)}, 2019, pp.
  1--5.

\bibitem{stratimirovic2018analysis}
D.~Stratimirovi{\'c}, D.~Sarvan, V.~Miljkovi{\'c}, and S.~Blesi{\'c},
  ``Analysis of cyclical behavior in time series of stock market returns,''
  \emph{Communications in Nonlinear Science and Numerical Simulation}, vol.~54,
  pp. 21--33, 2018.

\bibitem{otazu2002detection}
X.~Otazu, M.~Rib{\'o}, M.~Peracaula, J.~Paredes, and J.~Nunez, ``Detection of
  superimposed periodic signals using wavelets,'' \emph{Monthly Notices of the
  Royal Astronomical Society}, vol. 333, no.~2, pp. 365--372, 2002.

\bibitem{elfeky2005warp}
M.~G. Elfeky, W.~G. Aref, and A.~K. Elmagarmid, ``Warp: time warping for
  periodicity detection,'' in \emph{IEEE International Conference on Data
  Mining (ICDM)}, 2005, pp. 8--pp.

\bibitem{boulnemour2018qp}
I.~Boulnemour and B.~Boucheham, ``Qp-dtw: upgrading dynamic time warping to
  handle quasi periodic time series alignment,'' \emph{Journal of Information
  Processing Systems}, vol.~14, no.~4, pp. 851--876, 2018.

\bibitem{song2022robust}
X.~Song, Q.~Wen, Y.~Li, and L.~Sun, ``Robust time series dissimilarity measure
  for outlier detection and periodicity detection,'' in \emph{ACM International
  Conference on Information \& Knowledge Management}, 2022, pp. 4510--4514.

\bibitem{suh2019topolines}
P.~Rosen, A.~Suh, C.~Salgado, and M.~Hajij, ``{TopoLines: Topological Smoothing
  for Line Charts},'' in \emph{EuroVis (Short Papers)}, 2020.

\bibitem{rensink2010perception}
R.~A. Rensink and G.~Baldridge, ``The perception of correlation in
  scatterplots,'' \emph{Computer Graphics Forum}, vol.~29, no.~3, pp.
  1203--1210, 2010.

\bibitem{palshikar2009simple}
G.~Palshikar \emph{et~al.}, ``Simple algorithms for peak detection in
  time-series,'' in \emph{International Conference on Advanced Data Analysis,
  Business Analytics and Intelligence}, vol. 122, 2009.

\bibitem{zhou2022improved}
Y.~Zhou, J.~Ma, F.~Li, B.~Chen, T.~Xian, and X.~Wei, ``An improved algorithm
  for peak detection based on weighted continuous wavelet transform,''
  \emph{IEEE Access}, vol.~10, pp. 118\,779--118\,788, 2022.

\bibitem{du2006improved}
P.~Du, W.~A. Kibbe, and S.~M. Lin, ``Improved peak detection in mass spectrum
  by incorporating continuous wavelet transform-based pattern matching,''
  \emph{Bioinformatics}, vol.~22, no.~17, pp. 2059--2065, 2006.

\bibitem{price1993signals}
J.~Price and T.~Goble, ``Signals and noise,'' in \emph{Telecommunications
  Engineer's Reference Book}.\hskip 1em plus 0.5em minus 0.4em\relax Elsevier,
  1993, pp. 10--1.

\bibitem{butterworth1930theory}
S.~Butterworth \emph{et~al.}, ``On the theory of filter amplifiers,''
  \emph{Wireless Engineer}, vol.~7, no.~6, 1930.

\bibitem{rhodes1980generalized}
J.~D. Rhodes and S.~Alseyab, ``The generalized chebyshev low-pass prototype
  filter,'' \emph{International Journal of Circuit Theory and Applications},
  vol.~8, no.~2, pp. 113--125, 1980.

\bibitem{raj2007fft}
B.~Raj, L.~Turicchia, B.~Schmidt-Nielsen, and R.~Sarpeshkar, ``An fft-based
  companding front end for noise-robust automatic speech recognition,''
  \emph{EURASIP Journal on Audio, Speech, and Music Processing}, vol. 2007, pp.
  1--13, 2007.

\bibitem{karar1997apd}
A.~Karar, R.~Tanaka, and J.~C. Vanel, ``Apd's excess noise measurements using
  spectral analysis (fft),'' \emph{Nuclear Instruments and Methods in Physics
  Research Section A: Accelerators, Spectrometers, Detectors and Associated
  Equipment}, vol. 387, no. 1-2, pp. 205--210, 1997.

\bibitem{parekh2018investigating}
V.~Parekh, M.~Bilalpur, C.~Jawahar, S.~Kumar, S.~Winkler, and R.~Subramanian,
  ``Investigating the generalizability of eeg-based cognitive load estimation
  across visualizations,'' in \emph{International Conference on Multimodal
  Interaction: Adjunct}, 2018, pp. 1--5.

\bibitem{2018entropy}
G.~Ryan, A.~Mosca, R.~Chang, and E.~Wu, ``At a glance: Pixel approximate
  entropy as a measure of line chart complexity,'' \emph{IEEE Trans.\ on
  Visualization and Computer Graphics}, vol.~25, no.~1, pp. 872--881, 2019.

\bibitem{nagadia_2022}
\BIBentryALTinterwordspacing
M.~Nagadia, ``Apple stock price from 1980-2021,'' 2022. [Online]. Available:
  \url{https://www.kaggle.com/datasets/meetnagadia/apple-stock-price-from-19802021}
\BIBentrySTDinterwordspacing

\bibitem{alma}
``{Alma Science Archive},'' \url{http://almascience.nrao.edu/aq/}, 2020.

\bibitem{chicago_crime}
{Chicago Police Department}, ``{Crimes - 2001 to present: City of Chicago: Data
  Portal},''
  \url{https://data.cityofchicago.org/Public-Safety/Crimes-2001-to-present/ijzp-q8t2},
  2020.

\bibitem{noaa}
{National Centers for Environmental Information}, ``{Climate Data Online: Web
  Services Documentation},''
  \url{https://www.ncdc.noaa.gov/cdo-web/webservices/v2}, 2020.

\bibitem{dhruvil_dave_2021}
\BIBentryALTinterwordspacing
D.~Dave, ``Dogecoin historical data,'' 2021. [Online]. Available:
  \url{https://www.kaggle.com/ds/1179216}
\BIBentrySTDinterwordspacing

\bibitem{eeg_database}
A.~Delorme, ``{EEG / ERP Public Database},''
  \url{https://sccn.ucsd.edu/~arno/fam2data/publicly_available_EEG_data.html},
  2020.

\bibitem{d3_zoomable}
M.~Bostack, ``{Observable: Zoomable Area Chart},''
  \url{https://observablehq.com/@d3/zoomable-area-chart}, 2020.

\bibitem{new_zealand}
``{New Zealand Tourist Arrivals},''
  \url{https://tradingeconomics.com/new-zealand/tourist-arrivals}, 2020.

\bibitem{us_bls}
``{U.S. Bureau of Labor Statistics},'' \url{https://www.bls.gov/}, 2020.

\bibitem{kim2021adaptive}
E.~Kim, J.~Kim, H.~Lee, and S.~Kim, ``Adaptive data augmentation to achieve
  noise robustness and overcome data deficiency for deep learning,''
  \emph{Applied Sciences}, vol.~11, no.~12, p. 5586, 2021.

\bibitem{satyanarayan2016vega}
A.~Satyanarayan, D.~Moritz, K.~Wongsuphasawat, and J.~Heer, ``Vega-lite: A
  grammar of interactive graphics,'' \emph{IEEE Trans.\ on Visualization and
  Computer Graphics}, vol.~23, no.~1, pp. 341--350, 2016.

\bibitem{mchugh2012interrater}
M.~L. McHugh, ``Interrater reliability: the kappa statistic,'' \emph{Biochemia
  medica}, vol.~22, no.~3, pp. 276--282, 2012.

\bibitem{eichmann2015evaluating}
P.~Eichmann and E.~Zgraggen, ``Evaluating subjective accuracy in time series
  pattern-matching using human-annotated rankings,'' in \emph{International
  Conference on Intelligent User Interfaces (IUI)}, 2015, pp. 28--37.

\bibitem{ding2008querying}
H.~Ding, G.~Trajcevski, P.~Scheuermann, X.~Wang, and E.~Keogh, ``Querying and
  mining of time series data: experimental comparison of representations and
  distance measures,'' \emph{Proceedings of the VLDB Endowment}, vol.~1, no.~2,
  pp. 1542--1552, 2008.

\bibitem{EdelsbrunnerHarer2010}
H.~Edelsbrunner and J.~Harer, \emph{Computational Topology: An
  Introduction}.\hskip 1em plus 0.5em minus 0.4em\relax Providence, RI, USA:
  American Mathematical Society, 2010.

\bibitem{lex2014upset}
A.~Lex, N.~Gehlenborg, H.~Strobelt, R.~Vuillemot, and H.~Pfister, ``Upset:
  visualization of intersecting sets,'' \emph{IEEE Trans.\ on Visualization and
  Computer Graphics}, vol.~20, no.~12, pp. 1983--1992, 2014.

\bibitem{elhamdadi2022we}
H.~Elhamdadi, A.~Gaba, Y.-S. Kim, and C.~Xiong, ``How do we measure trust in
  visual data communication?'' in \emph{Evaluation and Beyond-Methodological
  Approaches for Visualization (BELIV)}, 2022, pp. 85--92.

\bibitem{kong2009perceptual}
N.~Kong and M.~Agrawala, ``Perceptual interpretation of ink annotations on line
  charts,'' in \emph{ACM Symposium on User Interface Software and Technology
  (UIST)}, 2009, pp. 233--236.

\bibitem{Rahman2025Annotate}
M.~D. Rahman, G.~J. Quadri, B.~Doppalapudi, D.~A. Szafir, and P.~Rosen, ``A
  qualitative analysis of common practices in annotations: A taxonomy and
  design space,'' \emph{IEEE Trans.\ on Visualization and Computer Graphics},
  vol.~31, no.~1, pp. 360--370, 2025.

\end{thebibliography}

\vspace{11pt}

\begin{IEEEbiography}[{\includegraphics[width=1in,height=1.25in,clip,keepaspectratio]{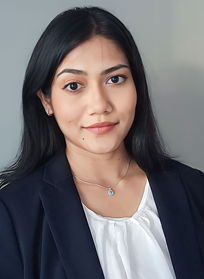}}]{Rifat Ara Proma}is a Ph.D. Candidate in the School of Computing at the University of Utah. Her dissertation focuses on developing strategies to mitigate visual clutter in line charts. She holds a Master's in Computer Science from the University of South Florida. Her research interests include information visualization, human-centric visualizations and human-computer interaction.
\end{IEEEbiography}

\begin{IEEEbiography}[{\includegraphics[width=1in,height=1.25in,clip,keepaspectratio]{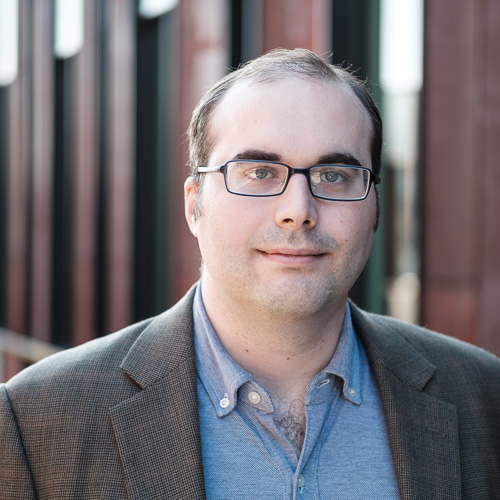}}]{Michael Correll} is a Research Associate Professor at Northeastern University's Roux Institute. He received the Ph.D. degree from the University of Wisconsin–Madison. His research interests include data ethics, communicating statistics to mass audiences, and investigating biased or misleading data visualizations.
\end{IEEEbiography}

\begin{IEEEbiography}[{\includegraphics[width=1in,height=1.25in,clip,keepaspectratio]{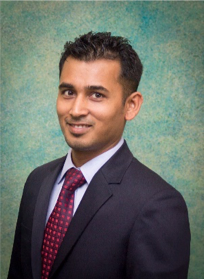}}]{Ghulam Jilani Quadri} is an Assistant Professor in the School of Computer Science at the University of Oklahoma. Prior to that, Dr. Quadri was a CIFellow postdoc at the University of North Carolina at Chapel Hill. He received his Ph.D. from the University of South Florida. Quadri received the 2021 Computing Innovation Fellow award. His research interests include creating human-centered frameworks to optimize visualization design and improve decision-making quality.
\end{IEEEbiography}

\begin{IEEEbiography}[{\includegraphics[width=1in,height=1.25in,clip,keepaspectratio]{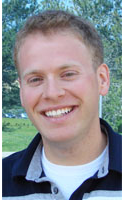}}]{Paul Rosen} is an Associate Professor at the University of Utah. He received his Ph.D.\ from Purdue University. His research interests include applying geometry- and topology-based approaches to problems in information visualization. Along with his collaborators, he has received best paper awards or honorable mentions at IEEE VIS, IEEE PacificVis, CG\&A, IVAPP, and SIBGRAPI.  Dr.\ Rosen received a National Science Foundation CAREER Award in 2019.
\end{IEEEbiography}

\vfill

\end{document}